\documentclass[preprint,authoryear,11pt]{elsarticle}

\usepackage{graphics}
\usepackage{graphicx}
\graphicspath{{Figures/}}
\usepackage{epsfig}
\usepackage{subfigure}
\usepackage{amssymb}
\usepackage{amsthm}
\usepackage{bm}
\usepackage{amsmath}
\usepackage{color}
\usepackage{geometry}
\usepackage{setspace}

\allowdisplaybreaks
\usepackage{enumerate}
\usepackage{booktabs}

\usepackage{hyperref}
\hypersetup{
	colorlinks=true,       
}

\biboptions{}

\journal{}

\begin{document}

\begin{frontmatter}



\title{Electrostatically tunable small-amplitude free vibrations \\ of pressurized electro-active spherical balloons}


\author[1]{Renwei Mao}
\author[2]{Bin Wu\corref{cor1}}
\ead{bin\_wu@polito.it}
\author[2]{Erasmo Carrera}
\author[1,3]{Weiqiu Chen}

\cortext[cor1]{Corresponding author. Tel.: +39-011-0906887; fax: +39-011-0906899.}
\address[1]{Key Laboratory of Soft Machines and Smart Devices of Zhejiang Province \\ and Department of Engineering Mechanics,\\ Zhejiang University, Hangzhou 310027, P.R. China;\\[6pt]}
\address[2]{Mul$^\text{2}$ Group, Department of Mechanical and Aerospace Engineering,\\Politecnico di Torino, Torino 10129, Italy; \\[6pt]}
\address[3]{Soft Matter Research Center, \\ Zhejiang University, Hangzhou 310027, P.R. China.}

\begin{abstract}
	
Designing tunable resonators is of practical importance in active/adaptive sound generation, noise control, vibration isolation and damping. In this paper, we propose to exploit the biasing fields (induced by internal pressure and radial electric voltage) to tune the three-dimensional and small-amplitude free vibration of a thick-walled soft electro-active (SEA) spherical balloon. The incompressible isotropic SEA balloon is characterized by both neo-Hookean and Gent ideal dielectric models. The equations governing small-amplitude vibrations under inhomogeneous biasing fields can be linearized and solved in spherical coordinates using the \textit{state-space formalism}, which establishes two separate transfer relations correlating the state vectors at the inner surface with those at the outer surface of the SEA balloon. By imposing the mechanical and electric boundary conditions, two separate analytical frequency equations are derived, which characterize two independent classes of vibration for torsional and spheroidal modes, respectively. Numerical examples are finally conducted to validate the theoretical derivation as well as to investigate the effects of both radial electric voltage and internal pressure on the resonant frequency of the SEA balloon. The reported analytical solution is truly and fully three-dimensional, covering from the purely radial breathing mode to torsional mode to any general spheroidal mode, and hence provides a more accurate prediction of the vibration characteristics of tunable resonant devices that incorporate the SEA spherical balloon as the tuning element.

\end{abstract}


\end{frontmatter}




\section{Introduction}

Soft electro-active (SEA) materials as one kind of promising smart materials have recently attracted considerable academic and industrial interest since they are equipped with many unique electromechanical properties such as low actuation voltage, rapid response and large deformation under electric stimuli, high fracture toughness and energy density. These superior advantages make SEA materials ideal candidates for broad application prospects as transducers, actuators, sensors, energy harvesters, biomedical devices as well as flexible electronics \citep{Pelrine836, carpi2011dielectric, anderson2012multi, zhao2014harnessing}. Moreover, the application of electric stimuli significantly and simultaneously modifies the effective electromechanical properties as well as the geometrical configurations, and thus results in changes of vibration and wave characteristics of SEA materials, prefiguring various potential applications including electrostatically tunable resonators and oscillators, waveguides, phononic crystals and metamaterials \citep{zhu2010resonant, zhao2016application, jin2017ratio, bortot2017tuning, galich2016manipulating, galich2017shear, wu2017guided, wu2018tuning}. A general theoretical framework of nonlinear electroelasticity has been well established \citep{mcmeeking2005electrostatic, dorfmann2006nonlinear, dorfmann2014nonlinear, suo2008nonlinear} to capture the nonlinear electromechanical behavior of SEA materials such as dielectric elastomers (DEs) that are high-speed electrically actuated elastomers with strain greater than 100\% in the pioneering work by \citet{Pelrine836}.

To explore how biasing fields (induced by for instances prestretch, internal pressure and electric stimuli) affect the small-amplitude dynamic characteristics of SEA materials, the linearized incremental theory based on the nonlinear electroelasticity theory is commonly adopted. With a particular attention focused on the SEA materials, a compact form of the linearized incremental theory in both Lagrangian and updated Lagrangian descriptions was developed by \citet{dorfmann2010electroelastic} to study the small-amplitude motions superimposed on finite biasing fields, which generalizes the pioneering work on the linearized incremental theory for piezoelectric materials by \citet{baumhauer1973nonlinear}. Recently, utilizing the concept of three configurations, \citet{wu2016} reviewed in detail different versions of nonlinear electroelasticity theories and relevant linearized incremental theories, and proved that all seemingly various theories in the literature are actually equivalent and have no substantive difference. Using the linearized incremental theory by \citet{dorfmann2010electroelastic}, much effort has been devoted to theoretically investigating the effects of biasing fields on the propagation of bulk waves (including longitudinal and transverse waves) in compressible DEs \citep{galich2016manipulating} and various types of guided waves in waveguides with different geometrical configurations, including the surface waves in a deformed SEA half-space \citep{dorfmann2010electroelastic}, the generalized Rayleigh-Lamb waves in an ideal DE layer \citep{shmuel2012rayleigh}, the axisymmetric and non-axisymmetric waves in a pre-stretched SEA cylinder and tube subjected to an axial electric displacement \citep{chen2012waves, su2016propagation}.

However, the aforementioned biasing fields are homogeneous such that exact dispersion relations can be obtained. In fact, radially inhomogeneous biasing fields may be generated by application of an radial electric potential difference (or electric voltage) to the compliant electrodes attached to the inner and outer surfaces of SEA spherical balloons or cylindrical tubes \citep{dorfmann2014nonlinear2, wu2017guided}. Thus, it is intractable to achieve exact analytical solutions for these loading cases. Based on the compound matrix method, \citet{shmuel2013axisymmetric} and \citet{shmuel2015manipulating} considered respectively the axisymmetric and torsional waves propagating in neo-Hookean ideal SEA tubes subjected to a radially applied electric voltage as well as an axial mechanical force. Note that for thick-walled tubes, \citet{shmuel2013axisymmetric} found that the compound matrix method failed to converge to a physically sound solution, which is probably attributed to its numerical difficulty when searching roots \citep{haughton1997eversion, amar2005growth}. More recently, the state-space method (SSM), which combines the state-space formalism with the approximate laminate technique, was put forward by \citet{wu2017guided} to effectively study the effect of inhomogeneous biasing fields on the circumferential waves in SEA tubes. For the aspect of finite-element implementation, the finite-element-based numerical simulation capability was presented by \citet{jandron2018numerical} for designing electrostatically tunable soft phononic crystals. Their finite-element computational procedures can deal with both the electro-mechanical coupling and the large-deformation capability as well as the inertial effects.

In addition to the wave characteristics, applying biasing fields can be also an efficient means to actively tune the free vibration behaviors (such as resonant frequencies) of SEA structures of finite size. Consequently, the SEA actuators have many potential applications in sound generation by designing tunable SEA loudspeakers, in noise control by using adaptive acoustic absorbers, and in the domain of active/adaptive vibration isolators and dampers by exploiting the stiffness tunability and viscoelasticity of SEA structures. The tunable SEA resonators as important elements for electromechanical devices play a key role in these applications. For instance, \citet{sugimoto2013lightweight} proposed a lightweight push-pull acoustic transducer composed of DE films for sound generation in advanced audio systems and demonstrated an advantage of push-pull driving in the suppression of harmonic distortion. Moreover, a hemispherical breathing mode loudspeaker activated by a DE actuator was fabricated, analyzed and measured by \citet{hosoya2015hemispherical} for its repeatability, sound pressure, vibration mode profiles, and acoustic radiation patterns. In addition, an electrostatically tunable duct silencer was developed by \citet{lu2015electronically} and their experimental results indicated that all the resonance peaks of this duct silencer could be adjusted using external control signals without any additional mechanical part. Earlier, voltage controllability of the resonant frequency for the SEA polymer membranes was studied experimentally and theoretically by \citet{dubois2008voltage} and a reduction in resonant frequency up to 77\% from the initial value was observed by adjusting the voltage. They also pointed out that the tuning way based on SEA structures requires no external actuators or variable elements compared with other resonance frequency tuning techniques. Therefore, the tunable SEA resonators are extremely appropriate for the next generation of acoustic treatment devices to replace the traditional acoustic treatment. Furthermore, \citet{bein2008smart} presented the design of active interfaces and semi-active vibration absorbers to meet the requirements of aerospace applications, and demonstrated the capability of DE actuators to set up control loops to suppress unwanted vibrations. 
A vibration damper based on DE actuators was developed by \citet{zhang2015tunable} in order to obtain a tunable and controllable vibration attenuation process. More recently, the development of SEA actuator technologies in acoustics and vibration control was reviewed in detail by \citet{zhao2016application}, to which the reader may be referred. 

\begin{figure}[htbp]
	\centering	
	\includegraphics[width=0.75\textwidth]{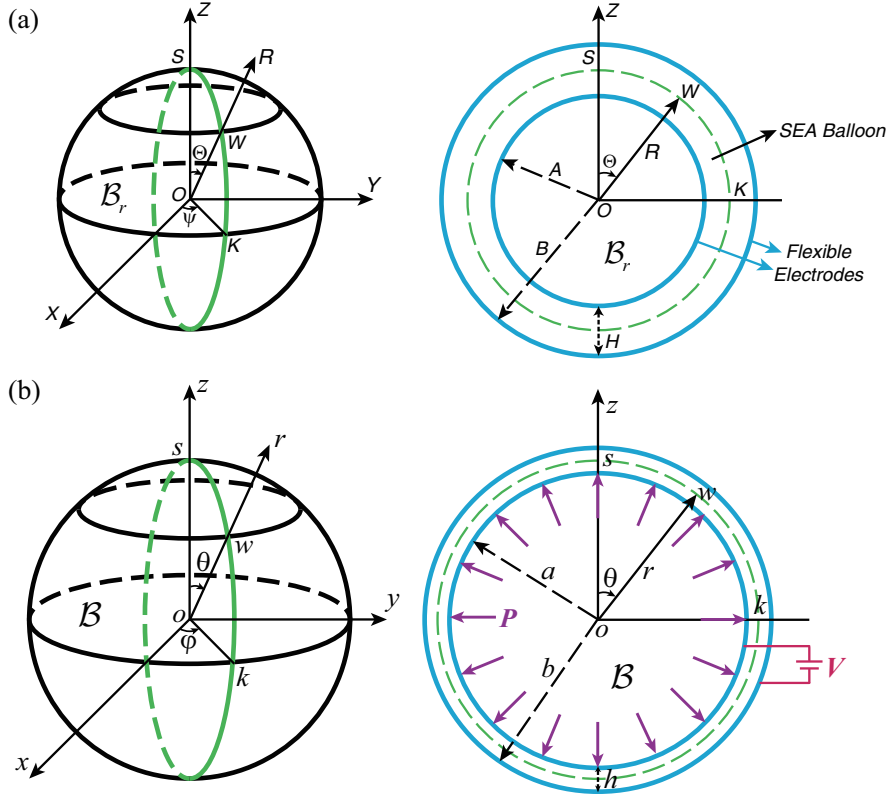}
	\caption{Schematic diagram of a SEA spherical balloon with flexible electrodes with related spherical coordinates and spherical surfaces: (a) undeformed configuration ${{\mathcal{B}}_{r}}$ before activation; (b) deformed configuration ${\mathcal{B}}$ after activation induced by a combined action of radial electric voltage $V$ and internal pressure $P$.}
	\label{Fig1}
\end{figure}

The primary objective of the present work is to exploit the SSM proposed by \citet{wu2017guided} for wave problems to investigate the 3D small-amplitude free vibrations of SEA balloons under radially inhomogeneous biasing fields, which are induced by the combined action of an internal pressure and an electric voltage applied to the electrodes on the inner and outer spherical surfaces (see Fig.~\ref{Fig1}(a)). Subjected to the biasing fields, the SEA balloon will reduce in thickness and inflate in area, as displayed in Fig.~\ref{Fig1}(b). As is well-known, the electromechanical instability occurs when the applied voltage reaches a threshold value, which is due to the positive feedback between an increasing electric field and a drastically decreasing thickness of balloons. Furthermore, the SEA spherical balloon with strain-stiffening effect (that can be described by the Gent model \citep{gent1996new}) may undergo snap-through instability and suddenly jump to another new equilibrium state with larger radius and smaller thickness. The snap-through instability of SEA balloons was used by \citet{rudykh2012snap} to design electrostatically controllable actuators and micropumps. Recently, the snap-through instability of phononic crystals made of SEA cylindrical structures was also utilized to realize sharp transitions in the bandgap width and position \citep{bortot2017tuning, wu2018tuning}. As a result, another goal of this paper is focused on how the resonant frequency of a SEA spherical balloon is influenced by the phenonenon of snap-through instability.

This paper is organized as follows: (i) First, some preliminary and introductory information concerning the nonlinear electroelasticity theory and the relevant linearized incremental theory \citep{dorfmann2010electroelastic} is briefly recalled in Sec.~\ref{section2}. (ii) Next, the spherically symmetric nonlinear inflation of an incompressible isotropic SEA balloon is considered in Subsec.~\ref{section3.1} for any form of energy function, which is specialized to the neo-Hookean and Gent ideal dielectric models in Subsec.~\ref{section3.2}. (iii) Subsequently, we derive the incremental governing equations in spherical coordinates and obtain two decoupled state equations by introducing appropriate displacement and stress functions in Sec.~\ref{section4}. (iv) Based on the state-space formalism for the incremental fields, the approximate laminate technique is employed in Sec.~\ref{section5} to efficiently derive two separate analytical frequency equations. (v) Then, numerical examples are conducted in Sec.~\ref{Sec6} to verify the effectiveness of the proposed SSM and elucidate the influence of biasing fields on the 3D free vibration characteristics. (vi) Finally, the main conclusions are drawn in Sec.~\ref{sec7}. Also, Appendices A-B provide some relevant mathematical expressions and derivations.


\section{Theoretical background}
\label{section2}

In this section, the general nonlinear electroelasticity theory and the relevant linearized incremental theory for small-amplitude motion superimposed on a finitely deformed configuration will be briefly outlined hereinafter for the sake of completeness. For more detailed discussions on the basic ideas and formulations, we refer to the papers of \citet{dorfmann2006nonlinear, dorfmann2010electroelastic} and the monograph of \citet{dorfmann2014nonlinear} as well as the references cited therein. 


\subsection{The equations of nonlinear electroelasticity}\label{section2.1}


Consider a deformable SEA continuum, which in the absence of mechanical loads and electric fields is unstressed at time ${{t}_{0}}$ and occupies in the Euclidean space the \emph{undeformed reference configuration} ${{\mathcal{B}}_{r}}$ with the boundary $\partial {{\mathcal{B}}_{r}}$ and the outward unit normal vector $\mathbf{N}$. An arbitrary material point in this state labelled as $X$ is identified by its position vector $\mathbf{X}$. At time $t$, the body is subjected to certain electric field and mechanical load, as a result of which it occupies the \emph{deformed current configuration} ${{\mathcal{B}}_{t}}$ with the boundary $\partial {{\mathcal{B}}_{t}}$ and the outward unit normal vector ${{\mathbf{n}}_{t}}$. Now the material point $X$ moves to a new position $\mathbf{x}=\mathbf{\chi }\text{(}\mathbf{X},t\text{)}$, where $\mathbf{\chi }$ is a vector function with a sufficiently regular property. The deformation gradient tensor is defined as $\mathbf{F}=\text{Grad }\mathbf{\chi }(\mathbf{X})=\text{Grad}\hskip 2pt\mathbf{x}$, where `$\text{Grad}$' is the gradient operator with respect to ${{\mathcal{B}}_{r}}$. Its Cartesian components are ${{\mathbf{F}}_{i\alpha }}=\partial {{x}_{i}}/\partial {{X}_{\alpha }}$, where in this paper Greek indices are associated with the reference configuration ${{\mathcal{B}}_{r}}$ and small Roman indices with the current configuration ${{\mathcal{B}}_{t}}$. The relations between the infinitesimal line element $\text{d}\mathbf{X}$, surface element $\text{d}A$ and volume element $\text{d}V$ in ${{\mathcal{B}}_{r}}$ and those in ${{\mathcal{B}}_{t}}$ are connected by $\text{d}\mathbf{x}=\mathbf{F}\text{d}\mathbf{X}$, the well-known Nanson's formula ${{\mathbf{n}}_{t}}\text{d}{{a}_{t}}=J{{\mathbf{F}}^{-\text{T}}}\mathbf{N}\text{d}A$, and $\text{d}v=J\text{d}V$, where the superscript $^{\text{T}}$ signifies the transpose of a second-order tensor if not otherwise stated and $J=\det \mathbf{F}$, the determinant of $\mathbf{F}$, denotes the local measure of the volume change. The incompressibility constraint results in $J=1$. The left and right Cauchy-Green strain tensors associated with $\mathbf{F}$ are defined as $\mathbf{b=F}{{\mathbf{F}}^{\text{T}}}$ and $\mathbf{C=}{{\mathbf{F}}^{\text{T}}}\mathbf{F}$ that will be used as the deformation measures. If not otherwise specified, the summation convention for repeated indices is adopted throughout this paper. 

In the absence of mechanical body forces, free charges and electric currents, and with the quasi-electrostatic approximation, the equations of motion, the Gauss's law and Faraday's law can be written as
\begin{equation} \label{governingEQ}
\mbox{div\hskip 1pt} \bm{\tau}=\rho {{\mathbf{x}}_{,tt}}, \quad \mbox{div\hskip 1pt}\mathbf{D}=0,\quad \mbox{curl\hskip 1pt} \mathbf{E}=\mathbf{0},
\end{equation}
where `$\text{curl}$' and `$\text{div}$' are the curl and divergence operators in ${{\mathcal{B}}_{t}}$, respectively; $\bm{\mathbf{\tau }}$ is the so-called \emph{total Cauchy stress tensor} incorporating the contribution of the electric body forces,  $\rho $ is the material mass density remaining unchanged during the motion due to the material incompressibility, and the subscript $t$ following a comma denotes the material time derivative; $\mathbf{D}$ and $\mathbf{E}$ are the electric displacement and electric field vectors in ${{\mathcal{B}}_{t}}$, respectively. The symmetry of $\bm{\tau }$ is ensured by the conservation of angular momentum.

According to the theory of nonlinear electroelasticity \citep{dorfmann2006nonlinear}, the nonlinear constitutive relations for incompressible SEA materials in terms of the total energy density function $\Omega^{\text{*}} (\mathbf{F},\bm{\mathcal{D}})$ (per unit reference volume rather than per unit mass) can be expressed as
\begin{equation}\label{s1-constitutiveL}
\mathbf{T}=\frac{\partial {{\Omega }^{\text{*}}}}{\partial \mathbf{F}}-p{{\mathbf{F}}^{-1}},\quad \bm{\mathcal{E}}=\frac{\partial {{\Omega }^{\text{*}}}}{\partial \bm{\mathcal{D}}},
\end{equation}
where $p$ is a Lagrange multiplier associated with the incompressibility constraint; $\mathbf{T}={{\mathbf{F}}^{-1}}\bm{\mathbf{\tau }}$, $\bm{\mathcal{D}}={{\mathbf{F}}^{-1}}\mathbf{D}$ and $\bm{\mathcal{E}}={{\mathbf{F}}^{\text{T}}}\mathbf{E}$ are the \emph{total nominal stress tensor}, the Lagrangian electric displacement and electric field vectors, respectively. Therefore, the corresponding expressions for $\bm{\mathbf{\tau }}$ and $\mathbf{E}$ are then given by 
\begin{equation}\label{s1-constitutiveE}
\bm{\tau }=\mathbf{F}\frac{\partial {{\Omega }^{\text{*}}}}{\partial \mathbf{F}}-p\mathbf{I},\quad \mathbf{E}={{\mathbf{F}}^{-\text{T}}}\frac{\partial {{\Omega }^{\text{*}}}}{\partial \bm{\mathcal{D}}},
\end{equation}
where $\mathbf{I}$ is the identity tensor. For an incompressible isotropic SEA material, the total energy density function can be expressed as $\Omega^{\text{*}} =\Omega^{\text{*}} ({{I}_{1}},{{I}_{2}},{{I}_{4}},{{I}_{5}},{{I}_{6}})$ depending on the following five invariants:
\begin{equation} \label{invariants}
{{I}_{1}}=\text{tr}\mathbf{C},\text{    }{{I}_{2}}={\left[ {{\left( \text{tr}\mathbf{C} \right)}^{2}}-\text{tr}\left( {{\mathbf{C}}^{2}} \right) \right]}/{2}\;,\text{    }{{I}_{4}}=\bm{\mathcal{D}}\cdot \bm{\mathcal{D}},\text{    }{{I}_{5}}=\bm{\mathcal{D}}\cdot \left( \mathbf{C}\bm{\mathcal{D}} \right),\text{    }{{I}_{6}}=\bm{\mathcal{D}}\cdot \left( {{\mathbf{C}}^{2}}\bm{\mathcal{D}} \right).
\end{equation}

Therefore, the total Cauchy stress tensor and the electric field vector can be derived from Eq.~\eqref{s1-constitutiveE} as \citep{dorfmann2010electroelastic}
\begin{equation} \label{initial-constitutive}
\begin{split}
& \bm{\tau }=2\Omega _{1}^{\text{*}}\mathbf{b}+2\Omega _{2}^{\text{*}}\left( {{I}_{1}}\mathbf{b}-{{\mathbf{b}}^{2}} \right)-p\mathbf{I}+2\Omega _{5}^{\text{*}}\mathbf{D}\otimes \mathbf{D}+2\Omega _{6}^{\text{*}}\left( \mathbf{D}\otimes \mathbf{bD}+\mathbf{bD}\otimes \mathbf{D} \right), \\ 
& \mathbf{E}=2\left( \Omega _{4}^{\text{*}}{{\mathbf{b}}^{-1}}\mathbf{D}+\Omega _{5}^{\text{*}}\mathbf{D}+\Omega _{6}^{\text{*}}\mathbf{bD} \right),
\end{split}
\end{equation}
where the shorthand notation ${{\Omega }^{\text{*}}_{m}}=\partial \Omega^{\text{*}} /\partial {{I}_{m}}\text{ }(m=1,2,4,5,6)$ is adopted hereafter.

In this paper, we will consider a SEA spherical balloon (coated with flexible electrodes on the inner and outer surfaces with equal and opposite free \textit{surface} charges) under the combined action of internal pressure and radial electric voltage. In this case, as will be shown in Subsec.~\ref{section3.1}, the electric field in the body is distributed radially in the deformed configuration and there is no electric field in the surrounding vacuum. Consequently,  the mechanical and electric boundary conditions to be satisfied on $\partial {{\mathcal{B}}_{t}}$ may be written in Eulerian form as
\begin{equation} \label{boundary}
\bm{\tau n}_t={{\mathbf{t}}^{a}}, \quad \mathbf{n}_t\times \mathbf{E=0},\quad \mathbf{n}_t\cdot \mathbf{D=}-{{\sigma }_{\text{f}}},
\end{equation}
where  ${{\mathbf{t}}^{a}}$ is the prescribed mechanical traction vector per unit area of $\partial {{\mathcal{B}}_{t}}$, which is associated with the applied mechanical traction vector ${{\mathbf{t}}^{A}}$ per unit area of $\partial {{\mathcal{B}}_{r}}$ through ${{\mathbf{t}}^{a}}\text{d}a_t={{\mathbf{t}}^{A}}\text{d}A$; ${{\sigma }_{\text{f}}}$ is the free surface charge density on $\partial {{\mathcal{B}}_{t}}$.

The basic formulations in Eulerian form described above can be easily transformed into the corresponding Lagrangian counterparts, which can be referred to the work of \citet{dorfmann2006nonlinear}.


\subsection{The linearized incremental formulations}\label{Sec2-2}


According to \citet{dorfmann2010electroelastic}, the governing equations for the time-dependent infinitesimal incremental motion $\mathbf{\dot{x}}(\mathbf{X},t)$ superimposed on a finitely deformed SEA body, occupying a static configuration $\mathcal{B}$ with the boundary $\partial \mathcal{B}$ and the outward unit normal vector $\mathbf{n}$, will be presented in this subsection. Since the incremental mechanical and electrical fields are assumed to be infinitesimal, a linearized theory can be established by using the perturbation method. Here and henceforth, a superposed dot indicates the increment in the quantity concerned.

In updated Lagrangian form, the incremental governing equations can be expressed as
\begin{equation} \label{incre-governEQ}
\mbox{div\hskip 1pt}{{\mathbf{\dot{T}}}_{0}}=\rho {{\mathbf{u}}_{,tt}}, \quad \mbox{div\hskip 1pt}{{\bm{\dot{\mathcal{D}}}}_{0}}=0,\quad \mbox{curl\hskip 1pt}{{\bm{\dot{\mathcal{E}}}}_{0}}=\mathbf{0}, 
\end{equation}
where $\mathbf{u}(\mathbf{x},t)=\mathbf{\dot{x}}(\mathbf{X},t)$ denotes the incremental displacement vector; ${{\mathbf{\dot{T}}}_{0}}$, ${{\bm{\dot{\mathcal{E}}}}_{0}}$ and ${{\bm{\dot{\mathcal{D}}}}_{0}}$ are the \emph{push-forward} versions of the corresponding Lagrangian increments, which means that the reference configuration is updated from the original unstressed reference configuration ${{\mathcal{B}}_{r}}$ to the initial deformed configuration $\mathcal{B}$ \citep{dorfmann2010electroelastic}. Note that the resulting \emph{push-forward} variables are identified by a subscript 0. The linearized incremental constitutive laws for incompressible SEA materials involving the increments ${{\mathbf{\dot{T}}}_{0}}$ and ${{\bm{\dot{\mathcal{E}}}}_{0}}$ can be derived from Eq.~\eqref{s1-constitutiveL} and the \emph{push-forward} operation as
\begin{equation} \label{increconsti}
{{\mathbf{\dot{T}}}_{0}}={{\bm{\mathcal{A}}}_{0}}\mathbf{H}+{{\bm{\mathcal{M}}}_{0}}{{\bm{\dot{\mathcal{D}}}}_{0}}+p\mathbf{H}-\dot{p}\mathbf{I},\quad {{\bm{\dot{\mathcal{E}}}}_{0}}=\bm{\mathcal{M}}_{0}^{\text{T}}\mathbf{H}+{{\bm{\mathcal{R}}}_{0}}{{\bm{\dot{\mathcal{D}}}}_{0}},
\end{equation}
where $\dot{p}$ is the incremental Lagrange multiplier and $\mathbf{H}=\mbox{grad\hskip 1pt}\mathbf{u}$ denotes the incremental displacement gradient tensor with `$\text{grad}$' being the gradient operator in $\mathcal{B}$. Note that the superscript $^{\text{T}}$ in Eq.~\eqref{increconsti} denotes the transpose of a third-order tensor between the pair of subscripts (the first two indices that always go together) and the single subscript (the third index), i.e., $\bm{\mathcal{M}}_{0}^{\text{T}}\mathbf{H}={\mathcal{M}}_{0piq}H_{pi}$. The \emph{instantaneous} electroelastic moduli tensors ${{\bm{\mathcal{A}}_{0}}}$, ${{\bm{\mathcal{M}}_{0}}}$ and ${{\bm{\mathcal{R}}_{0}}}$ are defined in component notation by
\begin{align}\label{s1-effective-material}
&{{\mathcal{A}}_{0piqj}}={{F}_{p\alpha }}{{F}_{q\beta }}{{\mathcal{A}}_{\alpha i\beta j}}={{\mathcal{A}}_{0qjpi}},\quad {{\mathcal{R}}_{0ij}}=F_{\alpha i}^{-1}F_{\beta j}^{-1}{{\mathcal{R}}_{\alpha \beta }}={{\mathcal{R}}_{0ji}},\notag\\ 
&{{\mathcal{M}}_{0piq}}={{F}_{p\alpha }}F_{\beta q}^{-1}{{\mathcal{M}}_{\alpha i\beta }}={{\mathcal{M}}_{0ipq}},
\end{align}
where $\bm{\mathcal{A}}$, $\bm{\mathcal{M}}$ and $\bm{\mathcal{R}}$ are the \emph{referential} electroelastic moduli tensors associated with $\Omega^{\text{*}} (\mathbf{F},\bm{\mathcal{D}})$, with their components defined by ${{\mathcal{A}}_{\alpha i\beta j}}={{\partial }^{2}}\Omega^{\text{*}} /(\partial {{F}_{i\alpha }}\partial {{F}_{j\beta }})$,  ${{\mathcal{M}}_{\alpha i\beta }}={{\partial }^{2}}\Omega^{\text{*}} /(\partial {{F}_{i\alpha }}\partial {{\mathcal{D}}_{\beta }})$, and ${{\mathcal{R}}_{\alpha \beta }}={{\partial }^{2}}\Omega^{\text{*}} /(\partial {{\mathcal{D}}_{\alpha }}\partial {{\mathcal{D}}_{\beta }})$. By using the incremental form of the symmetry condition of the Cauchy stress $\mathbf{FT}={{(\mathbf{FT})}^{\text{T}}}$, the connections between ${{\bm{\mathcal{A}}}_{0}}$ and $\bm{\mathbf{\tau }}$ for an incompressible material may be found as \citep{dorfmann2014nonlinear}
\begin{equation}
{{\mathcal{A}}_{0jisk}}-{{\mathcal{A}}_{0ijsk}}=\left( {{\tau }_{js}}+p{{\delta }_{js}} \right){{\delta }_{ik}}-\left( {{\tau }_{is}}+p{{\delta }_{is}} \right){{\delta }_{jk}}.
\end{equation}
The additional incremental incompressibility condition reads
\begin{equation}\label{incompre}
\mbox{div\hskip 1pt}\mathbf{u} =\mbox{tr\hskip 1pt}\mathbf{H}=0.
\end{equation}

When the increments of electrical variables  in the surrounding vacuum are disregarded, the updated Lagrangian incremental forms of the electric and mechanical boundary conditions, which are to be satisfied on $\partial \mathcal{B}$, can be expressed as
\begin{equation}\label{s1-incremental-boundary}
{{\bm{\dot{\mathcal{E}}}}_{0}}\times \mathbf{n}=0,\quad {{\bm{\dot{\mathcal{D}}}}_{0}}\cdot \mathbf{n}=-{{\dot{\sigma }}_{\text{F0}}},\quad \mathbf{\dot{T}}_{0}^{\text{T}}\mathbf{n}=\mathbf{\dot{t}}_{0}^{A},
\end{equation}
where ${{\dot{\sigma }}_{\text{F0}}}$ is the incremental surface charge density on $\partial \mathcal{B}$ and the relation $\mathbf{\dot{t}}_{0}^{A}\text{d}a={{\mathbf{\dot{t}}}^{A}}\text{d}A$ holds based on the Nanson's formula, with ${{\mathbf{\dot{t}}}^{A}}$ being the Lagrangian incremental traction vector per unit area of $\partial \mathcal{B}_r$. If there is a pressure $p_a$ applied to the boundary $\partial \mathcal{B}$, the incremental mechanical boundary condition \eqref{s1-incremental-boundary}$_3$ can be rewritten as \citep{bustamante2013axisymmetric, wu2018propagation}
\begin{equation} \label{Pcondition}
\mathbf{\dot{t}}_{0}^{A}={{p}_{a}}{{\mathbf{H}}^{\operatorname{T}}}\mathbf{n}-{{\dot{p}}_{a}}\mathbf{n},
\end{equation}
where $\dot{p}_a$ is the increment of the applied pressure.


\section{Spherically symmetric nonlinear deformation}\label{section3}


The nonlinear electroelasticity theory described in Subsec.~\ref{section2.1} is now specialized for the problem of spherically symmetric nonlinear inflation of a thick-walled SEA balloon with flexible electrodes on its inner and outer surfaces, which is subjected to an internal pressure and a radial electric biasing field. Actually, the same problem has been discussed by \citet{dorfmann2014nonlinear2}, who presented the spherically symmetric nonlinear response (especially the snap-through instability) of a single-layer SEA balloon for several particular material models such as the neo-Hookean, Gent, and Ogden models. Recently, the investigation of the nonlinear electromechanical response has been extended by \citet{bortot2017analysis} to a multilayered SEA balloon, formed by embedding the active membrane between two soft passive layers and described by neo-Hookean and Gent models. In this section, the formulations governing the spherically symmetric nonlinear response for an ideal dielectric elastic model will be briefly outlined in our notation for completeness. Moreover, we will specialize the results to the neo-Hookean and Gent ideal dielectric models and present their explicit expressions for both the nonlinear static response and the radially inhomogeneous biasing fields.

\subsection{Arbitrary ideal dielectric elastic model}\label{section3.1}

The schematic diagram of a SEA spherical balloon with flexible electrodes before and after activation are displayed in Figs.~\ref{Fig1}(a) and \ref{Fig1}(b), respectively. In the undeformed configuration, the inner and outer radii of the balloon are $A$ and $B$, respectively. When a radial electric voltage $V$ is applied to the electrodes of the balloon, which is also simultaneously subjected to an internal pressure $P$, the inner and outer radii of the deformed balloon become $a$ and $b$, respectively. For incompressible SEA balloons, the spherically symmetric deformation is described by
\begin{equation} \label{kinematic}
{{r}^{3}}-{{a}^{3}}={{R}^{3}}-{{A}^{3}},\quad \theta =\mit{\Theta},\quad \varphi =\Psi,
\end{equation}
where $(R,\mit{\Theta},\Psi)$ and $\left( r,\theta ,\varphi \right)$ are the spherical coordinate systems in the undeformed and deformed configurations, respectively. Therefore, with respect to spherical polar coordinate axes, the deformation gradient tensor may be represented by a diagonal matrix $\mathbf{F}=\text{diag}[\lambda^{-2},{{\lambda }},{{\lambda }}]$, where ${{\lambda }_{r}}=\lambda^{-2}$ is the radial principal stretch, while ${{\lambda }_{\theta }}={{\lambda }_{\varphi }}=r/R \equiv \lambda$ is the principal stretch in the $\theta$ and $\varphi$ directions. For convenience of notation, we define the following notational convention as 
\begin{equation} \label{notation}
H=B-A,\quad h=b-a,\quad {{\lambda }_{a}}=a/A,\quad {{\lambda }_{b}}=b/B,\quad \eta =B/A,\quad \overline{\eta }=b/a.
\end{equation}
It then follows from Eq.~\eqref{kinematic} and Eq.~\eqref{notation}$_{3-5}$ that the stretches ${\lambda }_{a}$ and ${\lambda }_{b}$ are connected by
\begin{equation} \label{lamdaab}
\lambda _{a}^{3}-1={{\eta }^{3}}\left( \lambda _{b}^{3}-1 \right).
\end{equation}

For the considered spherically symmetric deformation and the applied radial electric field, the radial component ${{D}_{r}}$ is the only non-zero component of the Eulerian electric displacement vector $\mathbf{D}$. Then the Lagrangian electric displacement vector, $\bm{\mathcal{D}}={{\mathbf{F}}^{-1}}\mathbf{D}$, also has only one non-zero component ${{\mathcal{D}}_{r}}={{\lambda}^{2}}{{D}_{r}}$. As a result, the five independent invariants in Eq.~\eqref{invariants} can be given now in the form
\begin{equation} \label{invariant-new}
{{I}_{1}}=2{{\lambda }^{2}}+{{\lambda }^{-4}},\quad {{I}_{2}}=2{{\lambda }^{-2}}+{{\lambda }^{4}},\quad {{I}_{4}}=\mathcal{D}_{r}^{2},\quad {{I}_{5}}={{\lambda }^{-4}}{{I}_{4}},\quad {{I}_{6}}={{\lambda }^{-8}}{{I}_{4}},
\end{equation}
substitution of which into the initial constitutive relations \eqref{initial-constitutive} yields the non-zero components of the total stress tensor $\bm{\tau}$ and the electric field vector $\mathbf{E}$ as
\begin{equation} \label{iniconstitutive}
\begin{split}
{{\tau }_{rr}}&=2\left[ \Omega _{1}^{\text{*}}{{\lambda }^{-4}}+2\Omega _{2}^{\text{*}}{{\lambda }^{-2}}+\left( \Omega _{5}^{\text{*}}{{\lambda }^{-4}}+2\Omega _{6}^{\text{*}}{{\lambda }^{-8}} \right){{I}_{4}} \right]-p, \\ 
{{\tau }_{\theta \theta }}&={{\tau }_{\varphi \varphi }}=2\left[ \Omega _{1}^{\text{*}}{{\lambda }^{2}}+\Omega _{2}^{\text{*}}\left( {{\lambda }^{4}}+{{\lambda }^{-2}} \right) \right]-p, \\ 
{{E}_{r}}&=2\left( \Omega _{4}^{\text{*}}{{\lambda }^{4}}+\Omega _{5}^{\text{*}}+\Omega _{6}^{\text{*}}{{\lambda }^{-4}} \right){{D}_{r}}. \\ 
\end{split}
\end{equation}
It is clear from Eq.~\eqref{invariant-new} that only two independent quantities ${{\lambda }}$ and ${{I}_{4}}$ remain and thus a reduced energy function ${{\Omega}}$ can be defined as
\begin{equation} \label{strainenergy}
\Omega \left( \lambda ,{{I}_{4}} \right)={{\Omega }^{\text{*}}}\left( {{I}_{1}},{{I}_{2}},{{I}_{4}},{{I}_{5}},{{I}_{6}} \right).
\end{equation}
Therefore, Eqs.~\eqref{invariant-new}-\eqref{strainenergy} lead to
\begin{equation} \label{Er}
2\left( {{\tau }_{\theta \theta }}-{{\tau }_{rr}} \right)=\lambda {{\Omega }_{\lambda }},\quad {{E}_{r}}=2{{\lambda }^{4}}{{\Omega }_{4}}{{D}_{r}},
\end{equation}
where ${{\Omega }_{\lambda }}=\partial \Omega /\partial \lambda$ and ${{\Omega }_{4}}=\partial \Omega /\partial {{I}_{4}}$.

Due to the spherically symmetric deformation, all initial physical quantities depend only on $r$. The Faraday's law \eqref{governingEQ}$_{3}$ is then satisfied automatically and the Gauss's law \eqref{governingEQ}$_{2}$ simplifies to
\begin{equation} \label{GaussEQ}
\frac{1}{{{r}^{2}}}\frac{\text{d}}{\text{d}r}\left( {{r}^{2}}{{D}_{r}} \right)=0,
\end{equation}
which yields that $r^{2}{{D}_{r}}$ is a constant with $r^{2}{{D}_{r}}=a^{2}{{D}_{r}}(a)=b^{2}{{D}_{r}}(b)$, where ${{D}_{r}}(a)$ and ${{D}_{r}}(b)$ are the radial electric displacement components at the inner and outer spherical surfaces. It is assumed that the compliant electrodes are perfectly attached to the inner and outer surfaces of the balloon, with equal and opposite free surface charges $Q(a)$ and $Q(b)$, respectively, such that $Q(a)+Q(b)=0$. Consequently, there is no electric field outside the balloon and the initial boundary condition \eqref{boundary}$_3$ provides ${{D}_{r}}(a)={{\sigma }_{\text{f}a}}$ and ${{D}_{r}}(b)=-{{\sigma }_{\text{f}b}}$, where ${{\sigma }_{\text{f}a}}$ and ${{\sigma }_{\text{f}b}}$ are the free surface charge densities per unit deformed area $\partial \mathcal{B}$ on the inner and outer surfaces, respectively, defined by
\begin{equation}
{{\sigma }_{\text{f}a}}=\frac{Q\left( a \right)}{4\pi {{a}^{2}}},\quad {{\sigma }_{\text{f}b}}=\frac{Q\left( b \right)}{4\pi {{b}^{2}}}.
\end{equation}
Thus, the solution of Eq.~\eqref{GaussEQ} may be obtained as
\begin{equation} \label{Dr}
{{D}_{r}}=\frac{Q\left( a \right)}{4\pi {{r}^{2}}}=-\frac{Q\left( b \right)}{4\pi {{r}^{2}}}.
\end{equation}

Furthermore, the curl-free electric field can be expressed as $\mathbf{E}=-\text{grad}\phi $ by introducing an electrostatic potential $\phi $, which gives the only non-zero electric field component ${{E}_{r}}=-\text{d}\phi /\text{d}r$. Hence, inserting Eq.~\eqref{Dr} into Eq.~\eqref{Er}$_2$, we have 
\begin{equation} \label{electricV}
\frac{\text{d}\phi }{\text{d}r}=-{{\lambda }^{4}}{{\Omega }_{4}}\frac{Q\left( a \right)}{2\pi {{r}^{2}}}={{\lambda }^{4}}{{\Omega }_{4}}\frac{Q\left( b \right)}{2\pi {{r}^{2}}}.
\end{equation}
Using $V=\phi (a)-\phi (b)$ to represent the electric potential difference between the inner and outer surfaces and integrating Eq.~\eqref{electricV} from the inner surface to the outer surface, we obtain
\begin{equation} \label{VQ}
V=-\frac{Q\left( a \right)}{2\pi }\int_{a}^{b}{{{\lambda }^{4}}{{\Omega }_{4}}\frac{\text{d}r}{{{r}^{2}}}}=\frac{Q\left( b \right)}{2\pi }\int_{a}^{b}{{{\lambda }^{4}}{{\Omega }_{4}}\frac{\text{d}r}{{{r}^{2}}}},
\end{equation}
which represents a general relationship between the electric voltage $V$ and the surface free charge $Q$, which depends on the initial deformation.

Since the deformation is spherically symmetric, the equilibrium equations $\text{div}\bm{\tau }=\mathbf{0}$, with the help of Eq.~\eqref{Er}$_1$, reduce to one equation only, i.e., 
\begin{equation} \label{EQstress}
\frac{\text{d}{{\tau }_{rr}}}{\text{d}r}=\frac{2\left( {{\tau }_{\theta \theta }}-{{\tau }_{rr}} \right)}{r}=\frac{\lambda {{\Omega }_{\lambda }}}{r},
\end{equation}
integrating which from $a$ to $b$ and utilizing a change of variable ${\text{d}r}/{r}=-{\text{d}\lambda }/[({{{\lambda }^{3}}-1}){\lambda}]$ (which is obtained from Eq.~\eqref{kinematic} with the definition $\lambda=r/R$), we have
\begin{equation} \label{radial-stress}
{{\tau }_{rr}}\left( b \right)-{{\tau }_{rr}}\left( a \right)=-\int_{{{\lambda }_{a}}}^{{{\lambda }_{b}}}{\frac{{{\Omega }_{\lambda }}}{{{\lambda }^{3}}-1}\text{d}\lambda }.
\end{equation}
Assuming that there is a pressure $P$ applied to the inner surface but the outer surface of the SEA balloon is traction-free, i.e., ${{\tau }_{rr}}(a)=-P$ and ${{\tau }_{rr}}(b)=0$, we can rewrite Eq.~\eqref{radial-stress} as 
\begin{equation} \label{POmega}
P=-\int_{{{\lambda }_{a}}}^{{{\lambda }_{b}}}{\frac{{{\Omega }_{\lambda }}}{{{\lambda }^{3}}-1}\text{d}\lambda }.
\end{equation}
Note that ${\lambda}_{b}$ may be expressed in terms of ${\lambda }_{a}$ by Eq.~\eqref{lamdaab}. Thus, Eq.~\eqref{POmega} establishes a general expression for the internal pressure $P$ in terms of the electrical variable ($Q$ or $V$, which is included in ${{\Omega }}$) and the inner radius $a$ (measured by ${{\lambda }_{a}}$) for any given initial geometry. Similarly, the radial normal stress can be derived by integrating Eq.~\eqref{EQstress} from $r$ to $b$ as 
\begin{equation} \label{radial-stress-new}
{{\tau }_{rr}}\left( r \right)=\int_{\lambda }^{{{\lambda }_{b}}}{\frac{{{\Omega }_{\lambda }}}{{{\lambda }^{3}}-1}\text{d}\lambda }.
\end{equation}

Therefore, once the reduced energy function ${{\Omega }}$ is given, one can analytically or numerically conduct the integrations \eqref{VQ}, \eqref{POmega} and \eqref{radial-stress-new}. Then the normal stresses ${{\tau }_{\theta \theta }}$ and ${{\tau }_{\varphi\varphi}}$ in the $\theta$ and $\varphi$ directions are obtained by Eq.~\eqref{Er}$_1$ after the radial normal stress ${{\tau }_{rr}}$ is determined from Eq.~\eqref{radial-stress-new}. Finally, one can derive the Lagrange multiplier $p$ by one of the two equations \eqref{iniconstitutive}$_{1,2}$.

\subsection{Two illustrative ideal dielectric elastic models}\label{section3.2}

The aforementioned formulations are completely general for an isotropic SEA balloon. For definiteness, we now specialize the previous results to the neo-Hookean and Gent ideal dielectric models, which are characterized by the following (reduced) energy density functions, respectively,
\begin{equation} \label{NH}
{{\Omega }^{\text{*}}}=\frac{\mu}{2} \left( {{I}_{1}}-3 \right)+\frac{{I}_{5}}{2\varepsilon},\quad \Omega \left( \lambda ,{{I}_{4}} \right)=\frac{\mu}{2} \left( 2{{\lambda }^{2}}+{{\lambda }^{-4}}-3 \right)+\frac{1}{2\varepsilon}{{\lambda }^{-4}}{{I}_{4}},
\end{equation} 
and
\begin{equation} \label{Gent}
\begin{split}
&{{\Omega }^{\text{*}}}=-\frac{\mu G}{2} \ln \left( 1-\frac{{{I}_{1}}-3}{G} \right)+\frac{{I}_{5}}{2\varepsilon},\quad \Omega \left( \lambda ,{{I}_{4}} \right)={{\Omega }_{G}}\left( \lambda  \right)+\frac{1}{2\varepsilon}{{\lambda }^{-4}}{{I}_{4}},\\ 
&{{\Omega }_{G}}\left( \lambda  \right)=-\frac{\mu G}{2}\ln \left( \frac{G+3-2{{\lambda }^{2}}-{{\lambda }^{-4}}}{G} \right),
\end{split}
\end{equation}
where $\mu$ denotes the shear modulus of the SEA material in the absence of biasing fields, $\varepsilon$ is the dielectric constant of the ideal dielectric material, and the parameter $G$ is the dimensionless Gent constant reflecting the limiting chain extensibility of rubber networks \citep{gent1996new} such that the elastic strain energy approaches infinity in the limit
$I_1-3 \rightarrow G$. Thus, the Gent model can account for the particular strain-stiffening effect in the inflation procedure of the balloon. Additionally, the Gent model \eqref{Gent} can reduce to the neo-Hookean model \eqref{NH} in the limit of $G \rightarrow \infty $.

Substituting Eq.~\eqref{NH} into Eqs.~\eqref{VQ} and \eqref{POmega} and carrying out the integrations, we obtain the spherically symmetric nonlinear response for the neo-Hookean model as
\begin{equation} \label{VQNH}
\overline{V}=\frac{\eta \lambda _{a}^{-1}-\lambda _{b}^{-1}}{\eta \left( \eta -1 \right)}\overline{Q}, 
\end{equation}
and
\begin{equation} \label{PQNH}
\overline{P}=\frac{\left( {{\lambda }_{a}}-{{\lambda }_{b}} \right)}{2{{\lambda }_{a}^{4}}{{\lambda }_{b}^{4}}}\left[ 4{{\lambda }_{a}^{3}}{{\lambda }_{b}^{3}}+\left( {{\lambda }_{a}}+{{\lambda }_{b}} \right)\left( {{\lambda }_{a}^{2}}+{{\lambda }_{b}^{2}} \right) \right]-\frac{\lambda_{a}^{-4}-{{\eta }^{-4}}\lambda_{b}^{-4}}{2}{{{\overline{Q}}}^{2}},
\end{equation}
where the dimensionless electric voltage, surface charge, and internal pressure are defined as $\overline{V}={V}\sqrt{\varepsilon/{\mu }}/H$, $\overline{Q}=Q\left( a \right)/\left({4\pi {{A}^{2}}\sqrt{\mu \varepsilon }}\right)$, and $\overline{P}=P/\mu$. Note that Eqs.~\eqref{VQNH} and \eqref{PQNH} are equivalent to the results obtained by \citet{dorfmann2014nonlinear2} and \citet{bortot2017analysis} but expressed in different notation.

For the Gent ideal dielectric model, the explicit expression between $\overline{V}$ and $\overline{Q}$ is the same as Eq.~\eqref{VQNH}. Making use of Eqs.~\eqref{POmega} and \eqref{Gent}$_{2,3}$, we can derive the explicit nonlinear relation for $\overline{P}$ in terms of $\overline{Q}$ and  ${{\lambda }_{a}}$ as follows:
\begin{equation} \label{PQGent}
\begin{split}
\overline{P}&=\int_{{{\lambda }_{b}}}^{{{\lambda }_{a}}}{\frac{{{{{\Omega_{G}^{'}}}}}\left( \lambda  \right)}{\mu \left( {{\lambda }^{3}}-1 \right)}\text{d}\lambda }-\frac{1}{2\mu \varepsilon }{{\left( \frac{Q\left( a \right)}{4\pi } \right)}^{2}}\frac{\left( {{b}^{4}}-{{a}^{4}} \right)}{{{a}^{4}}{{b}^{4}}}\\
&=\int_{{{\lambda }_{b}}}^{{{\lambda }_{a}}}{\frac{2G\left( {{\lambda }^{3}}+1 \right)}{\lambda \left[ \left( G+3 \right){{\lambda }^{4}}-2{{\lambda }^{6}}-1 \right]}}\text{d}\lambda -\frac{\lambda _{a}^{-4}-{{\eta }^{-4}}\lambda _{b}^{-4}}{2}{{{\overline{Q}}}^{2}} \\ 
&=G\int_{{{\lambda }_{a}}}^{{{\lambda }_{b}}}{\frac{{{\lambda }^{3}}+1}{\lambda \left( {{\lambda }^{2}}-{{\lambda }_{01}} \right)\left( {{\lambda }^{2}}-{{\lambda }_{02}} \right)\left( {{\lambda }^{2}}-{{\lambda }_{03}} \right)}}\text{d}\lambda -\frac{\lambda _{a}^{-4}-{{\eta }^{-4}}\lambda _{b}^{-4}}{2}{{{\overline{Q}}}^{2}},
\end{split}
\end{equation}
where $\Omega_{G}^{'} (\lambda)=\partial \Omega_G /\partial \lambda$ and $\lambda_{0j} \,(j=1,2,3)$ are the three roots of the following cubic polynomial equation of $\lambda^{2}$:
\begin{equation} \label{limiting}
\left( G+3 \right){{\lambda }^{4}}-2{{\lambda }^{6}}-1=0.
\end{equation}
Thus, integrating Eq.~\eqref{PQGent} finally yields
\begin{equation}
\overline{P}=\frac{G}{2}\left[ f\left( {{\lambda }_{b}} \right)-f\left( {{\lambda }_{a}} \right) \right]-\frac{\lambda _{a}^{-4}-{{\eta }^{-4}}\lambda _{b}^{-4}}{2}{{{\overline{Q}}}^{2}},
\end{equation}
where
\begin{align}
&f\left( \lambda  \right)=\sum\limits_{i=1}^{3}{\left[ \frac{\ln \left( {{\lambda }^{2}}-{{\lambda }_{0i}} \right)}{{{\lambda }_{0i}}\left( {{\lambda }_{0i}}-{{\lambda }_{0j}} \right)\left( {{\lambda }_{0i}}-{{\lambda }_{0k}} \right)}-\frac{2\sqrt{{{\lambda }_{0i}}}\operatorname{arctanh}\left( \lambda /\sqrt{{{\lambda }_{0i}}} \right)}{\left( {{\lambda }_{0i}}-{{\lambda }_{0j}} \right)\left( {{\lambda }_{0i}}-{{\lambda }_{0k}} \right)} \right]}-\frac{2\ln \left( \lambda  \right)}{{{\lambda }_{01}}{{\lambda }_{02}}{{\lambda }_{03}}},\notag \\
&\left( j,k\in \left\{ 1,2,3 \right\};\,\, i\ne j\ne k\ne i \right).
\end{align}

Furthermore, by introducing the dimensionless radial coordinates $\Lambda =R/A\in [1,\eta ]$ and $\kappa =r/a\in \left[ 1,\overline{\eta } \right]$ in the undeformed and deformed configurations, we have ${{\kappa }^{3}}-1=\left( {{\Lambda }^{3}}-1 \right)\lambda _{a}^{-3}$ from Eq.~\eqref{kinematic}$_1$. In order to calculate the frequency of electrostatically tunable small-amplitude free vibrations, we should first obtain the biasing fields that determine the effective material parameters (see Secs.~\ref{section4} and \ref{section5}). Hence, using Eqs.~\eqref{kinematic}$_1$, \eqref{Dr}, \eqref{radial-stress-new}, and \eqref{iniconstitutive}$_{1}$, we can derive the radially inhomogeneous biasing fields (including the initial stretch $\lambda$, the radial electric displacement $D_r$, the radial normal stress $\tau_{rr}$, and the Lagrange multiplier $p$) as
\begin{equation}
\begin{split}
& \lambda ={{\left( \frac{{{\Lambda }^{3}}-1+\lambda _{a}^{3}}{{{\Lambda }^{3}}} \right)}^{1/3}}={{\left( \frac{{{\kappa }^{3}}}{{{\kappa }^{3}}-1+\lambda _{a}^{-3}} \right)}^{1/3}},\quad {{{\overline{D}}}_{r}}=\frac{\overline{Q}\lambda _{a}^{-2}}{{{\kappa }^{2}}}, \\ 
& {{{\overline{\tau }}}_{rr}}=\frac{G}{2}\left[ f\left( \lambda  \right)-f\left( {{\lambda }_{b}} \right) \right]+\frac{\lambda _{a}^{-4}{\kappa }^{-4}-{{\eta }^{-4}}\lambda _{b}^{-4}}{2}{{{\overline{Q}}}^{2}}, \quad \overline{p}=\frac{G{{\lambda }^{-4}}}{G-{{I}_{1}}+3}+\overline{D}_{r}^{2}-{{{\overline{\tau }}}_{rr}}, 
\end{split}
\end{equation}
where the dimensionless quantities are defined as ${{{\overline{D}}}_{r}}={{{D}_{r}}}/{\sqrt{\mu \varepsilon}}$, ${{{\overline{\tau }}}_{rr}}={{{\tau }_{rr}}}/{\mu }$, and $\overline{p}={p}/{\mu }$.


\section{Incremental equations and state-space formalism in spherical coordinates}
\label{section4}


Without the need to specify the energy density function, the linearized incremental governing equations given in Subsec.~\ref{Sec2-2} can be first recast into their equivalent forms in the spherical coordinates $\left( r,\theta ,\varphi \right)$. Then the state-space formalism for incremental fields in the spherical coordinates will be derived to describe the time-dependent incremental motion and the accompanying incremental electric field superimposed on the radially inhomogeneous biasing fields that are determined in Sec.~\ref{section3} for the pressurized SEA spherical balloon.

\subsection{Incremental governing equations}

After denoting the components of the incremental displacement vector
$\mathbf{u}$ by ${{u}_{r}}, {{u}_{\theta }}$ and ${{u}_{\varphi}}$ along the $r$-, $\theta$- and $\varphi$-directions, respectively, the incremental displacement gradient tensor $\mathbf{H}$ has the matrix representation as
\begin{equation} \label{39}
\mathbf{H}=\left[ \begin{matrix}
\dfrac{\partial {{u}_{r}}}{\partial r} & \dfrac{1}{r}\left( \dfrac{\partial {{u}_{r}}}{\partial \theta }-{{u}_{\theta }} \right) & \dfrac{1}{r}\left( \dfrac{1}{\sin \theta }\dfrac{\partial {{u}_{r}}}{\partial \varphi }-{{u}_{\varphi }} \right)  \\[12pt]
\dfrac{\partial {{u}_{\theta }}}{\partial r} & \dfrac{1}{r}\left( \dfrac{\partial {{u}_{\theta }}}{\partial \theta }+{{u}_{r}} \right) & \dfrac{1}{r}\left( \dfrac{1}{\sin \theta }\dfrac{\partial {{u}_{\theta }}}{\partial \varphi }-{{u}_{\varphi }}\cot \theta  \right)  \\[12pt]
\dfrac{\partial {{u}_{\varphi }}}{\partial r} & \dfrac{1}{r}\dfrac{\partial {{u}_{\varphi }}}{\partial \theta } & \dfrac{1}{r}\left( \dfrac{1}{\sin \theta }\dfrac{\partial {{u}_{\varphi }}}{\partial \varphi }+{{u}_{r}}+{{u}_{\theta }}\cot \theta  \right)  \\
\end{matrix} \right],
\end{equation}
and the incremental incompressibility constraint \eqref{incompre} becomes
\begin{equation} \label{40}
\frac{1}{{{r}^{2}}}\frac{\partial }{\partial r}( {{r}^{2}}{{u}_{r}})+\frac{1}{r\sin \theta }\frac{\partial \left( {{u}_{\theta }}\sin \theta  \right)}{\partial \theta }+\frac{1}{r\sin \theta }\frac{\partial {{u}_{\varphi }}}{\partial \varphi }=0.
\end{equation}

The corresponding incremental equations of motion \eqref{incre-governEQ}$_1$ and incremental Gauss's law \eqref{incre-governEQ}$_2$ in the spherical coordinates can be specialized to the respective forms
\begin{equation}\label{41}
\begin{split}
& \frac{1}{r^2}\frac{\partial}{\partial r}(r^2{{{\dot{T}}}_{0rr}}) +\frac{1}{r \sin \theta} \frac{\partial ({{{\dot{T}}}_{0\theta r}}\sin \theta)}{\partial \theta } +\frac{1}{r\sin \theta }\frac{\partial {{{\dot{T}}}_{0\varphi r}}}{\partial \varphi } - \frac{{{{\dot{T}}}_{0\theta \theta }}+{{{\dot{T}}}_{0\varphi \varphi }}}{r} =\rho \frac{{{\partial }^{2}}{{u}_{r}}}{\partial {{t}^{2}}}, \\
& \frac{1}{r^2}\frac{\partial}{\partial r}(r^2{{{\dot{T}}}_{0r\theta }}) +\frac{1}{r\sin \theta}\frac{\partial ({{{\dot{T}}}_{0\theta \theta }}\sin \theta)}{\partial \theta } +\frac{1}{r\sin \theta }\frac{\partial {{{\dot{T}}}_{0\varphi \theta }}}{\partial \varphi } +\frac{{{{\dot{T}}}_{0\theta r}}-{{{\dot{T}}}_{0\varphi \varphi }}\cot \theta}{r}=\rho \frac{{{\partial }^{2}}{{u}_{\theta }}}{\partial {{t}^{2}}}, \\
& \frac{1}{r^2}\frac{\partial}{\partial r}(r^2{{{\dot{T}}}_{0r\varphi}})+\frac{1}{r \sin \theta} \frac{\partial ({{{\dot{T}}}_{0\theta \varphi}}\sin \theta)}{\partial \theta }+\frac{1}{r\sin \theta }\frac{\partial {{{\dot{T}}}_{0\varphi \varphi }}}{\partial \varphi }+\frac{{{{\dot{T}}}_{0\varphi r}}+{{{\dot{T}}}_{0\varphi \theta }} \cot \theta }{r}=\rho \frac{{{\partial }^{2}}{{u}_{\varphi }}}{\partial {{t}^{2}}},
\end{split} 
\end{equation}
with ${{\dot{T}}}_{0ij}\neq{{\dot{T}}}_{0ji} \,(i\neq j; i,j=r,\theta,\varphi)$, and
\begin{equation} \label{42}
\frac{1}{{{r}^{2}}}\frac{\partial }{\partial r}( {{r}^{2}}{{{\dot{\mathcal{D}}}}_{0r}} )+\frac{1}{r\sin \theta }\frac{\partial ( {{{\dot{\mathcal{D}}}}_{0\theta }}\sin \theta  )}{\partial \theta }+\frac{1}{r\sin \theta }\frac{\partial {{{\dot{\mathcal{D}}}}_{0\varphi }}}{\partial \varphi }=0.
\end{equation}

To satisfy the incremental Faraday's law \eqref{incre-governEQ}$_3$ identically, an incremental electric potential $\dot{\phi }$ is introduced such that ${{\dot{\mathcal{E}}}_{0}}=-\text{grad}\dot{\phi }$, the components of which are given by
\begin{equation} \label{43}
{{\dot{\mathcal{E}}}_{0r}}=-\frac{\partial \dot{\phi }}{\partial r},\quad {{\dot{\mathcal{E}}}_{0\theta }}=-\frac{1}{r}\frac{\partial \dot{\phi }}{\partial \theta },\quad {{\dot{\mathcal{E}}}_{0\varphi }}=-\frac{1}{r\sin \theta }\frac{\partial \dot{\phi }}{\partial \varphi }.
\end{equation}

For the spherically symmetric deformation \eqref{kinematic} of the incompressible isotropic SEA balloon subjected to an internal pressure and a radial electric displacement field, one can derive the instantaneous electroelastic moduli tensors ${{\mathcal{A}}_{0}}$, ${{\mathcal{M}}_{0}}$ and ${{\mathcal{R}}_{0}}$ following \citet{dorfmann2010electroelastic}. For the completeness of the presentation, their non-zero components are provided in \ref{AppeA} and, in addition, we have
\begin{equation} \label{instant}
\begin{split}
& {{\mathcal{A}}_{0iijk}}=0,\quad {{\mathcal{R}}_{0jk}}=0,\quad \text{for }j\ne k,\text{   }i\in \left\{ 1,2,3 \right\},\text{   no sum over }i, \\ 
& {{\mathcal{M}}_{0ii2}}={{\mathcal{M}}_{0ii3}}={{\mathcal{M}}_{02ii}}={{\mathcal{M}}_{03ii}}=0,\quad i\in \left\{ 1,2,3 \right\},\text{   no sum over }i, \\ 
& {{\mathcal{M}}_{0ijk}}=0,\quad\text{for   }i\ne j\ne k\ne i.
\end{split} 
\end{equation}
where here and in what follows the components of ${{\mathcal{A}}_{0}}$, ${{\mathcal{M}}_{0}}$ and ${{\mathcal{R}}_{0}}$ denoted by the subscripts 1, 2 and 3 correspond to those along the $r$-, $\theta$- and $\varphi$-directions, respectively (e.g. ${{\mathcal{A}}_{01122}} = {{\mathcal{A}}_{0rr\theta\theta}}$, ${{\mathcal{A}}_{02323}} = {{\mathcal{A}}_{0\theta\varphi\theta\varphi}}$, etc.). Similar convention is applicable to $\mathbf{H}$ in Eqs.~\eqref{45} and \eqref{46}.

It is noted from \ref{AppeA} that the instantaneous electroelastic moduli are determined by the applied prestretch ${{\lambda }}$, the electric biasing field ${{D}_{r}}$, and the specific form of the energy function $\Omega$. Thus, by adjusting the internal pressure $P$ and the electric voltage $V$, one can alter the instantaneous electromechanical properties of the SEA spherical balloon, which will in turn play an important role on the dynamic characteristics of the incremental motions.

Utilizing Eq.~\eqref{instant}, the incremental constitutive equations \eqref{increconsti} can be written as
\begin{equation} \label{45}
\begin{split}
& {{{\dot{T}}}_{0rr}}=({{\mathcal{A}}_{01111}}+p){{H}_{11}}+{{\mathcal{A}}_{01122}}\left( {{H}_{22}}+{{H}_{33}} \right)+{{\mathcal{M}}_{0111}}{{{\dot{\mathcal{D}}}}_{0r}}-\dot{p}, \\ 
& {{{\dot{T}}}_{0\theta \theta }}={{\mathcal{A}}_{01122}}{{H}_{11}}+({{\mathcal{A}}_{02222}}+p){{H}_{22}}+{{\mathcal{A}}_{02233}}{{H}_{33}}+{{\mathcal{M}}_{0221}}{{{\dot{\mathcal{D}}}}_{0r}}-\dot{p}, \\ 
& {{{\dot{T}}}_{0\varphi \varphi }}={{\mathcal{A}}_{01122}}{{H}_{11}}+{{\mathcal{A}}_{02233}}{{H}_{22}}+({{\mathcal{A}}_{02222}}+p){{H}_{33}}+{{\mathcal{M}}_{0221}}{{{\dot{\mathcal{D}}}}_{0r}}-\dot{p}, \\ 
& {{{\dot{T}}}_{0r\theta }}={{\mathcal{A}}_{01212}}{{H}_{21}}+({{\mathcal{A}}_{01221}}+p){{H}_{12}}+{{\mathcal{M}}_{0122}}{{{\dot{\mathcal{D}}}}_{0\theta }}, \\ 
& {{{\dot{T}}}_{0\theta r}}={{\mathcal{A}}_{02121}}{{H}_{12}}+({{\mathcal{A}}_{01221}}+p){{H}_{21}}+{{\mathcal{M}}_{0122}}{{{\dot{\mathcal{D}}}}_{0\theta }}, \\ 
& {{{\dot{T}}}_{0r\varphi }}={{\mathcal{A}}_{01212}}{{H}_{31}}+({{\mathcal{A}}_{01221}}+p){{H}_{13}}+{{\mathcal{M}}_{0122}}{{{\dot{\mathcal{D}}}}_{0\varphi }}, \\ 
& {{{\dot{T}}}_{0\varphi r}}={{\mathcal{A}}_{02121}}{{H}_{13}}+({{\mathcal{A}}_{01221}}+p){{H}_{31}}+{{\mathcal{M}}_{0122}}{{{\dot{\mathcal{D}}}}_{0\varphi }}, \\ 
& {{{\dot{T}}}_{0\theta \varphi }}={{\mathcal{A}}_{02323}}{{H}_{32}}+({{\mathcal{A}}_{02332}}+p){{H}_{23}},\text{ }{{{\dot{T}}}_{0\varphi \theta }}={{\mathcal{A}}_{02323}}{{H}_{23}}+({{\mathcal{A}}_{02332}}+p){{H}_{32}},
\end{split} 
\end{equation}
and
\begin{equation} \label{46}
\begin{split}
& {{{\dot{\mathcal{E}}}}_{0r}}={{\mathcal{M}}_{0111}}{{H}_{11}}+{{\mathcal{M}}_{0221}}\left( {{H}_{22}}+{{H}_{33}} \right)+{{\mathcal{R}}_{011}}{{{\dot{\mathcal{D}}}}_{0r}}, \\ 
& {{{\dot{\mathcal{E}}}}_{0\theta }}={{\mathcal{M}}_{0122}}\left( {{H}_{21}}+{{H}_{12}} \right)+{{\mathcal{R}}_{022}}{{{\dot{\mathcal{D}}}}_{0\theta }}, \\ 
& {{{\dot{\mathcal{E}}}}_{0\varphi }}={{\mathcal{M}}_{0122}}\left( {{H}_{31}}+{{H}_{13}} \right)+{{\mathcal{R}}_{022}}{{{\dot{\mathcal{D}}}}_{0\varphi }}.
\end{split} 
\end{equation}

After solving for ${{\bm{\dot{\mathcal{D}}}}_{0}}$ in Eq.~\eqref{46} in terms of ${{\bm{\dot{\mathcal{E}}}}_{0}}$, substituting the resulting expressions into Eq.~\eqref{45} and accounting for the relations \eqref{39} and \eqref{43}, the incremental constitutive equations \eqref{45} and \eqref{46} can be transformed into those in terms of the incremental electric potential $\dot{\phi }$ and incremental displacement vector $\mathbf{u}$ as
\begin{equation} \label{47}
\begin{split}
& {{{\dot{\mathcal{D}}}}_{0r}}={{e}_{11}}\frac{\partial {{u}_{r}}}{\partial r}+{{e}_{12}}\frac{1}{r}\left( \frac{\partial {{u}_{\theta }}}{\partial \theta }+2{{u}_{r}}+\frac{1}{\sin \theta }\frac{\partial {{u}_{\varphi }}}{\partial \varphi }+{{u}_{\theta }}\cot \theta  \right)-{{\varepsilon }_{11}}\frac{\partial \dot{\phi }}{\partial r}, \\ 
& {{{\dot{\mathcal{D}}}}_{0\theta }}={{e}_{26}}\left[ \frac{1}{r}\left( \frac{\partial {{u}_{r}}}{\partial \theta }-{{u}_{\theta }} \right)+\frac{\partial {{u}_{\theta }}}{\partial r} \right]-{{\varepsilon }_{22}}\frac{1}{r}\frac{\partial \dot{\phi }}{\partial \theta }, \\ 
& {{{\dot{\mathcal{D}}}}_{0\varphi }}={{e}_{26}}\left[ \frac{1}{r}\left( \frac{1}{\sin \theta }\frac{\partial {{u}_{r}}}{\partial \varphi }-{{u}_{\varphi }} \right)+\frac{\partial {{u}_{\varphi }}}{\partial r} \right]-{{\varepsilon }_{22}}\frac{1}{r\sin \theta }\frac{\partial \dot{\phi }}{\partial \varphi }, 
\end{split} 
\end{equation}
and
\begin{align} \label{48}
& {{{\dot{T}}}_{0rr}}={{c}_{11}}\frac{\partial {{u}_{r}}}{\partial r}+{{c}_{12}}\frac{1}{r}\left( \frac{\partial {{u}_{\theta }}}{\partial \theta }+2{{u}_{r}}+\frac{1}{\sin \theta }\frac{\partial {{u}_{\varphi }}}{\partial \varphi }+{{u}_{\theta }}\cot \theta  \right)+{{e}_{11}}\frac{\partial \dot{\phi }}{\partial r}-\dot{p}, \notag \\ 
& {{{\dot{T}}}_{0\theta \theta }}={{c}_{12}}\frac{\partial {{u}_{r}}}{\partial r}+{{c}_{22}}\frac{1}{r}\left( \frac{\partial {{u}_{\theta }}}{\partial \theta }+{{u}_{r}} \right)+{{c}_{23}}\frac{1}{r}\left( \frac{1}{\sin \theta }\frac{\partial {{u}_{\varphi }}}{\partial \varphi }+{{u}_{r}}+{{u}_{\theta }}\cot \theta  \right)+{{e}_{12}}\frac{\partial \dot{\phi }}{\partial r}-\dot{p}, \notag \\ 
& {{{\dot{T}}}_{0\varphi \varphi }}={{c}_{12}}\frac{\partial {{u}_{r}}}{\partial r}+{{c}_{23}}\frac{1}{r}\left( \frac{\partial {{u}_{\theta }}}{\partial \theta }+{{u}_{r}} \right)+{{c}_{22}}\frac{1}{r}\left( \frac{1}{\sin \theta }\frac{\partial {{u}_{\varphi }}}{\partial \varphi }+{{u}_{r}}+{{u}_{\theta }}\cot \theta  \right)+{{e}_{12}}\frac{\partial \dot{\phi }}{\partial r}-\dot{p}, \notag \\ 
& {{{\dot{T}}}_{0r\theta }}={{c}_{66}}\frac{\partial {{u}_{\theta }}}{\partial r}+{{c}_{69}}\frac{1}{r}\left( \frac{\partial {{u}_{r}}}{\partial \theta }-{{u}_{\theta }} \right)+{{e}_{26}}\frac{1}{r}\frac{\partial \dot{\phi }}{\partial \theta }, \notag \\
&{{{\dot{T}}}_{0\theta r}}={{c}_{69}}\frac{\partial {{u}_{\theta }}}{\partial r}+{{c}_{99}}\frac{1}{r}\left( \frac{\partial {{u}_{r}}}{\partial \theta }-{{u}_{\theta }} \right)+{{e}_{26}}\frac{1}{r}\frac{\partial \dot{\phi }}{\partial \theta }, \notag \\ 
& {{{\dot{T}}}_{0r\varphi }}={{c}_{66}}\frac{\partial {{u}_{\varphi }}}{\partial r}+{{c}_{69}}\frac{1}{r}\left( \frac{1}{\sin \theta }\frac{\partial {{u}_{r}}}{\partial \varphi }-{{u}_{\varphi }} \right)+{{e}_{26}}\frac{1}{r\sin \theta }\frac{\partial \dot{\phi }}{\partial \varphi }, \notag \\ 
& {{{\dot{T}}}_{0\varphi r}}={{c}_{69}}\frac{\partial {{u}_{\varphi }}}{\partial r}+{{c}_{99}}\frac{1}{r}\left( \frac{1}{\sin \theta }\frac{\partial {{u}_{r}}}{\partial \varphi }-{{u}_{\varphi }} \right)+{{e}_{26}}\frac{1}{r\sin \theta }\frac{\partial \dot{\phi }}{\partial \varphi }, \notag \\ 
& {{{\dot{T}}}_{0\theta \varphi }}={{c}_{44}}\frac{1}{r}\frac{\partial {{u}_{\varphi }}}{\partial \theta }+{{c}_{47}}\frac{1}{r}\left( \frac{1}{\sin \theta }\frac{\partial {{u}_{\theta }}}{\partial \varphi }-{{u}_{\varphi }}\cot \theta  \right), \notag \\
&{{{\dot{T}}}_{0\varphi \theta }}={{c}_{44}}\frac{1}{r}\left( \frac{1}{\sin \theta }\frac{\partial {{u}_{\theta }}}{\partial \varphi }-{{u}_{\varphi }}\cot \theta  \right)+{{c}_{47}}\frac{1}{r}\frac{\partial {{u}_{\varphi }}}{\partial \theta },  
\end{align}
where
\begin{align} \label{materialparameter1}
& {{\varepsilon }_{11}}=1/{{\mathcal{R}}_{011}},\,\,{{\varepsilon }_{22}}=1/{{\mathcal{R}}_{022}},\,\,{{e}_{11}}=-{{\mathcal{M}}_{0111}}{{\varepsilon}_{11}},\,\,{{e}_{12}}=-{{\mathcal{M}}_{0221}}{{\varepsilon }_{11}},\,\,{{e}_{26}}=-{{\mathcal{M}}_{0122}}{{\varepsilon }_{22}}, \notag \\ 
& {{c}_{11}}={{\mathcal{A}}_{01111}}+{{\mathcal{M}}_{0111}}{{e}_{11}}+p,\,\,{{c}_{12}}={{\mathcal{A}}_{01122}}+{{\mathcal{M}}_{0111}}{{e}_{12}},\,{{c}_{22}}={{\mathcal{A}}_{02222}}+{{\mathcal{M}}_{0221}}{{e}_{12}}+p, \notag \\ 
& {{c}_{23}}={{\mathcal{A}}_{02233}}+{{\mathcal{M}}_{0221}}{{e}_{12}},\,\,{{c}_{66}}={{\mathcal{A}}_{01212}}+{{\mathcal{M}}_{0122}}{{e}_{26}},\,\,{{c}_{69}}={{\mathcal{A}}_{01221}}+{{\mathcal{M}}_{0122}}{{e}_{26}}+p, \notag \\ 
& {{c}_{99}}={{\mathcal{A}}_{02121}}+{{\mathcal{M}}_{0122}}{{e}_{26}},\,\,{{c}_{47}}={{\mathcal{A}}_{02332}}+p,\,\,{{c}_{44}}={{\mathcal{A}}_{02323}}.
\end{align}

Therefore, the application of internal pressure $P$ and radial electric voltage $V$ makes the originally isotropic SEA spherical balloon exhibit the spherically isotropic behavior (a special kind of transverse isotropy) with respect to the underlying deformed configuration, which can be seen from the incremental constitutive Eqs.~\eqref{47}-\eqref{materialparameter1}. This symmetry-breaking due to the biasing fields is well-known as the \textit{deformation-induced anisotropy} \citep{wu2016, wu2017guided}. Furthermore, from the constitutive point of view, an isotropic SEA balloon with a radial voltage behaves like a conventional spherically isotropic piezoelectric shell since the applied electric displacement vector plays a similar role to that of the preferred direction for the spherically isotropic piezoelectric shell in the reference configuration.

As described in Sec.~\ref{section3}, the initial physical variables for the spherically symmetric deformation in the pressurized SEA balloon subjected to the radial electric voltage are radially inhomogeneous, resulting in the $r$-dependence of the instantaneous electroelastic moduli. In general, the resulting incremental displacement equations are a system of coupled partial differential equations with variable coefficients, which are intractable to solve analytically or even numerically via the conventional displacement-based method. It is noted that in order to circumvent this difficulty, the \textit{state-space method} (SSM), which combines the incremental state-space formalism with the approximate laminate or multi-layer technique, has been developed by \citet{wu2017guided, wu2018propagation} to investigate the effects of inhomogeneous biasing fields on the circumferential waves in a homogeneous SEA tube and on the axisymmetric waves in a pressurized functionally graded elastomeric hollow cylinder, respectively. As a result, the SSM will be employed in this work to identify the small-amplitude free vibration characteristics of SEA balloons under the inhomogeneous biasing fields.

\subsection{State-space formalism for incremental fields}
\label{4.2}

To obtain the decoupled state-space formalism for incremental fields (which will be shown below), we rewrite Eqs.~\eqref{41}, \eqref{42}, \eqref{40} and \eqref{48}, \eqref{47} as follows:
\begin{align} \label{50}
& {{\nabla }_{2}}{{\Sigma }_{rr}}+\frac{\partial {{\Sigma }_{\theta r}}}{\partial \theta }+\frac{1}{\sin \theta }\frac{\partial {{\Sigma }_{\varphi r}}}{\partial \varphi }+{{\Sigma }_{rr}}-{{\Sigma }_{\theta \theta }}-{{\Sigma }_{\varphi \varphi }}+{{\Sigma }_{\theta r}}\cot \theta =\rho {{r}^{2}}\frac{{{\partial }^{2}}{{u}_{r}}}{\partial {{t}^{2}}}, \notag \\ 
& {{\nabla }_{2}}{{\Sigma }_{r\theta }}+\frac{\partial {{\Sigma }_{\theta \theta }}}{\partial \theta }+\frac{1}{\sin \theta }\frac{\partial {{\Sigma }_{\varphi \theta }}}{\partial \varphi }+{{\Sigma }_{r\theta }}+{{\Sigma }_{\theta r}}+\left( {{\Sigma }_{\theta \theta }}-{{\Sigma }_{\varphi \varphi }} \right)\cot \theta =\rho {{r}^{2}}\frac{{{\partial }^{2}}{{u}_{\theta }}}{\partial {{t}^{2}}}, \notag \\ 
& {{\nabla }_{2}}{{\Sigma }_{r\varphi }}+\frac{\partial {{\Sigma }_{\theta \varphi }}}{\partial \theta }+\frac{1}{\sin \theta }\frac{\partial {{\Sigma }_{\varphi \varphi }}}{\partial \varphi }+{{\Sigma }_{r\varphi }}+{{\Sigma }_{\varphi r}}+\left( {{\Sigma }_{\theta \varphi }}+{{\Sigma }_{\varphi \theta }} \right)\cot \theta =\rho {{r}^{2}}\frac{{{\partial }^{2}}{{u}_{\varphi }}}{\partial {{t}^{2}}}, \notag \\ 
& {{\nabla }_{2}}{{\Delta }_{r}}+{{\Delta }_{r}}+\frac{1}{\sin \theta }\frac{\partial \left( {{\Delta }_{\theta }}\sin \theta  \right)}{\partial \theta }+\frac{1}{\sin \theta }\frac{\partial {{\Delta }_{\varphi }}}{\partial \varphi }=0, \notag \\
&{{\nabla }_{2}}{{u}_{r}}+2{{u}_{r}}+\frac{\partial {{u}_{\theta }}}{\partial \theta }+{{u}_{\theta }}\cot \theta+\frac{1}{\sin \theta }\frac{\partial {{u}_{\varphi }}}{\partial \varphi }=0, 
\end{align}
and
\begin{align} \label{51}
& {{\Sigma }_{rr}}={{c}_{11}}{{\nabla }_{2}}{{u}_{r}}+{{c}_{12}}\left( \frac{\partial {{u}_{\theta }}}{\partial \theta }+2{{u}_{r}}+\frac{1}{\sin \theta }\frac{\partial {{u}_{\varphi }}}{\partial \varphi }+{{u}_{\theta }}\cot \theta  \right)+{{e}_{11}}{{\nabla }_{2}}\dot{\phi }-r\dot{p}, \notag \\ 
& {{\Sigma }_{\theta \theta }}={{c}_{12}}{{\nabla }_{2}}{{u}_{r}}+{{c}_{22}}\left( \frac{\partial {{u}_{\theta }}}{\partial \theta }+{{u}_{r}} \right)+{{c}_{23}}\left( \frac{1}{\sin \theta }\frac{\partial {{u}_{\varphi }}}{\partial \varphi }+{{u}_{r}}+{{u}_{\theta }}\cot \theta  \right)+{{e}_{12}}{{\nabla }_{2}}\dot{\phi }-r\dot{p}, \notag \\ 
& {{\Sigma }_{\varphi \varphi }}={{c}_{12}}{{\nabla }_{2}}{{u}_{r}}+{{c}_{23}}\left( \frac{\partial {{u}_{\theta }}}{\partial \theta }+{{u}_{r}} \right)+{{c}_{22}}\left( \frac{1}{\sin \theta }\frac{\partial {{u}_{\varphi }}}{\partial \varphi }+{{u}_{r}}+{{u}_{\theta }}\cot \theta  \right)+{{e}_{12}}{{\nabla }_{2}}\dot{\phi }-r\dot{p}, \notag \\ 
& {{\Sigma }_{r\theta }}={{c}_{66}}{{\nabla }_{2}}{{u}_{\theta }}+{{c}_{69}}\left( \frac{\partial {{u}_{r}}}{\partial \theta }-{{u}_{\theta }} \right)+{{e}_{26}}\frac{\partial \dot{\phi }}{\partial \theta },\notag \\
& {{\Sigma }_{\theta r}}={{c}_{69}}{{\nabla }_{2}}{{u}_{\theta }}+{{c}_{99}}\left( \frac{\partial {{u}_{r}}}{\partial \theta }-{{u}_{\theta }} \right)+{{e}_{26}}\frac{\partial \dot{\phi }}{\partial \theta },\notag \\
&{{\Sigma }_{r\varphi }}={{c}_{66}}{{\nabla }_{2}}{{u}_{\varphi }}+{{c}_{69}}\left( \frac{1}{\sin \theta }\frac{\partial {{u}_{r}}}{\partial \varphi }-{{u}_{\varphi }} \right)+\frac{{{e}_{26}}}{\sin \theta }\frac{\partial \dot{\phi }}{\partial \varphi },\notag \\
&{{\Sigma }_{\varphi r}}={{c}_{69}}{{\nabla }_{2}}{{u}_{\varphi }}+{{c}_{99}}\left( \frac{1}{\sin \theta }\frac{\partial {{u}_{r}}}{\partial \varphi }-{{u}_{\varphi }} \right)+\frac{{{e}_{26}}}{\sin \theta }\frac{\partial \dot{\phi }}{\partial \varphi },\notag \\
&{{\Sigma }_{\theta \varphi }}={{c}_{44}}\frac{\partial {{u}_{\varphi }}}{\partial \theta }+{{c}_{47}}\left( \frac{1}{\sin \theta }\frac{\partial {{u}_{\theta }}}{\partial \varphi }-{{u}_{\varphi }}\cot \theta  \right), \notag \\ 
&{{\Sigma }_{\varphi \theta }}={{c}_{47}}\frac{\partial {{u}_{\varphi }}}{\partial \theta }+{{c}_{44}}\left( \frac{1}{\sin \theta }\frac{\partial {{u}_{\theta }}}{\partial \varphi }-{{u}_{\varphi }}\cot \theta  \right), \notag \\ 
& {{\Delta }_{r}}={{e}_{11}}{{\nabla }_{2}}{{u}_{r}}+{{e}_{12}}\left( \frac{\partial {{u}_{\theta }}}{\partial \theta }+2{{u}_{r}}+\frac{1}{\sin \theta }\frac{\partial {{u}_{\varphi }}}{\partial \varphi }+{{u}_{\theta }}\cot \theta  \right)-{{\varepsilon }_{11}}{{\nabla }_{2}}\dot{\phi }, \notag \\ 
& {{\Delta }_{\theta }}={{e}_{26}}\left( \frac{\partial {{u}_{r}}}{\partial \theta }+{{\nabla }_{2}}{{u}_{\theta }}-{{u}_{\theta }} \right)-{{\varepsilon }_{22}}\frac{\partial \dot{\phi }}{\partial \theta },\notag \\
&{{\Delta }_{\varphi }}={{e}_{26}}\left( \frac{1}{\sin \theta }\frac{\partial {{u}_{r}}}{\partial \varphi }+{{\nabla }_{2}}{{u}_{\varphi }}-{{u}_{\varphi }} \right)-\frac{{{\varepsilon }_{22}}}{\sin \theta }\frac{\partial \dot{\phi }}{\partial \varphi }, 
\end{align}
where ${\nabla }_{2}=r\partial/\partial{r}$, ${\Sigma }_{ij}=r{{\dot{T}}}_{0ij}$, and ${{\Delta }_{i }}=r{{{\dot{\mathcal{D}}}}_{0i}} \,(i,j=r,\theta,\varphi)$.

In the linear piezoelectricity scenario, it has been shown that by adopting a proper means, not only can the basic governing equations be decoupled with the order reduced, but also the subsequent solving procedure becomes simpler \citep{chen2001free, wu2018free}. To do so, we can introduce the following three displacement functions ($\psi$, $G$, $w$), two stress functions (${\Sigma }_{1}$, ${\Sigma }_{2}$) and one electric potential function $\Phi$ to express $(u_r, u_\theta, u_\varphi)$ and  $({{\Sigma }_{r\theta }}, {{\Sigma }_{r\varphi }}, \dot{\phi }) $ as
\begin{equation} \label{52}
\begin{split}
& {{u}_{\theta }}=-\frac{1}{\sin \theta }\frac{\partial \psi }{\partial \varphi }-\frac{\partial G}{\partial \theta },\quad {{u}_{\varphi }}=\frac{\partial \psi }{\partial \theta }-\frac{1}{\sin \theta }\frac{\partial G}{\partial \varphi },\quad {{u}_{r}}=w, \\ 
& {{\Sigma }_{r\theta }}=-\frac{1}{\sin \theta }\frac{\partial {{\Sigma }_{1}}}{\partial \varphi }-\frac{\partial {{\Sigma }_{2}}}{\partial \theta },\quad {{\Sigma }_{r\varphi }}=\frac{\partial {{\Sigma }_{1}}}{\partial \theta }-\frac{1}{\sin \theta }\frac{\partial {{\Sigma }_{2}}}{\partial \varphi },\quad \dot{\phi }=\Phi.
\end{split}
\end{equation}

The specific derivation procedure for the decoupled state equations is demonstrated in \ref{AppeB}. Collecting and rearranging Eqs.~\eqref{B5}$_1$ and \eqref{B16} as well as Eqs.~\eqref{B2}, \eqref{B3}, \eqref{B5}$_2$, \eqref{B8}, \eqref{B10} and \eqref{B17}, we can transform the original incremental governing equations into the following two decoupled incremental state equations:
\begin{align} \label{53}
{{\nabla }_{2}}{{\mathbf{Y}}_{k}}={{\mathbf{K}}_{k}}\left( r,\nabla _{1}^{2} \right){{\mathbf{Y}}_{k}},\quad k\in \left[ 1,2 \right], 
\end{align}
where $\nabla _{1}^{2}={{{\partial }^{2}}}/{\partial {{\theta }^{2}}}+({1}/{{{\sin }^{2}}\theta}){{{\partial }^{2}}}/{\partial {{\varphi }^{2}}}+(\cot \theta){\partial}/{\partial \theta }$ is the 2D Laplacian on a spherical surface, the incremental state vectors $\mathbf{Y_1}$ and $\mathbf{Y_2}$ are defined as
\begin{align} \label{54}
{{\mathbf{Y}}_{1}}={{\left[ {{\Sigma }_{1}},\psi  \right]}^{\text{T}}},\quad {{\mathbf{Y}}_{2}}={{\left[ {{\Sigma }_{rr}},{{\Sigma }_{2}},G,w,{{\Delta }_{r}},\Phi  \right]}^{\text{T}}},
\end{align}
with the elements being the state variables, and the $2\times2$ and $4 \times 4$ system matrices ${\mathbf{K}}_{1}$ and ${\mathbf{K}}_{2}$ are given by
\begin{equation} \label{55}
{{\mathbf{K}}_{1}}=\left[ \begin{matrix}
-{{q}_{7}} & \rho {{r}^{2}}\dfrac{{{\partial }^{2}}}{\partial {{t}^{2}}}-{{c}_{44}}\nabla _{1}^{2}-{{q}_{1}}-{{q}_{2}}  \\[6pt]
\dfrac{1}{{{c}_{66}}} & \dfrac{{{c}_{69}}}{{{c}_{66}}}  \\
\end{matrix} \right],
\end{equation}
\begin{equation} \label{56}
\setlength{\arraycolsep}{2.5pt}
{{\mathbf{K}}_{2}}=\left[ \begin{matrix}
1 & \dfrac{{{c}_{69}}}{{{c}_{66}}}\nabla _{1}^{2} & \left( {{q}_{1}}+{{q}_{2}}+2{{q}_{3}} \right)\nabla _{1}^{2} & {{q}_{1}}\nabla _{1}^{2}-2\left( {{q}_{2}}+2{{q}_{3}} \right)+\rho {{r}^{2}}\dfrac{{{\partial }^{2}}}{\partial {{t}^{2}}} & 2{{q}_{4}} & {{q}_{5}}\nabla _{1}^{2}  \\
1 & -{{q}_{7}} & {{q}_{3}}\nabla _{1}^{2}+\rho {{r}^{2}}\dfrac{{{\partial }^{2}}}{\partial {{t}^{2}}}-{{q}_{1}}-{{q}_{2}} & -\left( {{q}_{1}}+{{q}_{2}}+2{{q}_{3}} \right) & {{q}_{4}} & -{{q}_{5}}  \\[10pt]
0 & \dfrac{1}{{{c}_{66}}} & \dfrac{{{c}_{69}}}{{{c}_{66}}} & \dfrac{{{c}_{69}}}{{{c}_{66}}} & 0 & \dfrac{{{e}_{26}}}{{{c}_{66}}}  \\[15pt]
0 & 0 & \nabla _{1}^{2} & -2 & 0 & 0  \\[6pt]
0 & \dfrac{{{e}_{26}}}{{{c}_{66}}}\nabla _{1}^{2} & {{q}_{5}}\nabla _{1}^{2} & {{q}_{5}}\nabla _{1}^{2} & -1 & {{q}_{6}}\nabla _{1}^{2}  \\
0 & 0 & {{q}_{4}}\nabla _{1}^{2} & -2{{q}_{4}} & -\dfrac{1}{{{\varepsilon }_{11}}} & 0  \\
\end{matrix}  \right],
\end{equation}
in which the effective material parameters associated with the biasing fields are expressed by
\begin{equation}
\begin{split}
&{{q}_{1}}=\frac{c_{69}^{2}}{{{c}_{66}}}-{{c}_{99}},\quad {{q}_{2}}={{c}_{44}}+{{c}_{47}},\quad {{q}_{3}}=2{{c}_{12}}-{{c}_{11}}-{{c}_{22}}-\frac{{{\left( {{e}_{11}}-{{e}_{12}} \right)}^{2}}}{{{\varepsilon }_{11}}},\\
&{{q}_{4}}=\frac{{{e}_{11}}-{{e}_{12}}}{{{\varepsilon }_{11}}},\quad
{{q}_{5}}=\left( \frac{{{c}_{69}}}{{{c}_{66}}}-1 \right){{e}_{26}},\quad {{q}_{6}}=\frac{e_{26}^{2}}{{{c}_{66}}}+{{\varepsilon }_{22}},\quad {{q}_{7}}=1+\frac{{{c}_{69}}}{{{c}_{66}}}.
\end{split}
\end{equation}

It is clear from Eqs.~\eqref{53}-\eqref{56} that these two unknown functions $\psi$ and ${{\Sigma }_{1}}$ are uncoupled from the other six unknown functions $G$, $w$, ${{\Sigma }_{2}}$, ${{\Sigma }_{rr}}$, ${{\Delta }_{r}}$ and $\Phi$, which means that there exist two classes of incremental free vibration superimposed on the underlying deformed configuration, which will be discussed in Sec.~\ref{section5} in detail. 

In contrast to the conventional displacement-based method, the SSM as a special mixed-variables method in general transforms the basic governing equations into a set of first-order ordinary differential equations with respect to one particular coordinate variable, the radial coordinate $r$ here. It should be emphasized that the state equation \eqref{53} can be applicable to arbitrary energy density function for incompressible isotropic SEA materials. After taking account of the appropriate incremental boundary conditions, the state equation \eqref{53} can be efficiently solved as demonstrated in Sec.~\ref{section5}. For the sake of brevity, we have discarded the output equations that are usually required for the determination of the physical variables other than the state variables in Eq.~\eqref{54}.


\section{Frequency equations of two classes of free vibration}
\label{section5}


In this section, we will consider the small-amplitude free vibration of a SEA spherical balloon subjected to the radially inhomogeneous biasing fields determined in Sec.~\ref{section3}. Since it is difficult to directly solve the incremental state equations \eqref{53} with variable coefficients, we first intend to adopt the approximate laminate technique \citep{wu2017guided, wu2018propagation} to obtain the relation between the state vectors at the inner and outer surfaces of the deformed SEA balloon in Subsec.~\ref{5.1}. Then, by properly accounting for the incremental mechanical and electric boundary conditions, we derive in Subsec.~\ref{5.2} the frequency equations of two classes of free vibration in the SEA balloon.

\subsection{Approximate laminate technique} \label{5.1} 

For the harmonic vibration of a closed or complete SEA balloon, the state variables in Eq.~\eqref{54} can be expanded in series in the following form:
\begin{equation} \label{series-solution}
\begin{split}
& {{\Sigma }_{1}}=\sum\limits_{m=0}^{n}{\sum\limits_{n=0}^{\infty }{{{\Sigma }_{1n}}(r)}S_{n}^{m}(\theta ,\varphi ){{e}^{\text{i}\omega t}}},\quad \psi =\sum\limits_{m=0}^{n}{\sum\limits_{n=0}^{\infty }{{{\psi }_{n}}(r)}S_{n}^{m}(\theta ,\varphi ){{e}^{\text{i}\omega t}}}, \\ 
& {{\Sigma }_{rr}}=\sum\limits_{m=0}^{n}{\sum\limits_{n=0}^{\infty }{{{\Sigma }_{rn}}(r)}S_{n}^{m}(\theta ,\varphi ){{e}^{\text{i}\omega t}}},\quad {{\Sigma }_{2}}=\sum\limits_{m=0}^{n}{\sum\limits_{n=0}^{\infty }{{{\Sigma }_{2n}}(r)}S_{n}^{m}(\theta ,\varphi ){{e}^{\text{i}\omega t}}},\\
& G=\sum\limits_{m=0}^{n}{\sum\limits_{n=0}^{\infty }{{{G}_{n}}(r)}S_{n}^{m}(\theta ,\varphi ){{e}^{\text{i}\omega t}}},\quad w=\sum\limits_{m=0}^{n}{\sum\limits_{n=0}^{\infty }{{{w}_{n}}(r)}S_{n}^{m}(\theta ,\varphi ){{e}^{\text{i}\omega t}}}, \\ 
& {{\Delta }_{r}}=\sum\limits_{m=0}^{n}{\sum\limits_{n=0}^{\infty }{{{\Delta }_{rn}}(r)}S_{n}^{m}(\theta ,\varphi ){{e}^{\text{i}\omega t}}},\quad \Phi =\sum\limits_{m=0}^{n}{\sum\limits_{n=0}^{\infty }{{{\Phi }_{n}}(r)}S_{n}^{m}(\theta ,\varphi ){{e}^{\text{i}\omega t}}}, 
\end{split}
\end{equation}
where  $S_{n}^{m}(\theta ,\varphi)=P_{n}^{m}(\cos\theta ){{e}^{\text{i}m\varphi }}$ are the spherical harmonic functions, $P_{n}^{m}(\cos\theta )$ are the associated Legendre polynomials, $\omega$ is the circular frequency of vibration, and the integers $n$ and $m$ specify the \emph{angular} and \emph{azimuthal} dependences of the vibration modes, respectively. Since the integer $m$ will not appear in the resulting ordinary differential equations, there is no need to indicate $m$ in the subscript of the unknown functions (${\Sigma }_{1n}(r)$, $\psi_n(r)$, etc.) in Eq.~\eqref{series-solution}, which will be shown in the following derivations.

Substituting Eq.~\eqref{series-solution} into Eqs.~\eqref{53}-\eqref{56} and utilizing the orthogonality of the associated Legendre polynomials with respect to the weight $\sin\theta$ yields
\begin{equation} \label{state-equation}
r\frac{\text{d}}{\text{d}r}{{\mathbf{V}}_{kn}}={{\mathbf{M}}_{kn}}\left( r \right){{\mathbf{V}}_{kn}},\quad k\in \left[ 1,2 \right],
\end{equation}
where ${{\mathbf{V}}_{1n}}={{\left[ {{\Sigma }_{1n}},{{\psi }_{n}} \right]}^{\text{T}}}$, ${{\mathbf{V}}_{2n}}={{\left[ {{\Sigma }_{rn}},{{\Sigma }_{2n}},{{G}_{n}},{{w}_{n}},{{\Delta }_{rn}},{{\Phi }_{n}} \right]}^{\text{T}}},$ and
\begin{equation}
{{\mathbf{M}}_{1n}}=\left[ \begin{matrix}
-{{q}_{7}} & {{c}_{44}}l-{{q}_{1}}-{{q}_{2}}-\rho {{\omega }^{2}}{{r}^{2}}  \\[6pt]
\dfrac{1}{{{c}_{66}}} & \dfrac{{{c}_{69}}}{{{c}_{66}}}  \\
\end{matrix} \right],
\end{equation}
\begin{equation} \label{M2n}
\setlength{\arraycolsep}{3.0pt}
{{\mathbf{M}}_{2n}}=\left[ \begin{matrix}
1 & -\dfrac{{{c}_{69}}}{{{c}_{66}}}l & -\left( {{q}_{1}}+{{q}_{2}}+2{{q}_{3}} \right)l & -{{q}_{1}}l-2\left( {{q}_{2}}+2{{q}_{3}} \right)-\rho {{\omega }^{2}}{{r}^{2}} & 2{{q}_{4}} & -{{q}_{5}}l  \\
1 & -{{q}_{7}} & -{{q}_{3}}l-\rho {{\omega }^{2}}{{r}^{2}}-{{q}_{1}}-{{q}_{2}} & -\left( {{q}_{1}}+{{q}_{2}}+2{{q}_{3}} \right) & {{q}_{4}} & -{{q}_{5}}  \\[4pt]
0 & \dfrac{1}{{{c}_{66}}} & \dfrac{{{c}_{69}}}{{{c}_{66}}} & \dfrac{{{c}_{69}}}{{{c}_{66}}} & 0 & \dfrac{{{e}_{26}}}{{{c}_{66}}}  \\[10pt]
0 & 0 & -l & -2 & 0 & 0  \\[3pt]
0 & -\dfrac{{{e}_{26}}}{{{c}_{66}}}l & -{{q}_{5}}l & -{{q}_{5}}l & -1 & -{{q}_{6}}l  \\
0 & 0 & -{{q}_{4}}l & -2{{q}_{4}} & -\dfrac{1}{{{\varepsilon }_{11}}} & 0  \\
\end{matrix} \right],
\end{equation}
in which $l=n\left( n+1 \right)$. Note that the identity $\nabla _{1}^{2}S_{n}^{m}(\theta ,\varphi )+n\left( n+1 \right)S_{n}^{m}(\theta ,\varphi )=0$ has been used to derive Eq.~\eqref{state-equation}.

Now we divide the deformed SEA balloon (shown in Fig.~\ref{Fig1}(b)) into $N$ equal and thin sublayers, and use $r_{j-1}$ and $r_{j}$ to represent the inner and outer radii of the $j$th sublayer, respectively. Thus, the thickness of the $j$th sublayer is $h_j=r_j-r_{j-1}=h/N$, and the inner and outer radii of the deformed SEA balloon are $a=r_0$ and $b=r_N$. If the thickness of each sublayer is sufficiently small (i.e., the number of sublayers $N$ is sufficiently large), the system matrices $\mathbf{M}_{1n}$ and $\mathbf{M}_{2n}$ within each sublayer may be
envisioned approximately as constant rather than variable. Specifically, we will calculate the values of the material parameters and the (dimensionless) radial coordinate itself at the intermediate surface of each sublayer.

Before proceeding, the following variable substitution for the $j$th sublayer is taken:
\begin{align} \label{transform}
r={{r}_{j-1}}{{e}^{\xi }},\quad \left( j=1,2,\cdots ,N;\text{  } r \in \left[ r_{j-1},{{r}_{j}} \right];\text{  }\xi \in \left[ 0,{{\xi }_{j}} \right] \right),
\end{align}
where  ${\xi }$ denotes the dimensionless logarithm radial coordinate and ${\xi }_{j}=\ln \left( {{r }_{j}}/{{r }_{j-1}} \right)$. From Eq.~\eqref{transform} we have ${\text{d}}/{\text{d}\xi }=r{\text{d}}/{\text{d}r} $. Thus, by introducing the dimensionless incremental state vectors ${{{\mathbf{\overline{V}}}}_{1nj}}=\left[ {{t}_{1n1}},{{t}_{1n2}} \right]_{j}^{\text{T}}$ and ${{{\mathbf{\overline{V}}}}_{2nj}}=\left[ {{t}_{2n1}},{{t}_{2n2}},{{t}_{2n3}},{{t}_{2n4}},{{t}_{2n5}},{{t}_{2n6}} \right]_{j}^{\text{T}}$ with their components defined as
\begin{equation} \label{state-component}
\begin{split}
& {{t}_{1n1}}={{\Sigma }_{1n}}/(a\mu),\quad {{t}_{1n2}}={{\psi }_{n}}/a,\quad {{t}_{2n1}}={{\Sigma }_{rn}}/( a\mu ),\quad{{t}_{2n2}}={{\Sigma }_{2n}}/( a\mu),\\
&{{t}_{2n3}}={{G}_{n}}/a,\quad {{t}_{2n4}}={{w}_{n}}/a,\quad {{t}_{2n5}}={{\Delta }_{rn}}/( a\sqrt{\mu \varepsilon } ),\quad {{t}_{2n6}}={{\Phi }_{n}}/( a\sqrt{\mu /\varepsilon } ),
\end{split}
\end{equation}
and inserting Eqs.~\eqref{transform} and \eqref{state-component} into Eqs.~\eqref{state-equation}-\eqref{M2n}, we obtain the incremental state equations of the  $j$th sublayer as
\begin{equation} \label{dimensionaless-SE}
\frac{\text{d}}{\text{d}\xi }{{{\mathbf{\overline{V}}}}_{knj}}={{{\mathbf{\overline{M}}}}_{knj}}\left( \xi  \right){{{\mathbf{\overline{V}}}}_{knj}},\quad k\in \left[ 1,2 \right], 
\end{equation}
where the dimensionless system matrices are given by
\begin{equation} \label{dimensionlessM12}
\begin{split}
&{{{\mathbf{\overline{M}}}}_{1nj}}=\left[ \begin{matrix}
-{{q}_{7}} & {{{\overline{c}}}_{44}}l-{{{\overline{q}}}_{1}}-{{{\overline{q}}}_{2}}-\varpi _{0}^{2}\kappa _{j-1}^{2}{{e}^{2\xi }}  \\[6pt]
\dfrac{1}{{{{\overline{c}}}_{66}}} & {{\beta }_{1}}  \\
\end{matrix} \right], \quad {{{\mathbf{\overline{M}}}}_{2nj}}= \\[4pt]
& \setlength{\arraycolsep}{2.0pt} \left[ \begin{matrix}
1 & -{{\beta }_{1}}l & -\left( {{{\overline{q}}}_{1}}+{{{\overline{q}}}_{2}}+2{{{\overline{q}}}_{3}} \right)l & -\varpi _{0}^{2}\kappa _{j-1}^{2}{{e}^{2\xi }}-\left[ {{{\overline{q}}}_{1}}l+2\left( {{{\overline{q}}}_{2}}+2{{{\overline{q}}}_{3}} \right) \right] & 2{{{\overline{q}}}_{4}} & -{{{\overline{q}}}_{5}}l  \\[6pt]
1 & -{{q}_{7}} & -\varpi _{0}^{2}\kappa _{j-1}^{2}{{e}^{2\xi }}-\left( {{{\overline{q}}}_{3}}l+{{{\overline{q}}}_{1}}+{{{\overline{q}}}_{2}} \right) & -\left( {{{\overline{q}}}_{1}}+{{{\overline{q}}}_{2}}+2{{{\overline{q}}}_{3}} \right) & {{{\overline{q}}}_{4}} & -{{{\overline{q}}}_{5}}  \\[6pt]
0 & \frac{1}{{{{\overline{c}}}_{66}}} & {{\beta }_{1}} & {{\beta }_{1}} & 0 & {{\beta }_{2}}  \\[6pt]
0 & 0 & -l & -2 & 0 & 0  \\[6pt]
0 & -{{\beta }_{2}}l & -{{{\overline{q}}}_{5}}l & -{{{\overline{q}}}_{5}}l & -1 & -{{{\overline{q}}}_{6}}l  \\[6pt]
0 & 0 & -{{{\overline{q}}}_{4}}l & -2{{{\overline{q}}}_{4}} & -\frac{1}{{{{\overline{\varepsilon }}}_{11}}} & 0  \\
\end{matrix} \right], 
\end{split}
\end{equation}
in which
\begin{equation} \label{66}
\begin{split}
& {{\kappa }_{j-1}}=\frac{{{r}_{j-1}}}{a},\quad {{{\overline{c}}}_{ij}}=\frac{{{c}_{ij}}}{\mu },\quad {{{\overline{\varepsilon }}}_{11}}=\frac{{{\varepsilon }_{11}}}{\varepsilon },\quad {{\beta }_{1}}=\frac{{{c}_{69}}}{{{c}_{66}}},\quad {{\beta }_{2}}=\frac{{{e}_{26}}}{{{c}_{66}}}\sqrt{\frac{\mu }{\varepsilon }}, \\ 
& {{{\overline{q}}}_{1}}=\frac{{{q}_{1}}}{\mu },\quad {{{\overline{q}}}_{2}}=\frac{{{q}_{2}}}{\mu },\quad {{{\overline{q}}}_{3}}=\frac{{{q}_{3}}}{\mu },\quad {{{\overline{q}}}_{4}}={{q}_{4}}\sqrt{\frac{\varepsilon }{\mu }},\quad {{{\overline{q}}}_{5}}=\frac{{{q}_{5}}}{\sqrt{\mu \varepsilon }},\quad {{{\overline{q}}}_{6}}=\frac{{{q}_{6}}}{\varepsilon },
\end{split}
\end{equation}
and
\begin{equation}
{{\varpi }_{0}}=\varpi {{\lambda }_{a}}/\left( \eta -1 \right), 
\end{equation}
with $\varpi =\omega H/\sqrt{\mu /\rho }$ being the dimensionless vibration frequency. Now in each lamina, we have obtained two separated incremental state equations \eqref{dimensionaless-SE} with varying coefficients in a dimensionless form.

The values of the dimensionless radial coordinate $\kappa=r/a$ in the deformed configuration at the inner, outer and intermediate surfaces of the $j$th sublayer are defined, respectively, as
\begin{equation}
\begin{split}
& {{\kappa }_{j-1}}=1+\frac{(j-1)(\overline{\eta }-1)}{N}, \quad {{\kappa }_{j}}=1+\frac{j(\overline{\eta }-1)}{N}, \quad {{\kappa }_{jm}}=1+\frac{(2j-1)(\overline{\eta }-1)}{2N}, 
\end{split}
\end{equation}
the latter two of which, according to Eq.~\eqref{transform}, correspond to the following dimensionless logarithm radial coordinates at the outer and intermediate surfaces of the $j$th sublayer:
\begin{equation}
\begin{split}
& {{\xi }_{j}}=\ln \left( {{\kappa }_{j}}/{{\kappa }_{j-1}} \right)=\ln \frac{1+j(\overline{\eta }-1)/N}{1+(j-1)(\overline{\eta }-1)/N},\\[5pt]
&{{\xi }_{jm}}=\ln \left( {{\kappa }_{jm}}/{{\kappa }_{j-1}} \right)=\ln \frac{1+(2j-1)(\overline{\eta }-1)/(2N)}{1+(j-1)(\overline{\eta }-1)/N}.
\end{split}
\end{equation}

For the $j$th sufficiently thin sublayer, we have ${\text{d}}{{{\mathbf{\overline{V}}}}_{knj}}/{\text{d}\xi }={{{\mathbf{\overline{M}}}}_{knj}}\left( {{\xi }_{jm}} \right){{{\mathbf{\overline{V}}}}_{knj}}$ by taking $\xi={\xi }_{jm}$ in Eq.~\eqref{dimensionaless-SE}, where ${{{\mathbf{\overline{M}}}}_{knj}}({{\xi }_{jm}})$ is the approximated system matrix which is constant within the $j$th sublayer. Consequently, the formal solution can be written as
\begin{equation} \label{formal-solu}
\begin{split}
& {{{\mathbf{\overline{V}}}}_{knj}}(\xi )=\exp \left[ \xi {{{\mathbf{\overline{M}}}}_{knj}}\left( {{\xi }_{jm}} \right) \right]{{{\mathbf{\overline{V}}}}_{knj}}(0),\quad k\in \left[ 1,2 \right], \\ 
& \left( 0\le \xi \le {{\xi }_{j}};\text{   }j=1,2,\cdots N \right).
\end{split}
\end{equation}
Setting $\xi={\xi }_{j}$ in Eq.~\eqref{formal-solu} leads to the relation between the incremental state vectors at the inner and outer surfaces of the $j$th sublayer as
\begin{equation} \label{transfer}
{{\mathbf{\overline{V}}}_{knj}}({{\xi }_{j}})=\exp \left[ {{\xi }_{j}}{{{\mathbf{\overline{M}}}}_{knj}}\left( {{\xi }_{jm}} \right) \right]{{\mathbf{\overline{V}}}_{knj}}(0),\quad k\in \left[ 1,2 \right].
\end{equation}
Taking account of the continuity conditions at the fictitious interfaces between equally divided sublayers that require the eight state variables be continuous, we can finally obtain from Eq.~\eqref{transfer}
\begin{equation} \label{GTM}
\mathbf{\overline{V}}_{kn}^{\text{ou}}={{\mathbf{S}}_{kn}}\mathbf{\overline{V}}_{kn}^{\text{in}},\quad \left( k\in \left[ 1,2 \right];\text{   }n\ge 1 \right), 
\end{equation}
where $\mathbf{\overline{V}}_{kn}^{\text{in}}$ and $\mathbf{\overline{V}}_{kn}^{\text{ou}}$ are the incremental state vectors at the inner and outer surfaces of the deformed SEA balloon, respectively, and ${{\mathbf{S}}_{kn}}=\prod{_{j=N}^{1}\exp \left[ {{\xi }_{j}}{{{\mathbf{\overline{M}}}}_{knj}}\left( {{\xi }_{jm}} \right) \right]}$ are the global transfer matrices of second-order
($k=1$) and sixth-order ($k=2$) through which the boundary state variables at the inner and outer surfaces are connected.

It should be pointed out that, for the breathing mode $n=0$ corresponding to the purely radial vibration, the final expressions of the incremental displacements $u_\theta$, $u_\varphi$ and the generalized incremental
stresses ${{\Sigma }_{r\theta}}$, ${{\Sigma }_{r\varphi}}$ all vanish, as can be seen from Eqs.~\eqref{52} and \eqref{series-solution}. As a result, $\psi_0$, $G_0$, ${\Sigma }_{10}$ and ${\Sigma }_{20}$  can be assumed zero and Eqs.~\eqref{dimensionaless-SE} and \eqref{dimensionlessM12} degenerate to the following incremental state equation:
\begin{equation} \label{breath}
\begin{split}
& \frac{\text{d}}{\text{d}\xi }{{{\mathbf{\overline{V}}}}_{20j}}={{{\mathbf{\overline{M}}}}_{20j}}\left( \xi  \right){{{\mathbf{\overline{V}}}}_{20j}},\quad {{{\mathbf{\overline{V}}}}_{20j}}=\left[ {{t}_{201}},{{t}_{204}},{{t}_{205}},{{t}_{206}} \right]_{j}^{\text{T}}, \\ 
& {{{\mathbf{\overline{M}}}}_{20j}}\left( \xi  \right)=\left[ \begin{matrix}
1 & -\varpi _{0}^{2}\kappa _{j-1}^{2}{{e}^{2\xi }}-2\left( {{{\overline{q}}}_{2}}+2{{{\overline{q}}}_{3}} \right) & 2{{{\overline{q}}}_{4}} & 0  \\[6pt]
0 & -2 & 0 & 0  \\[6pt]
0 & 0 & -1 & 0  \\[6pt]
0 & -2{{{\overline{q}}}_{4}} & -\frac{1}{{{{\overline{\varepsilon }}}_{11}}} & 0 
\end{matrix} \right], 
\end{split}
\end{equation}
Following the previous derivation, from Eq.~\eqref{breath} we can obtain
\begin{equation} \label{breathing}
\begin{split}
& {{\left[ t_{201}^{\text{ou}},t_{204}^{\text{ou}},t_{205}^{\text{ou}},t_{206}^{\text{ou}} \right]}^{\text{T}}}={{\mathbf{S}}_{20}}{{\left[ t_{201}^{\text{in}},t_{204}^{\text{in}},t_{205}^{\text{in}},t_{206}^{\text{in}} \right]}^{\text{T}}},\quad \left( n=0 \right),
\end{split}
\end{equation}
where ${{\mathbf{S}}_{20}}=\prod{_{j=N}^{1}\exp \left[ {{\xi }_{j}}{{{\mathbf{\overline{M}}}}_{20j}}\left( {{\xi }_{jm}} \right) \right]}$ is a fourth-order global transfer matrix of the deformed SEA balloon.

\subsection{Incremental boundary conditions and frequency equations}
\label{5.2}

For the incremental free vibrations of the SEA spherical balloon, there are also no incremental electric fields outside the balloon and its outer surface is also free from mechanical traction. Moreover, if we assume that the applied internal pressure $p_a=P$ and electric potential difference keep unchanged during the incremental motion, the corresponding incremental mechanical and electric boundary conditions \eqref{s1-incremental-boundary}$_{1,3}$ and \eqref{Pcondition} become
\begin{equation} \label{MEBC}
\begin{split}
& \left\{
\begin{array}{lr}
{{{\dot{T}}}_{0rr}}=P\dfrac{\partial {{u}_{r}}}{\partial r},\text{ } {{{\dot{T}}}_{0r\theta }}=P\dfrac{1}{r}\left( \dfrac{\partial {{u}_{r}}}{\partial \theta }-{{u}_{\theta }} \right), & \\[10pt] {{{\dot{T}}}_{0r\varphi }}=P\dfrac{1}{r}\left( \dfrac{1}{\sin \theta }\dfrac{\partial {{u}_{r}}}{\partial \varphi }-{{u}_{\varphi }} \right), \quad \dot{\phi }=0, & \left( r=a \right);
\end{array}
\right. \\ 
& \quad \,\,\,{{{\dot{T}}}_{0rr}}={{{\dot{T}}}_{0r\theta }}={{{\dot{T}}}_{0r\varphi }}=0,\quad \dot{\phi }=0,\quad \left( r=b \right), 
\end{split}
\end{equation}
which, combined with Eq.~\eqref{52} and the relation ${\Sigma }_{rr}=r{{\dot{T}}}_{0rr}$, yields
\begin{equation} \label{IncreBound}
\begin{split}
& {{\Sigma }_{rr}}=P\left( \nabla _{1}^{2}G-2w \right),\text{ } {{\Sigma }_{1}}=-P\psi,\text{ }{{\Sigma }_{2}}=-P\left( w+G \right),\text{ } \Phi =0, \quad \left( r=a \right); \\ 
& {{\Sigma }_{rr}}={{\Sigma }_{1}}={{\Sigma }_{2}}=\Phi =0, \quad \left( r=b \right)
\end{split}
\end{equation}
where we have utilized the incremental incompressibility constraint \eqref{40}. Substituting Eq.~\eqref{series-solution} into Eq.~\eqref{IncreBound} and using Eq.~\eqref{state-component}, we have the dimensionless form of the incremental boundary conditions as follows:
\begin{align} \label{IBC}
& {{t}_{2n1}}=-\overline{P}\left( l{{t}_{2n3}}+2{{t}_{2n4}} \right),\text{ }{{t}_{1n1}}=-\overline{P}{{t}_{1n2}},\text{ }{{t}_{2n2}}=-\overline{P}\left( {{t}_{2n4}}+{{t}_{2n3}} \right),\text{ }{{t}_{2n6}}=0, \quad \left( r=a \right); \notag \\ 
& {{t}_{2n1}}={{t}_{1n1}}={{t}_{2n2}}={{t}_{2n6}}=0,\quad \left( r=b \right). 
\end{align}

Utilizing the incremental boundary conditions \eqref{IBC} in Eq. \eqref{GTM}, we acquire two sets of independent linear algebraic equations as
\begin{equation} \label{LFE}
\left( {{S}_{1n12}}-\overline{P}{{S}_{1n11}} \right)t_{1n2}^{\text{in}}=0,\quad {{\mathbf{S}}^{*}}{{\left[ t_{2n3}^{\text{in}},t_{2n4}^{\text{in}},t_{2n5}^{\text{in}} \right]}^{\text{T}}}=\mathbf{0},\quad \left( n\ge 1 \right),
\end{equation}
where ${\mathbf{S}}^{*}$ is the coefficient matrix of third-order, whose elements are given by
\begin{align}
& S_{j1}^{*}={{S}_{2nj3}}-l\overline{P}{{S}_{2nj1}}-\overline{P}{{S}_{2nj2}},\ S_{j2}^{*}={{S}_{2nj4}}-2\overline{P}{{S}_{2nj1}}-\overline{P}{{S}_{2nj2}},\ S_{j3}^{*}={{S}_{2nj5}},\ (j=1,2), \notag \\ 
& S_{31}^{*}={{S}_{2n63}}-l\overline{P}{{S}_{2n61}}-\overline{P}{{S}_{2n62}},\ S_{32}^{*}={{S}_{2n64}}-2\overline{P}{{S}_{2n61}}-\overline{P}{{S}_{2n62}},\ S_{33}^{*}={{S}_{2n65}},
\end{align}
in which ${{S}_{knij}}$ are the elements of the global transfer matrix ${{\mathbf{S}}_{kn}}$. The non-trivial solutions of Eq.~\eqref{LFE} give
\begin{equation} \label{freEQ}
{{S}_{1n12}}-\overline{P}{{S}_{1n11}}=0,\quad \left| {{\mathbf{S}}^{*}} \right|=0,\quad \left( n\ge 1 \right),
\end{equation}
which provide the characteristic frequency equations of two independent classes of free vibration of the deformed SEA balloon under radially inhomogeneous biasing fields for angular mode numbers $n\ge 1$.

For the breathing mode $n=0$, the necessary incremental boundary conditions \eqref{MEBC} and \eqref{IBC} reduce to 
\begin{equation}
\begin{split}
& {{{\dot{T}}}_{0rr}}=P\frac{\partial {{u}_{r}}}{\partial r},\text{ }\dot{\phi }=0;\quad {{t}_{201}}=-2\overline{P}{{t}_{204}},\text{ } {{t}_{206}}=0, \quad \left( r=a \right); \\ 
& {{{\dot{T}}}_{0rr}}=\dot{\phi }=0;\quad {{t}_{201}}={{t}_{206}}=0, \quad \left( r=b \right),
\end{split}
\end{equation}
which, together with Eq.~\eqref{breathing}, results in the following frequency equation:
\begin{equation} \label{freEQ-breath}
\left[ \begin{matrix}
{{S}_{2012}}-2\overline{P}{{S}_{2011}} & {{S}_{2013}}  \\
{{S}_{2042}}-2\overline{P}{{S}_{2041}} & {{S}_{2043}}  \\
\end{matrix} \right]\left[ \begin{matrix}
{t_{204}^{\text{in}}}  \\ {t_{205}^{\text{in}}}   \\
\end{matrix} \right]=0, \quad
\left| \begin{matrix}
{{S}_{2012}}-2\overline{P}{{S}_{2011}} & {{S}_{2013}}  \\
{{S}_{2042}}-2\overline{P}{{S}_{2041}} & {{S}_{2043}}  \\
\end{matrix} \right|=0,\quad \left( n=0 \right).
\end{equation}

In view of the state equations and the boundary conditions for the incremental fields obtained above, the 3D small-amplitude free vibrations of a SEA balloon under biasing fields may be divided into two independent classes, just like the classical linear elastic and piezoelectric cases without biasing fields \citep{lamb1881vibrations,chen2001free, wu2018free}: the first class ($n\ge1$) with the frequency equation \eqref{freEQ}$_1$ is described by ${{{\mathbf{\overline{V}}}}_{1n}}$ and ${{{\mathbf{\overline{M}}}}_{1n}}$, corresponds to torsional (or toroidal) modes, and belongs to an equi-volumetric motion of the SEA balloon without the radial displacement component, while the second class ($n\ge0$) with the frequency equations \eqref{freEQ}$_2$ and \eqref{freEQ-breath}$_2$ is governed by ${{{\mathbf{\overline{V}}}}_{2n}}$ and ${{{\mathbf{\overline{M}}}}_{2n}}$, whose mechanical displacements generally possess both transverse and radial components except for the breathing mode $n=0$, for which there exists only the radial component. The second class of vibration corresponds to the spheroidal modes including, for instances, the breathing ($n=0$), dipolar ($n=1$) and quadrupolar ($n=2$) modes.

In addition, it is emphasized here that, the resulting characteristic frequency equations \eqref{freEQ} and \eqref{freEQ-breath}$_2$ do not include the azimuthal integer $m$ representing the non-axisymmetric vibration modes, which is physically understandable since any non-axisymmetric vibration mode can be obtained by the superposition of different axisymmetric ones with an identical resonant frequency with respect to differently oriented spherical coordinates \citep{chen2001free}. Thus, we can only consider the axisymmetric modes with $m=0$ in the numerical calculations to be conducted in Sec.~\ref{Sec6}.

Once the vibration frequencies for a given angular mode number $n$ are solved numerically from Eqs.~\eqref{freEQ} and \eqref{freEQ-breath}$_2$, the incremental state vectors at the inner spherical surface can be obtained from Eqs.~\eqref{LFE} and \eqref{IBC}$_1$. Then we can calculate the incremental state vectors at any interior point in the SEA balloon by using the following formula:
\begin{equation}
\begin{split}
& {{{\mathbf{\overline{V}}}}_{knj}}(\xi ) =\exp \left[\xi {{\mathbf{\overline{M}}}_{knj}}\left( {{\xi }_{jm}} \right)\right] \prod\limits_{i=j-1}^{1}\exp \left[{{\xi }_{i}}{{{\mathbf{\overline{M}}}}_{kni}}\left( {{\xi }_{im}} \right)\right]\mathbf{\overline{V}}_{kn}^{\text{in}}, \\ 
& \left(k\in \left[ 1,2 \right];\quad 0\le \xi \le {{\xi }_{j}}; \quad j=1,2,\cdots N \right).
\end{split}
\end{equation}

The previous formulations in this section are valid for an incompressible isotropic SEA balloon modelled by a general form of energy density function. For the Gent ideal dielectric model \eqref{Gent}, we have non-zero derivatives of the energy function with respect to invariants as
\begin{equation}
{{\Omega }_{1}}=\frac{\mu G}{2\left( G-{{I}_{1}}+3 \right)},\quad {{\Omega }_{11}}=\frac{\mu G}{2{{\left( G-{{I}_{1}}+3 \right)}^{2}}},\quad {{\Omega }_{5}}=\frac{1}{2\varepsilon }, 
\end{equation}
which are substituted into \ref{AppeA} to evaluate the non-zero components of the instantaneous electroelastic moduli tensors as
\begin{align} \label{85}
& {{\mathcal{A}}_{01111}}=\frac{\mu G}{{{\lambda }^{8}}\left( G-{{I}_{1}}+3 \right)}\left( \frac{2}{G-{{I}_{1}}+3}+{{\lambda }^{4}} \right)+\frac{D_{r}^{2}}{\varepsilon },\quad {{\mathcal{A}}_{02233}}=\frac{2\mu G{{\lambda }^{4}}}{{{\left( G-{{I}_{1}}+3 \right)}^{2}}}, \notag \\
&{{\mathcal{A}}_{02222}}={{\mathcal{A}}_{03333}}=\frac{\mu G{{\lambda }^{2}}}{G-{{I}_{1}}+3}\left( 1+\frac{2}{G-{{I}_{1}}+3}{{\lambda }^{2}} \right), \quad {{\mathcal{A}}_{01221}}={{\mathcal{A}}_{01331}}={{\mathcal{A}}_{02332}}=0, \notag \\ 
& {{\mathcal{A}}_{01122}}={{\mathcal{A}}_{01133}}=\frac{2\mu G}{{{\lambda }^{2}}{{\left( G-{{I}_{1}}+3 \right)}^{2}}}, \quad {{\mathcal{A}}_{01212}}={{\mathcal{A}}_{01313}}=\frac{\mu G}{{{\lambda }^{4}}\left( G-{{I}_{1}}+3 \right)}+\frac{D_{r}^{2}}{\varepsilon }, \notag \\ 
&{{\mathcal{A}}_{02121}}={{\mathcal{A}}_{03131}}={{\mathcal{A}}_{02323}}={{\mathcal{A}}_{03232}}=\frac{\mu G{{\lambda }^{2}}}{G-{{I}_{1}}+3}, \quad {{\mathcal{M}}_{0221}}={{\mathcal{M}}_{0331}}=0, \notag \\ 
& {{\mathcal{M}}_{0111}}=2{{\varepsilon }^{-1}}{{D}_{r}},\quad {{\mathcal{M}}_{0122}}={{\mathcal{M}}_{0133}}={{\varepsilon }^{-1}}{{D}_{r}},\quad {{\mathcal{R}}_{011}}={{\mathcal{R}}_{022}}={{\mathcal{R}}_{033}}={{\varepsilon }^{-1}}. 
\end{align}
Thus, by defining the dimensionless quantities ${{{\overline{\varepsilon }}}_{ij}}={{{\varepsilon }_{ij}}}/{\varepsilon }$, ${{{\overline{e}}}_{ij}}={{{e}_{ij}}}/{\sqrt{\mu \varepsilon }}$ and ${{{\overline{c}}}_{ij}}={{{c}_{ij}}}/{\mu },$ and using Eq.~\eqref{85}, we can obtain the material parameters defined in Eqs.~\eqref{materialparameter1} and \eqref{66} as
\begin{equation}
\begin{split}
& {{{\overline{\varepsilon }}}_{11}}=1,\text{ } {{{\overline{\varepsilon }}}_{22}}=1, \text{ } {{{\overline{e}}}_{11}}=-2{{{\overline{D}}}_{r}},\text{ } {{{\overline{e}}}_{12}}=0,\text{ } {{{\overline{e}}}_{26}}=-{{{\overline{D}}}_{r}},\text{ } {{{\overline{c}}}_{12}}=\frac{2G}{{{\lambda }^{2}}{{\left( G-{{I}_{1}}+3 \right)}^{2}}}, \\
& {{{\overline{c}}}_{11}}=\frac{G}{{{\lambda }^{8}}\left( G-{{I}_{1}}+3 \right)}\left( \frac{2}{G-{{I}_{1}}+3}+{{\lambda }^{4}} \right)-3\overline{D}_{r}^{2}+\overline{p}, \quad {{{\overline{c}}}_{23}}=\frac{2G{{\lambda }^{4}}}{{{\left( G-{{I}_{1}}+3 \right)}^{2}}}, \\ 
& {{{\overline{c}}}_{22}}=\frac{G{{\lambda }^{2}}}{G-{{I}_{1}}+3}\left( 1+\frac{2{{\lambda }^{2}}}{G-{{I}_{1}}+3} \right)+\overline{p}, \quad {{{\overline{c}}}_{66}}=\frac{G}{{{\lambda }^{4}}\left( G-{{I}_{1}}+3 \right)},  \\ 
& {{{\overline{c}}}_{69}}=-\overline{D}_{r}^{2}+\overline{p},\quad {{{\overline{c}}}_{99}}=\frac{G{{\lambda }^{2}}}{G-{{I}_{1}}+3}-\overline{D}_{r}^{2},\quad {{{\overline{c}}}_{47}}=\overline{p},\quad {{{\overline{c}}}_{44}}=\frac{G{{\lambda }^{2}}}{G-{{I}_{1}}+3},
\end{split}
\end{equation}
and
\begin{equation}
\begin{split}
&{{{\overline{q}}}_{1}}=\frac{\overline{c}_{69}^{2}}{{{{\overline{c}}}_{66}}}-{{{\overline{c}}}_{99}}, \text{ } {{{\overline{q}}}_{2}}={{{\overline{c}}}_{44}}+{{{\overline{c}}}_{47}},\text{ } {{{\overline{q}}}_{3}}=2{{{\overline{c}}}_{12}}-{{{\overline{c}}}_{11}}-{{{\overline{c}}}_{22}}-\frac{{{\left( {{{\overline{e}}}_{11}}-{{{\overline{e}}}_{12}} \right)}^{2}}}{{{{\overline{\varepsilon }}}_{11}}}, \text{ } {{\beta }_{1}}=\frac{{{{\overline{c}}}_{69}}}{{{{\overline{c}}}_{66}}}, \\
&{{{\overline{q}}}_{4}}=\frac{{{{\overline{e}}}_{11}}-{{{\overline{e}}}_{12}}}{{{{\overline{\varepsilon }}}_{11}}},\text{ } {{{\overline{q}}}_{5}}=\left( \frac{{{{\overline{c}}}_{69}}}{{{{\overline{c}}}_{66}}}-1 \right){{{\overline{e}}}_{26}}, \text{ } {{{\overline{q}}}_{6}}=\frac{\overline{e}_{26}^{2}}{{{{\overline{c}}}_{66}}}+{{{\overline{\varepsilon }}}_{22}},\text{ } {{q}_{7}}=1+\frac{{{{\overline{c}}}_{69}}}{{{{\overline{c}}}_{66}}},\text{ } {{\beta }_{2}}=\frac{{{{\overline{e}}}_{26}}}{{{{\overline{c}}}_{66}}}. 
\end{split}
\end{equation}

As mentioned earlier, the above expressions can be degenerated to those for the neo‐Hookean ideal dielectric model by simply taking the limit of $G \rightarrow \infty $.


\section{Numerical results and discussions} \label{Sec6}


\subsection{Nonlinear static response}

Before studying the 3D free vibration characteristics of SEA spherical balloons, we first show the spherically symmetric nonlinear static response under the combined action of electric voltage and internal pressure. The results presented in this subsection will be used to clearly reveal some unique correlations between the resonant frequency and the electromechanical biasing fields.

\begin{figure}[htbp]
	\centering	
	\includegraphics[width=0.49\textwidth]{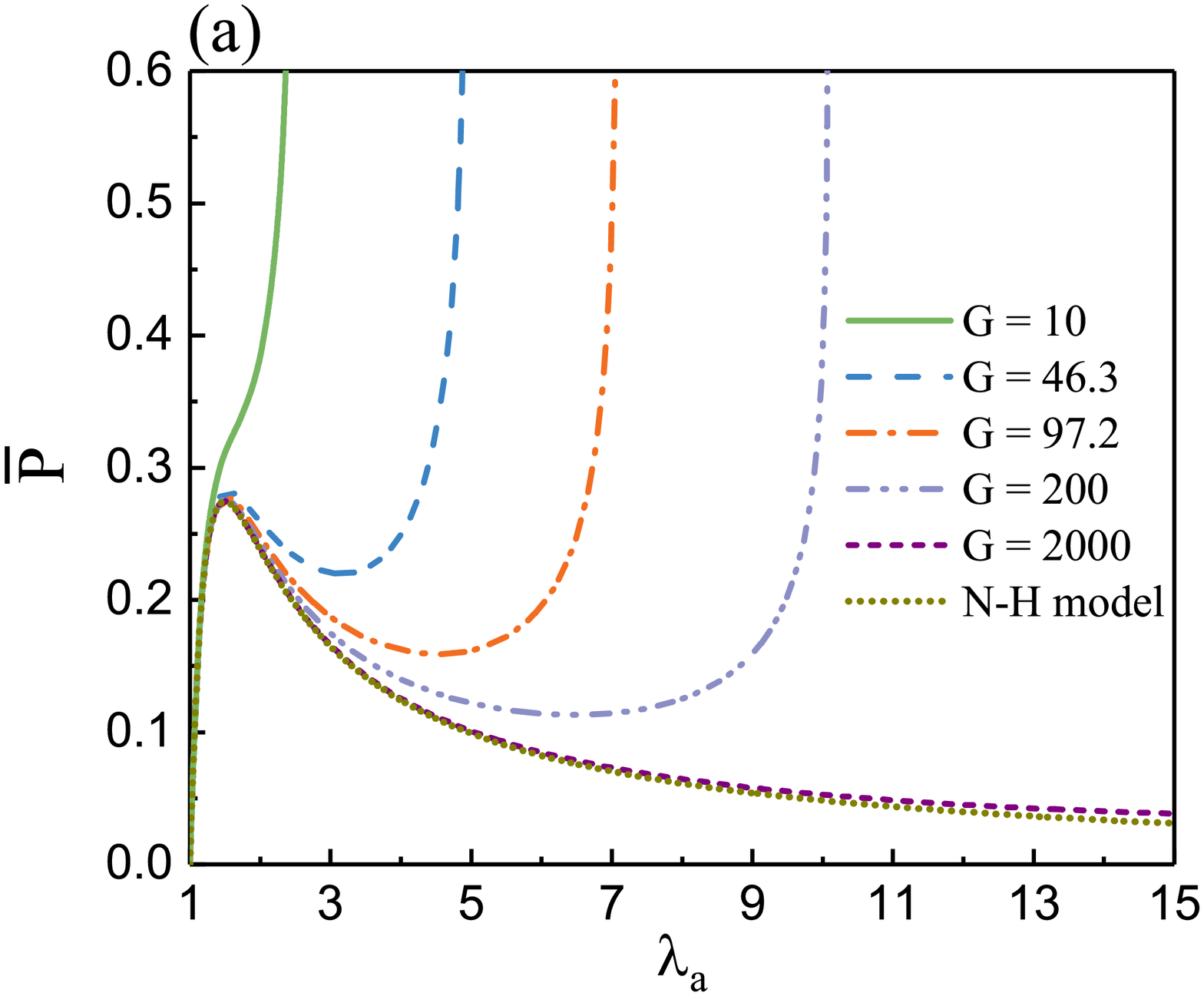}
	\includegraphics[width=0.49\textwidth]{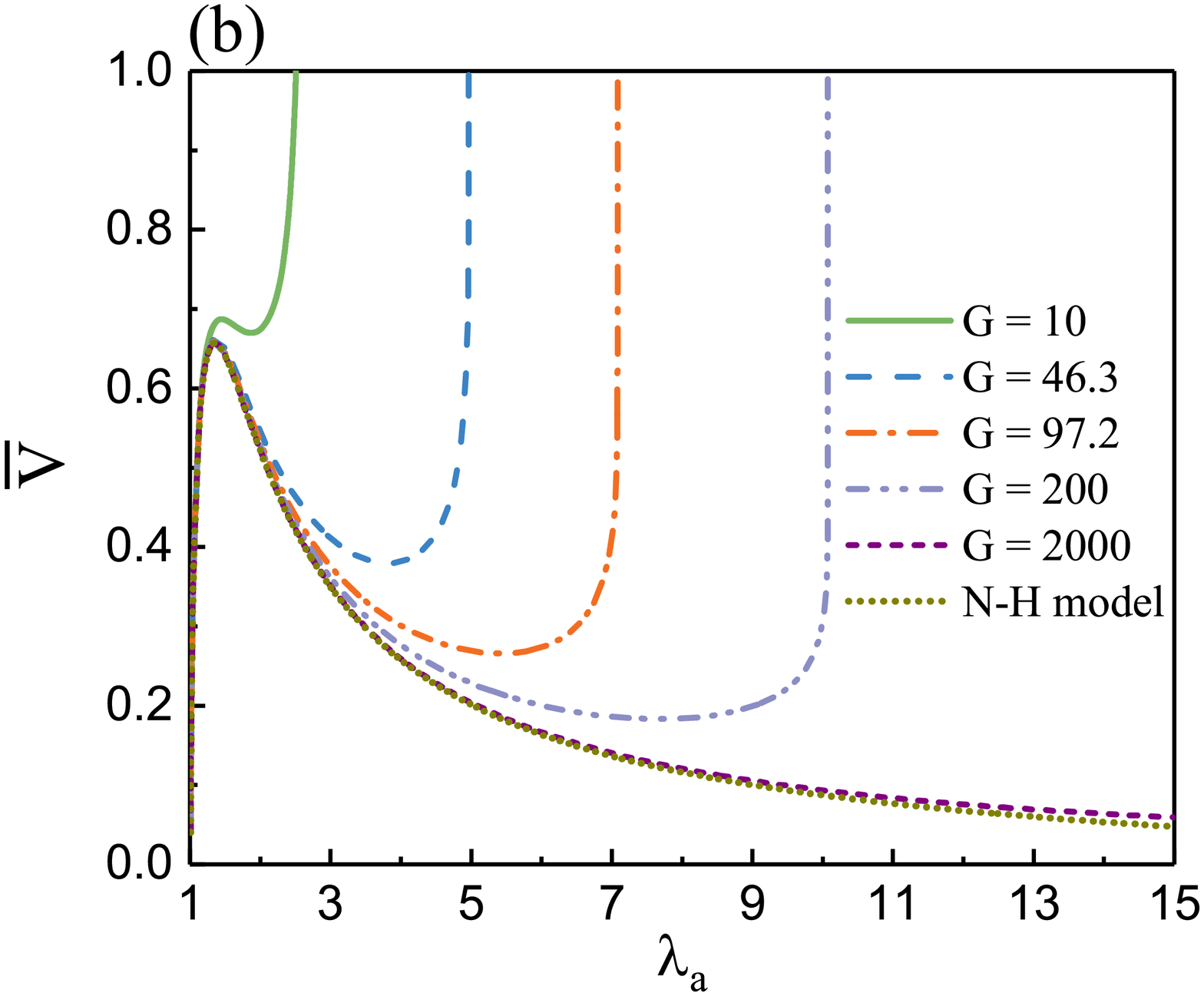}
    \caption{Variations of the dimensionless internal pressure $\overline{P}$ for a fixed electric voltage $\overline{V}=0.02$ (a) and electric voltage $\overline{V}$ for a prescribed internal pressure $\overline{P}=0.02$ (b) with the stretch ratio $\lambda_a$ in a  moderately thick SEA balloon ($\eta=1.25$) for different values of Gent constant $G$.}
	\label{Fig2}
\end{figure}

For a \textit{moderately thick} balloon ($\eta=1.25$) characterized by the Gent ideal dielectric model with different Gent constants $G$, Fig.~\ref{Fig2}(a) displays the dimensionless inflation pressure $\overline{P}$ versus the stretch ratio $\lambda_a$ for a fixed electric voltage $ \overline{V}=0.02 $, while Fig.~\ref{Fig2}(b) shows the dimensionless electric voltage $\overline{V}$ versus the stretch ratio $\lambda_a$ for a prescribed internal pressure $ \overline{P}=0.02 $. The results for the neo-Hookean model are also shown in Fig.~\ref{Fig2} for comparison purpose. Note that the specific value of $G$ found by \citet{gent1996new} to be suitable for the \textit{unfilled rubber vulcanizate} is $G=97.2$, which was also adopted by \citet{dorfmann2014nonlinear2}; for \textit{Silicone CF19-2186} from the manufacturer Nusil$^{\textrm{TM}}$ (Nusil$^{\textrm{TM}}$ Technology LLC, Carpinterio, USA), \citet{GETZ201724} utilized the experimental elongation at failure to first estimate the limiting stretch and then calculate the Gent constant as $G=46.3$. Therefore, the chosen $G=97.2$ and $46.3$ are assumed to be reasonable for our numerical calculations. As described in Subsec.~\ref{section3.2}, the Gent model accounts for the strain-stiffening effect in the inflation procedure of the spherical balloon and can degenerate to the neo-Hookean model when $G \rightarrow \infty $, which is clearly indicated in Fig.~\ref{Fig2}. Note that the snap-through phenomenon is observed for the Gent model with $G \le 200$, i.e., after a critical value of internal pressure or electric voltage is reached with the inflation of the SEA balloon, a sudden transition to a new equilibrium state occurs. In cases of fixed internal pressure and electric voltage, the critical values will not have an obvious change when the Gent constant decreases from 2000 to 46.3. But for small values of $G$, the critical voltage value inceases with the decreasing $G$ since the strain-stiffening effect is enhanced. Interestingly, the curve of $\overline{P}$ versus $\lambda_a$ for $G = 10$ in Fig.~\ref{Fig2}(a) exhibits a monotonically increasing variation owing to extremely strong strain-stiffening effect. In addition, the Gent model has an asymptote ($\overline{P} \rightarrow \infty $ or $\overline{V}\rightarrow \infty$) at limiting stretch ratios that can be determined by Eq.~\eqref{limiting}.

\begin{figure}[htbp]
	\centering	
	\includegraphics[width=0.49\textwidth]{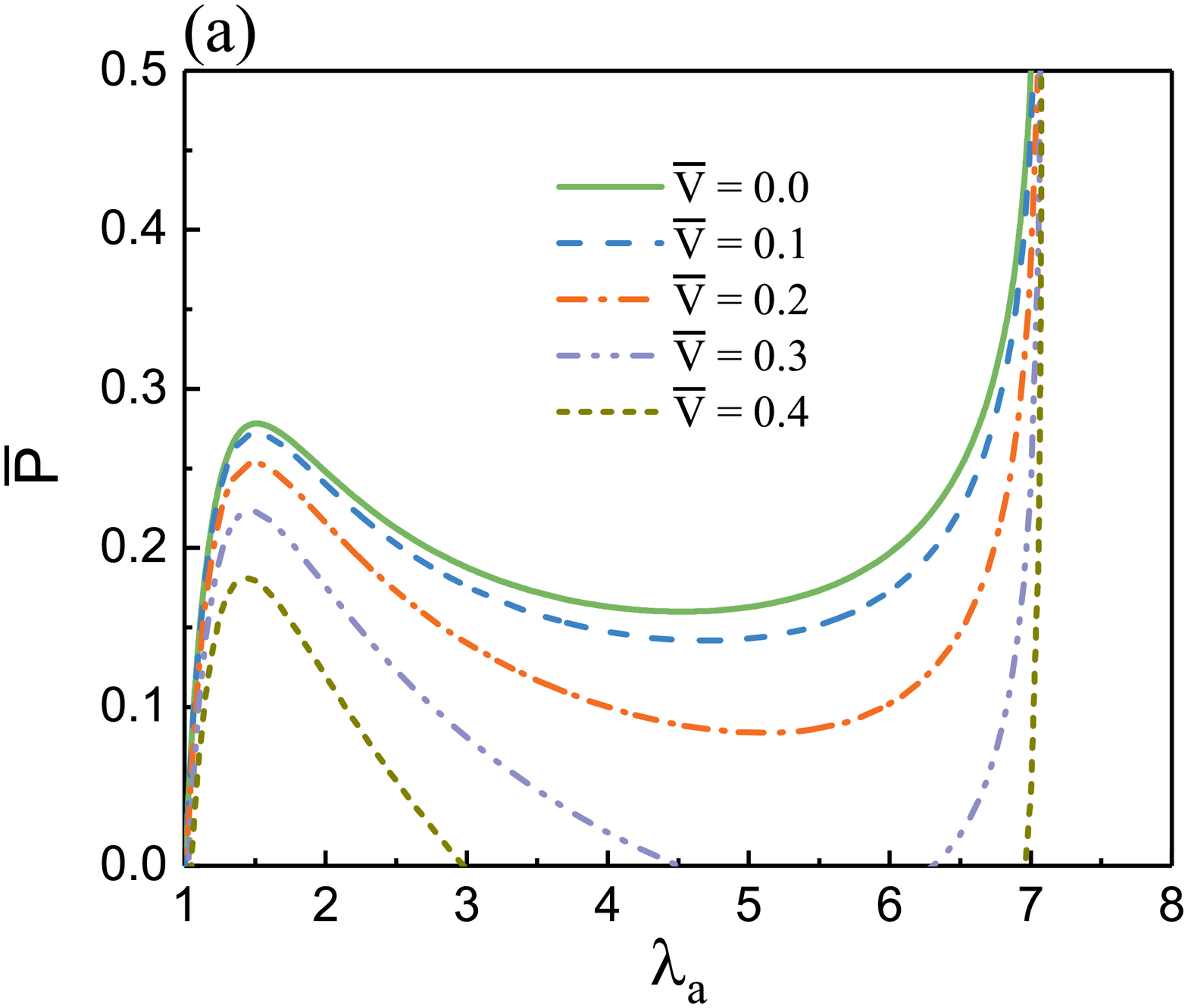}
	\includegraphics[width=0.49\textwidth]{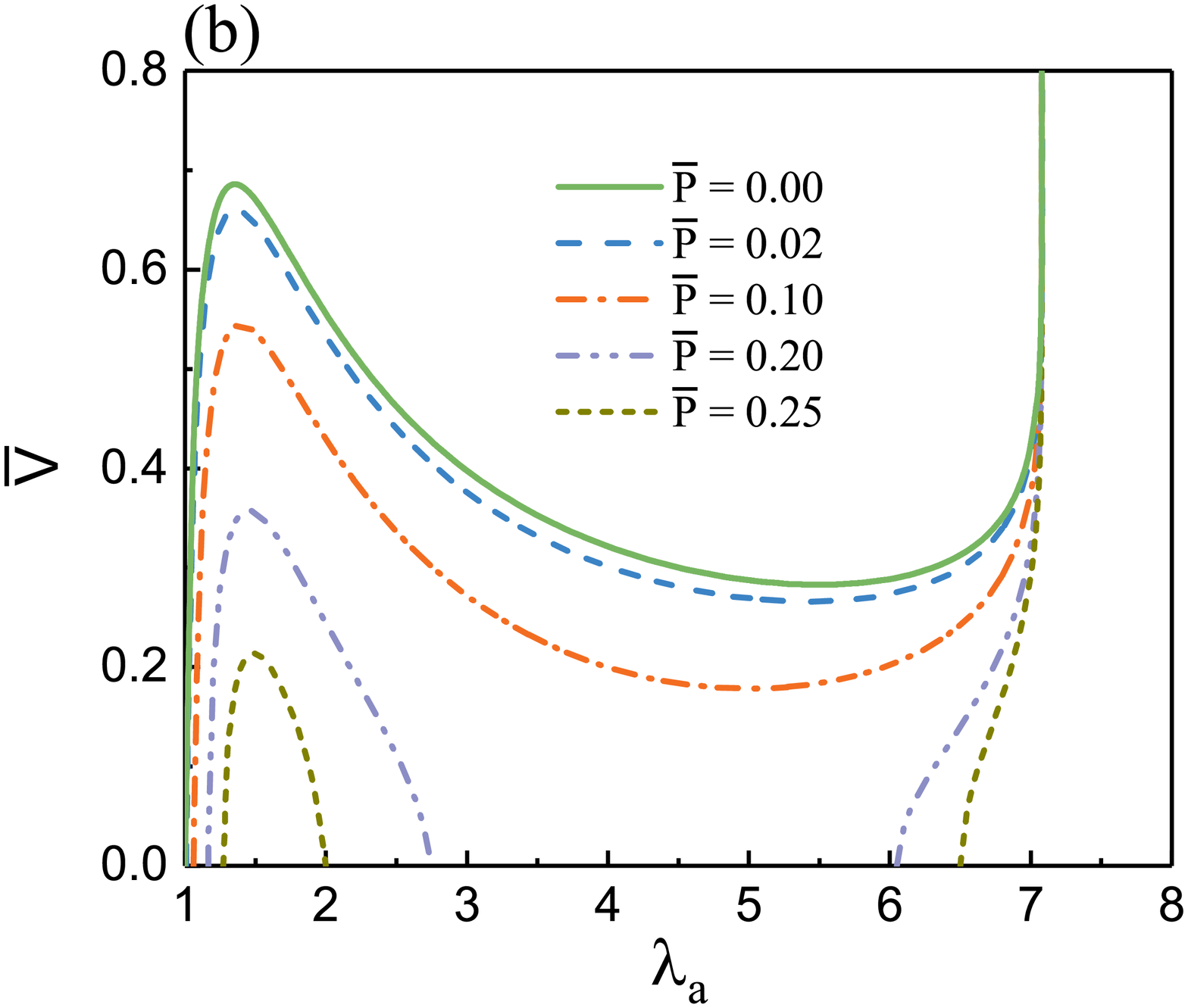}
    \caption{Variations of the dimensionless internal pressure $\overline{P}$ (a) or electric voltage $\overline{V}$ (b) with the stretch ratio $\lambda_a$ in a moderately thick SEA balloon ($\eta=1.25$) with Gent constant $G=97.2$ for different values of electric voltage (a) or of internal pressure (b).}
	\label{Fig3}
\end{figure}

For a moderately thick balloon with a given Gent constant $G=97.2$, results for the variations of internal pressure (or electric voltage) with the stretch ratio are depicted in Fig.~\ref{Fig3}(a) (or Fig.~\ref{Fig3}(b)) by adjusting the values of electric voltage (or internal pressure). As we can see from Fig.~\ref{Fig3}(a) that, for the SEA balloon pre-expanded electrically, the critical pressure value at which snap-through is triggered decreases when the applied electric voltage has a rise. Similarly, Fig.~\ref{Fig3}(b) shows that the critical voltage value at which the pre-inflated balloon snaps decreases with the increasing internal pressure. However, the fixed Gent constant $G=97.2$ leads to the same limiting stretch ratio $\lambda_a=7.08$ for all cases. These phenomena are qualitatively the same as those observed by \citet{rudykh2012snap} and \citet{dorfmann2014nonlinear2}.

\subsection{Small-amplitude free vibration analysis}

We now conduct numerical calculations to quantitatively investigate the effects of electromechanical biasing fields and structural parameters on the 3D small-amplitude free vibration characteristics of the SEA balloon. It should be mentioned that since the frequency equations \eqref{freEQ} and \eqref{freEQ-breath}$_2$ are derived for 3D motions, there exist an infinite number of resonant frequencies (i.e., the number of radial nodes or the harmonic order in Lamb's theory \citep{lamb1881vibrations} varies) for each angular mode number $n$. Furthermore, there are two zero frequencies when $n=1$, of which one is due to a rigid-body rotation in the first class of vibration and the other is caused by a rigid-body translation in the second class of vibration \citep{ding2006elasticity}. In what follows, we will only pay attention to the smallest nonzero root (i.e., the first-order radial mode) for each angular mode number $n$ and the higher order ones can be readily obtained if needed.

\subsubsection{Validation of the proposed approach}

The effectiveness of the proposed SSM will be first verified in terms of its accuracy and convergence for the free vibration analysis of the SEA balloon.

\begin{table}[htb]
	\centering
	\begin{tabular}{ccccccccc}
		\hline
		\hline \specialrule{0em}{0pt}{4pt}
		$ \eta=1.25$ & $n$ & 0 & 1 & 2 & 3 & 4 & 5 & 6   \\
		\specialrule{0em}{0pt}{2pt} \hline \specialrule{0em}{0pt}{8pt}
		SSM & First  & --- & 3.18862 & 0.43989 & 0.69537 & 0.93265 & 1.16278 & 1.38916  \\
		($N=40$) & Second  & 0.78102 & 0.93144 & 0.28260 & 0.39519 & 0.53107 & 0.69951 & 0.89153 \\
		\specialrule{0em}{0pt}{2pt} \hline \specialrule{0em}{0pt}{8pt}
		SSM & First  & --- & 3.18861 & 0.43989 & 0.69536 & 0.93264 & 1.16277 & 1.38915  \\
		($N=50$) & Second  & 0.78102 & 0.93144 & 0.28260 & 0.39519 & 0.53107 & 0.69951 & 0.89153  \\
		\specialrule{0em}{0pt}{4pt} \hline \specialrule{0em}{0pt}{6pt}
		Exact & First  & --- & 3.18860 & 0.43989 & 0.69536 & 0.93264 & 1.16277 & 1.38914 \\
		Solution & Second  & 0.78102 & 0.93144 & 0.28260 & 0.39519 & 0.53107 & 0.69951 & 0.89153  \\
		\specialrule{0em}{0pt}{4pt}
		\hline
		\hline
	\end{tabular}
	\caption{Accuracy and convergency analysis of the dimensionless resonant frequencies $\varpi =\omega H/\sqrt{\mu /\rho }$ of two classes of vibration predicted by the SSM for a \textit{moderately thick} SEA balloon ($\eta=1.25$) compared with the exact theoretical solutions made by \citet{shah1969three1, shah1969three2}.}
	\label{Table1}
\end{table}

\begin{table}[htb]
	\centering
	\begin{tabular}{ccccccccc}
		\hline
		\hline \specialrule{0em}{0pt}{4pt}
		$ \eta=3.00 $ & $n$ & 0 & 1 & 2 & 3 & 4 & 5 & 6   \\
		\specialrule{0em}{0pt}{2pt} \hline \specialrule{0em}{0pt}{8pt}
		SSM & First  & --- & 4.04127 & 1.65962 & 2.57348 & 3.39555 & 4.17704 & 4.93582  \\
		($N=90$) & Second  & 4.80717 & 2.79937 & 1.58438 & 2.55895 & 3.40048 & 4.15251 & 4.85991 \\
		\specialrule{0em}{0pt}{2pt} \hline \specialrule{0em}{0pt}{8pt}
		SSM & First  & --- & 4.04130 & 1.65960 & 2.57344 & 3.39548 & 4.17694 & 4.93571  \\
		($N=100$) & Second  & 4.80724 & 2.79938 & 1.58438 & 2.55895 & 3.40046 & 4.15249 & 4.85988  \\
		\specialrule{0em}{0pt}{4pt} \hline \specialrule{0em}{0pt}{6pt}
		Exact & First  & --- & 4.04132 & 1.65958 & 2.57342 & 3.39546 & 4.17691 & 4.93566 \\
		Solution & Second  & 4.80730 & 2.79939 & 1.58438 & 2.55894 & 3.40045 & 4.15247 & 4.85985  \\
		\specialrule{0em}{0pt}{4pt}
		\hline
		\hline
	\end{tabular}
	\caption{Accuracy and convergency analysis of the dimensionless resonant frequencies $\varpi =\omega H/\sqrt{\mu /\rho }$ of two classes of vibration predicted by the SSM for a \textit{thick} SEA balloon ($\eta=3.00$) compared with the exact theoretical solutions made by \citet{shah1969three1, shah1969three2}.}
	\label{Table2}
\end{table}

If there is no applied internal pressure and electric voltage, the SEA balloon has no deformation and behaves like an incompressible isotropic material. For this case, the analytical frequency equations for the two classes of free vibration were derived by \citet{shah1969three1, shah1969three2}. Therefore, for some angular mode numbers $n$, Tables \ref{Table1} and \ref{Table2} compare the numerical results of SSM based on different sublayer numbers $N$ with the exact solutions by \citet{shah1969three1, shah1969three2} for moderately thick ($\eta=1.25$) and thick ($\eta=3.00$) SEA balloons, respectively. Note that the Poisson's ratio should be set to approach $0.5$ in these exact solutions for incompressible materials. Table \ref{Table1} for $\eta=1.25$ shows that the SSM prediction results for $N=40$ and $50$ are very close and both agree exceptionally well with the 3D exact solutions \citep{shah1969three1, shah1969three2}. Similar analysis can be conducted for the thick-walled SEA balloon ($\eta=3.00$) in Table \ref{Table2}. Overall, the present SSM can be utilized to obtain accurate resonant frequencies of the free vibration with a high precision. In Subsecs.~\ref{6.2.2} and \ref{6.2.3}, only the moderately thick SEA balloon ($\eta=1.25$) is considered as an illustrative example, and we will take the number of the equally discretized thin layer $N$ as 50. The predicted numerical results can be assumed to be highly accurate (see Table \ref{Table1}).

\begin{figure}[htbp]
	\centering	
	\includegraphics[width=0.485\textwidth]{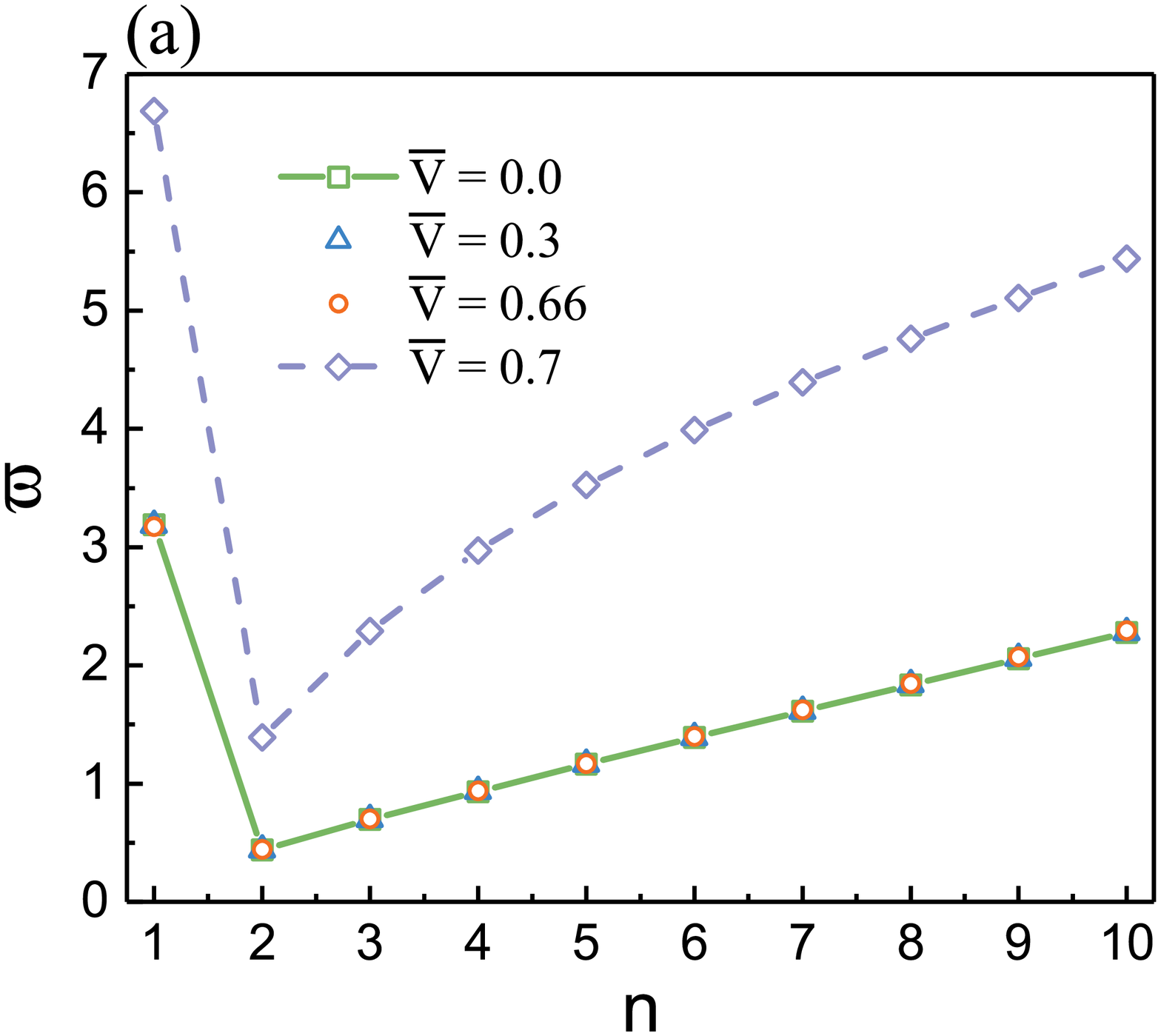}
	\includegraphics[width=0.50\textwidth]{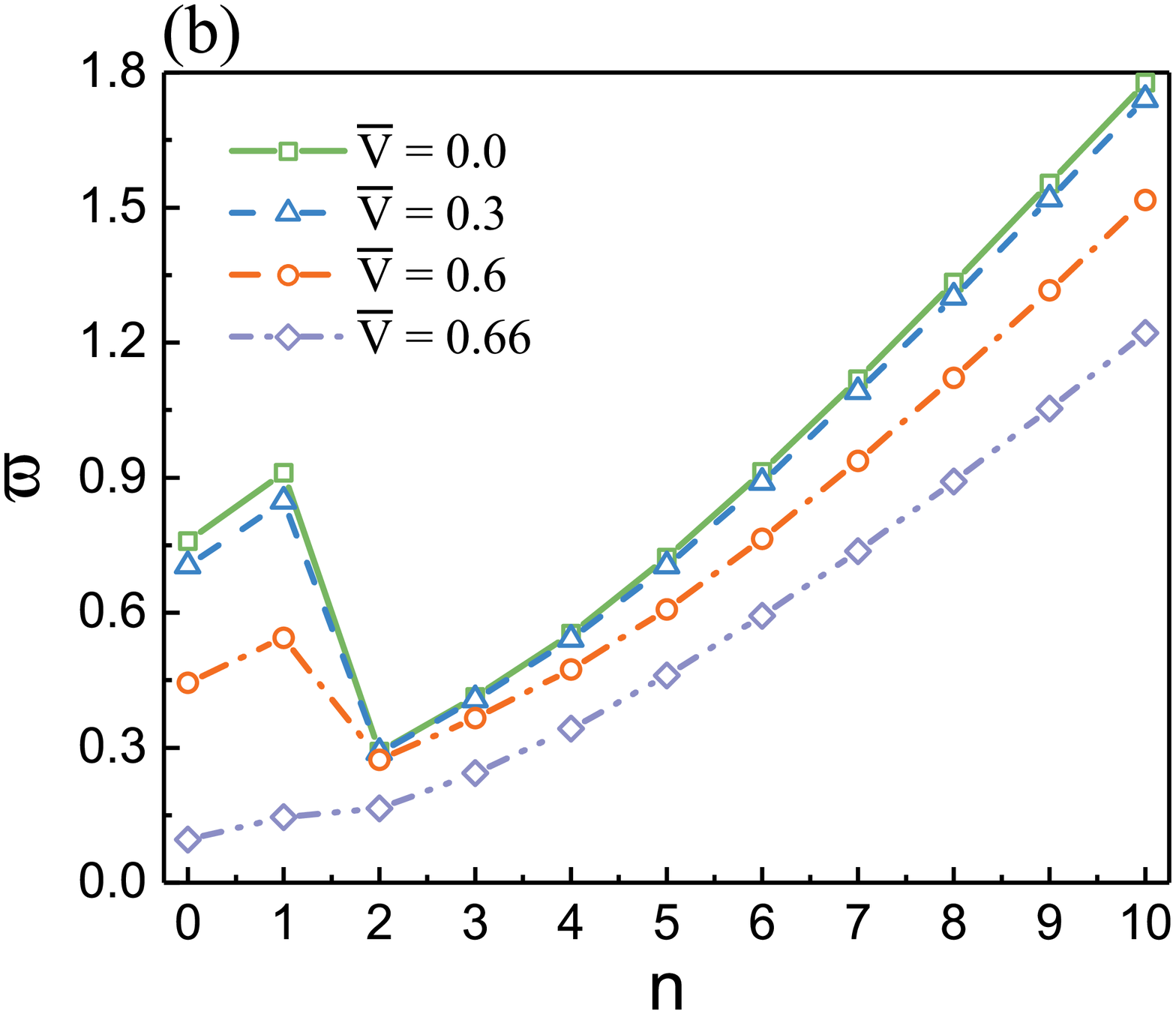}
	\caption{Dimensionless vibration frequencies $\varpi$ versus the angular mode number $n$ of a moderately thick SEA balloon ($\eta=1.25$) for 
		$\overline{P}=0.02$, $G=97.2$ and different values of electric voltage: (a) the first class of vibration (torsional modes); (b) the second class of vibration (spheroidal modes).}
	\label{Fig4}
\end{figure}


\subsubsection{Effect of the electric voltage}\label{6.2.2}

First of all, how the applied electric voltage influences the resonant frequency of these two classes of vibration of the SEA balloon is studied. For a fixed internal pressure ($\overline{P}=0.02$) and Gent constant ($G=97.2$), the curves of the lowest dimensionless vibration frequency versus the angular mode number $n$ are depicted in Fig.~\ref{Fig4} for different values of the electric voltage, wherein Fig.~\ref{Fig4}(a) displays the results of the first class of vibration (torsional modes) while Fig.~\ref{Fig4}(b) shows those of the second class (spheroidal modes).

One can observe from Fig.~\ref{Fig4}(a) that, for the first class of vibration, the vibration frequency for the purely torsional or rotary mode ($n=1$) is larger than those of the higher-order torsional modes ($n\ge2$), and the frequency of the torsional mode has a gradual rise with the increasing angular mode number for $n\ge2$. Furthermore, when the applied electric voltage is below the critical snap-through value ($\overline{V}_{cr} \simeq 0.67$, as shown in Fig.~\ref{Fig3}(b)), the resonant frequency is hardly affected by the electric voltage, which is physically understandable that the applied direction of the electric voltage is perpendicular to the vibration direction of the torsional modes and thus the work of the biasing electric fields vanishes. Nonetheless, for an electric voltage above the critical value, for example $\overline{V}= 0.7$, the resonant frequency has a dramatic increase owing to the snap-through phenomenon.

For the second class of vibration shown in Fig.~\ref{Fig4}(b), the lowest vibration frequency of the spheroidal modes is taken at $n=2$ corresponding to the quadrupolar mode if the applied voltage ($\overline{V}\le0.6$) is much lower than the critical value. On the contrary, for an electric voltage ($\overline{V}=0.66$) approaching $\overline{V}_{cr}$, the lowest vibration frequency occurs at the breathing mode ($n=0$) and the frequency exhibits a monotonically increasing trend in the whole range of the angular mode number. Moreover, the resonant frequency gradually decreases with the increasing electric voltage below $\overline{V}_{cr}$ for all the angular mode numbers. This is because the global stiffness of the balloon decreases when the electric voltage grows but still keeps smaller than the critical value.

\begin{figure}[htbp]
	\centering	
	\includegraphics[width=0.49\textwidth]{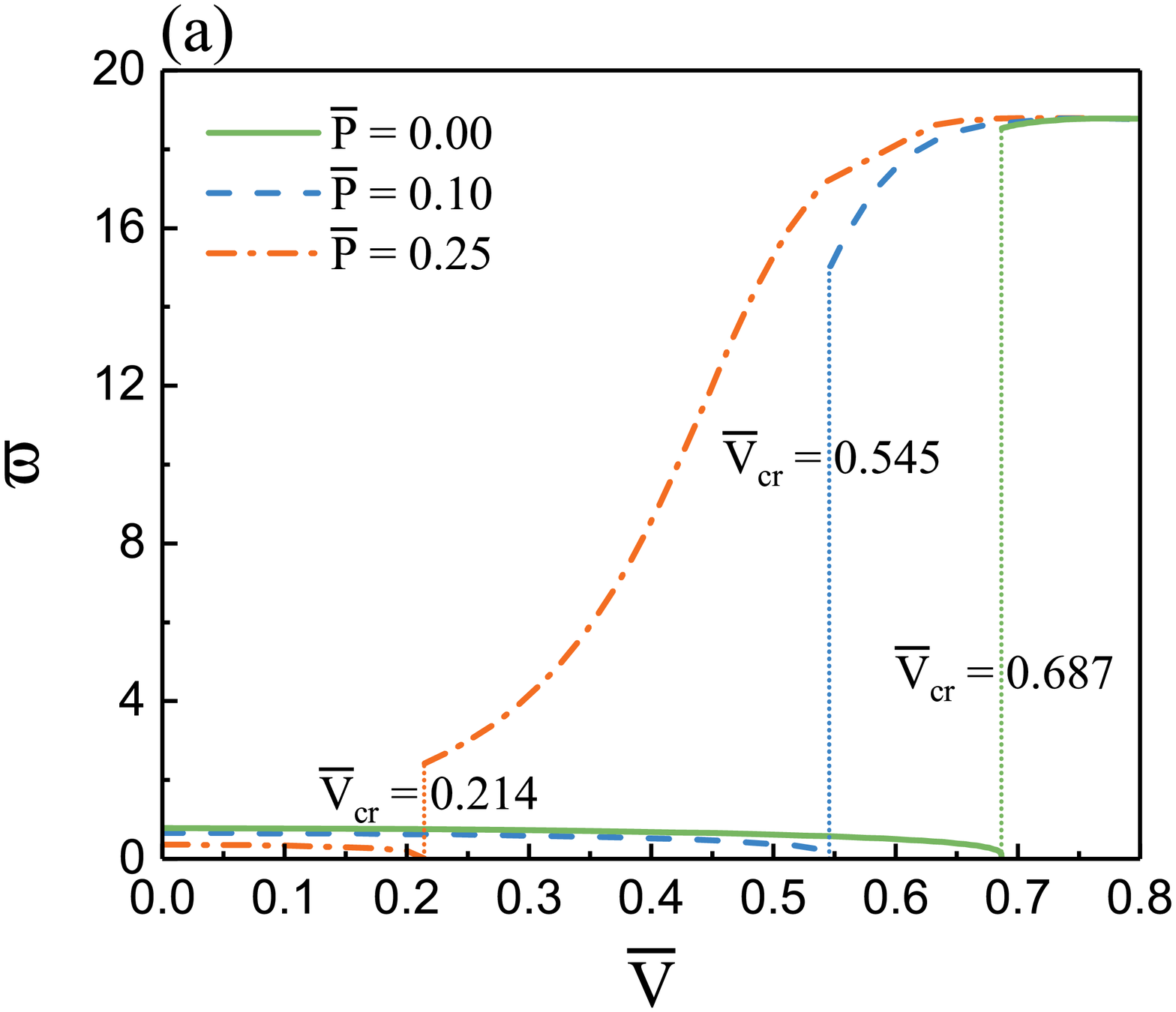}
	\includegraphics[width=0.49\textwidth]{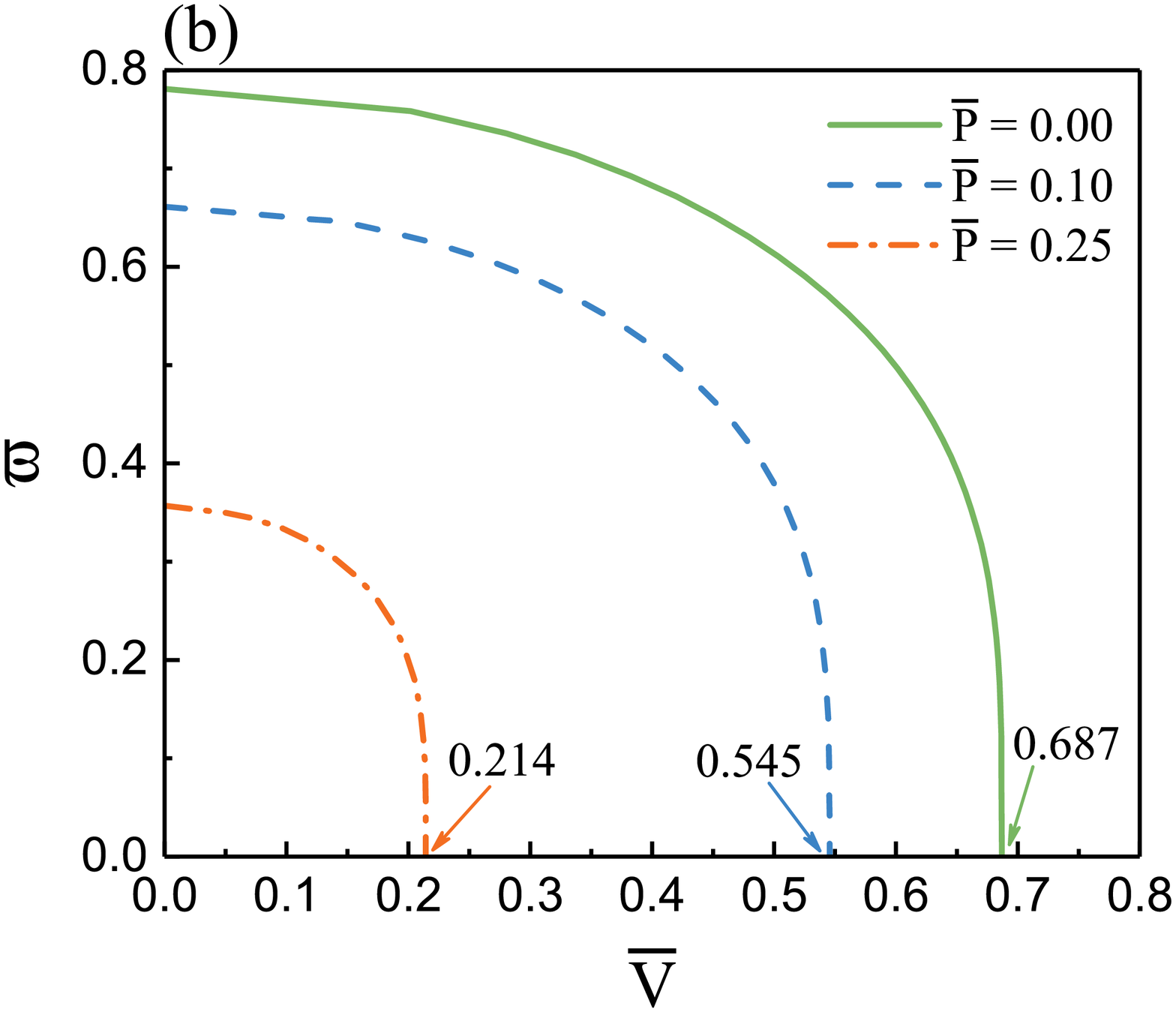}
	\includegraphics[width=0.49\textwidth]{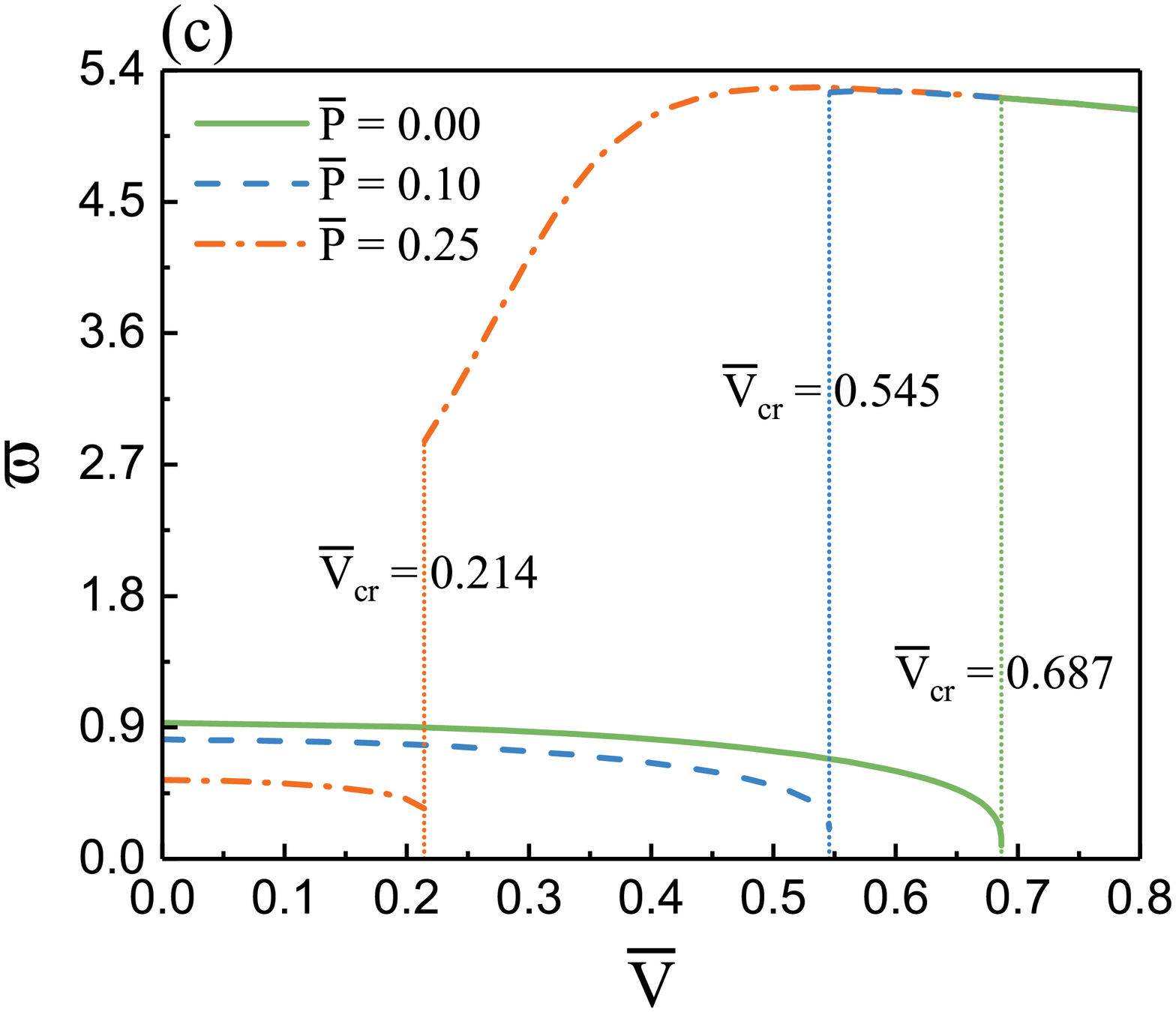}
	\includegraphics[width=0.49\textwidth]{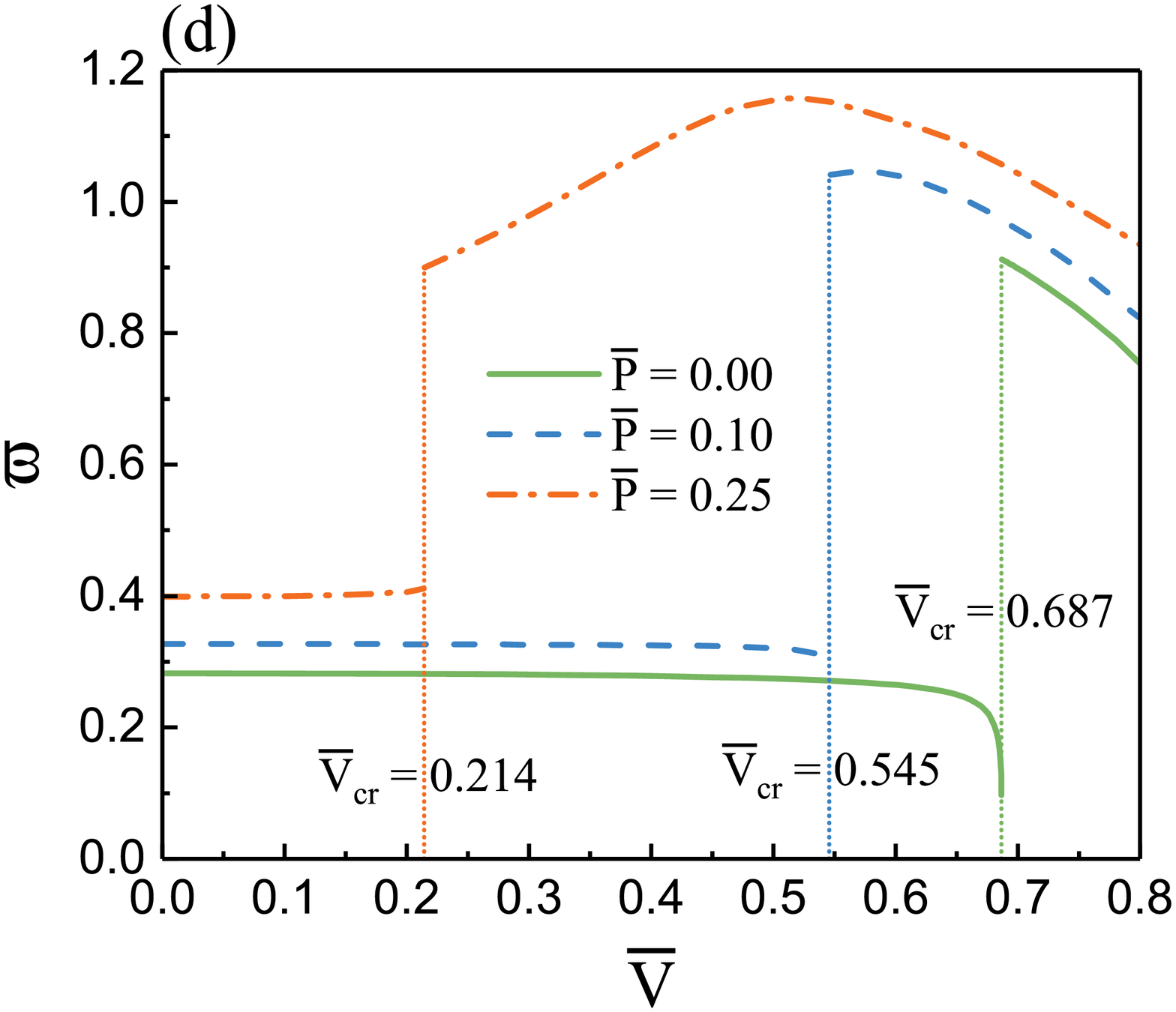}
	\caption{Dimensionless vibration frequencies $\varpi$ of the second class of vibration (spheroidal modes) as a function of the dimensionless electric voltage $\overline{V}$ for a moderately thick SEA balloon with $G=97.2$ under different values of internal pressure: (a) breathing mode ($n=0$) with snap-through instability; (b) breathing mode ($n=0$) without snap-through instability; (c) dipolar mode ($n=1$); (d) quadrupolar mode ($n=2$).}
	\label{Fig5}
\end{figure}

In order to clearly show the dependence of vibration characteristics on the electric biasing fields, Fig.~\ref{Fig5} displays the variations of the resonant frequency with the applied electric voltage for the breathing ($n=0$), dipolar ($n=1$), and quadrupolar modes ($n=2$) of the second class under three different internal pressures. Obviously, as the applied electric voltage increases from
zero to the critical value $\overline{V}_{cr}$ which depends on the internal pressure, the resonant frequency of these three modes will decrease to a minimum. Note that a higher internal pressure leads to a lower critical value $\overline{V}_{cr}$ (see Fig.~\ref{Fig3}(b)). Interestingly, when the electric voltage reaches $\overline{V}_{cr}$, the frequency of the breathing mode becomes zero, which corresponds to the snap-through instability.

In addition, Fig.~\ref{Fig5} also demonstrates the effect of the snap-through phenomenon on the vibration frequency of the breathing, dipolar and quadrupolar modes. Following the snap-through instability, the resonant frequency of these three modes increases considerably. With the increasing electric voltage above $\overline{V}_{cr}$, the frequency of the breathing mode increases monotonically for all the applied internal pressures. Nevertheless, the frequency variation of the dipolar and quadrupolar modes with the electric voltage depends on the internal pressure. Specifically, for a small pressure ($\overline{P}=0.10$), their resonant frequency will have a monotonous decrease, while for a large one ($\overline{P}=0.25$), the frequency will arrive at a maximum and then decrease reversely. It is worth pointing out that increasing the internal pressure enhances the adjustable range of the electric voltage for the resonant frequency. To conclude, all the above-mentioned results confirm the capability to electrostatically tune the small-amplitude free vibrations of SEA balloons.


\subsubsection{Effect of the internal pressure}\label{6.2.3}

\begin{figure}[htbp]
	\centering	
	\includegraphics[width=0.485\textwidth]{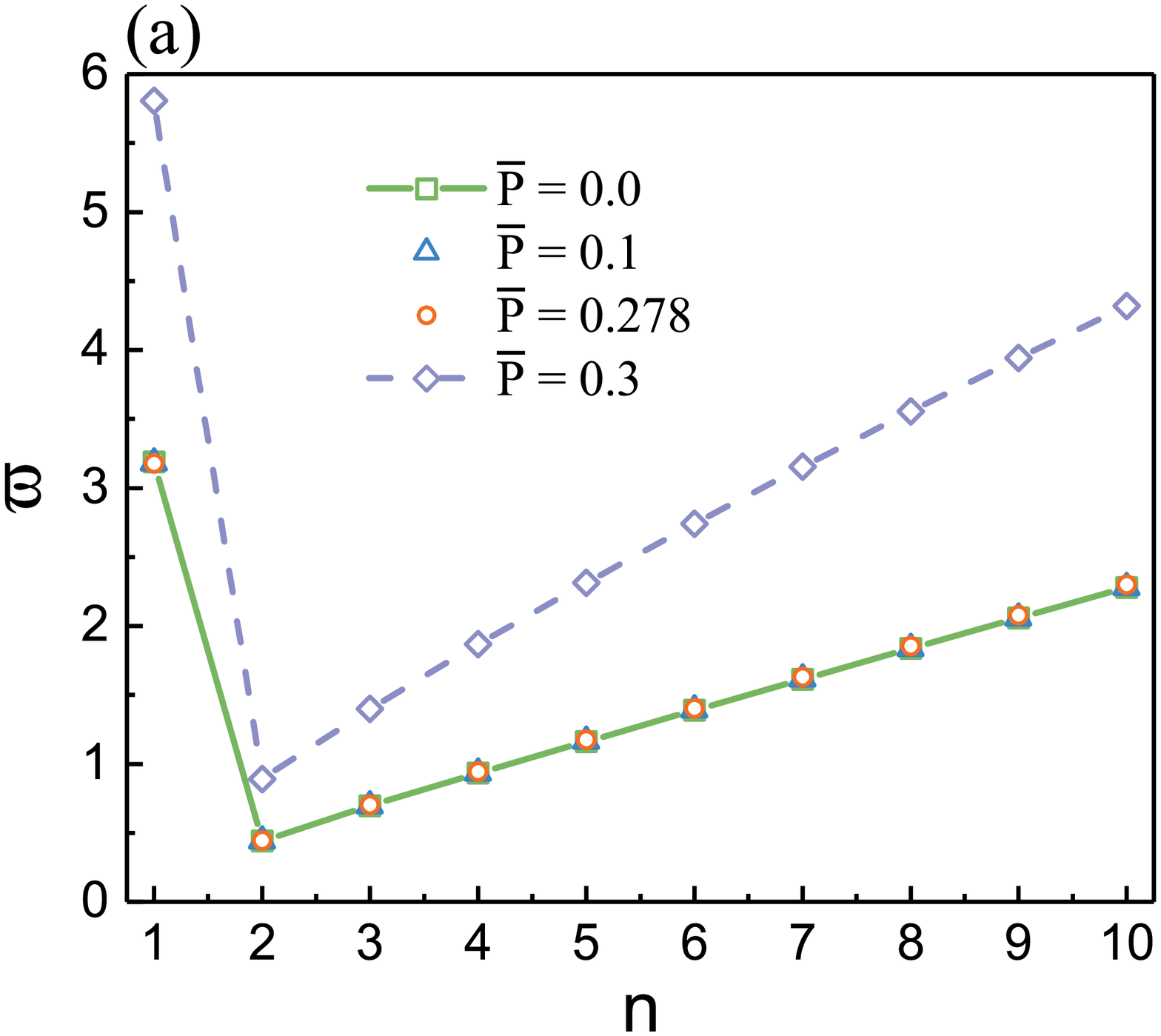}
	\includegraphics[width=0.50\textwidth]{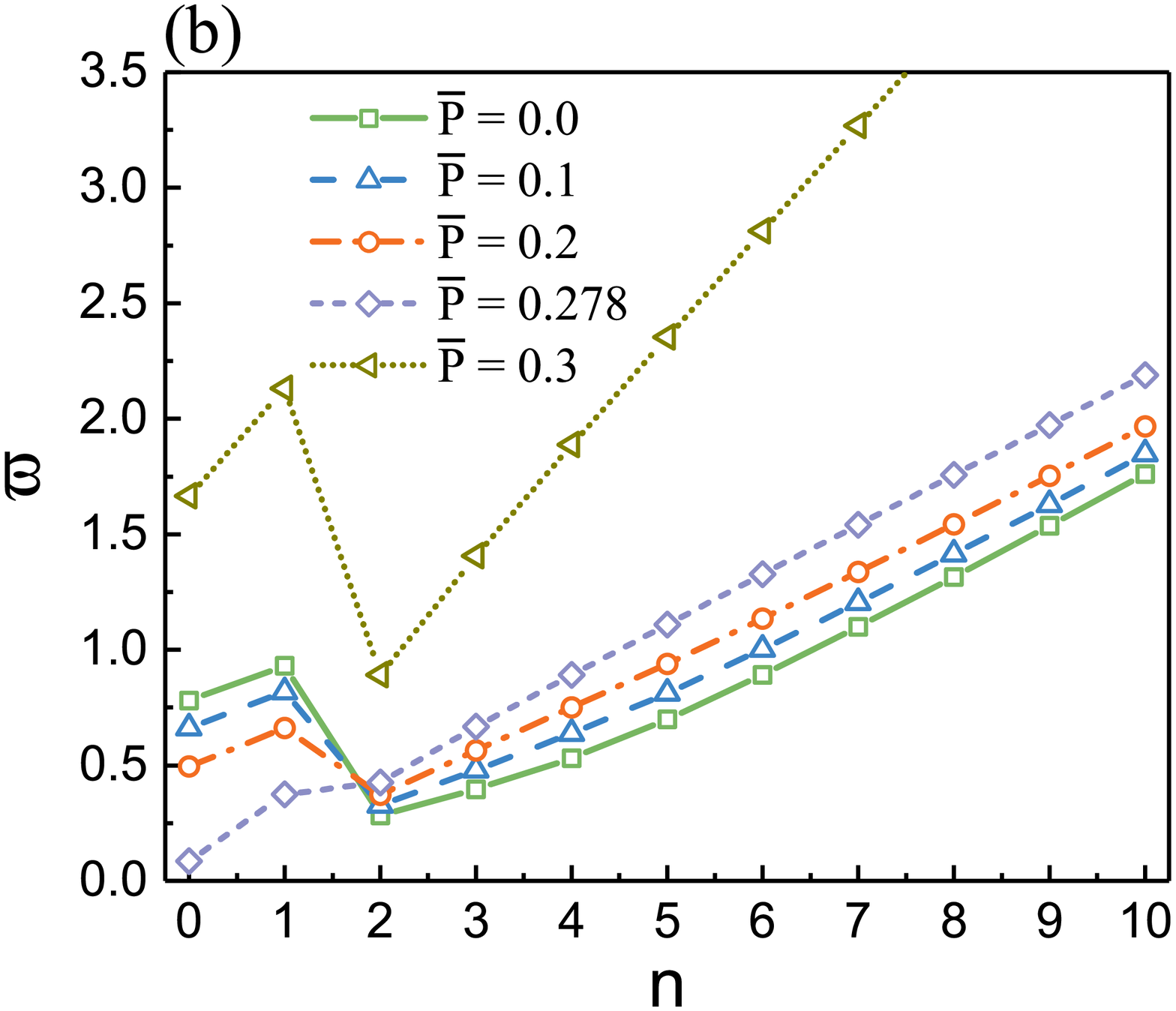}
	\caption{Dimensionless vibration frequencies $\varpi$ versus the angular mode number $n$ of a moderately thick SEA balloon ($\eta=1.25$) for 
		$\overline{V}=0.02$, $G=97.2$ and different values of internal pressure: (a) the first class of vibration (torsional modes); (b) the second class of vibration (spheroidal modes).}
	\label{Fig6}
\end{figure}

Now we turn to the investigation of the effect of internal pressure on the free vibrations. For a fixed electric voltage ($\overline{V}=0.02$) and Gent constant ($G=97.2$), Figs.~\ref{Fig6}(a) and \ref{Fig6}(b) display the variations of the resonant frequency with the angular mode number $n$ for the torsional and spheroidal modes, respectively, under different values of internal pressure. Note that the critical pressure value at which the snap-through is triggered for this case is $\overline{P}_{cr} \simeq 0.28$ (see Fig.~\ref{Fig3}(a)). The analysis of the results is qualitatively similar to that in Fig. 4 and thus not reported here for brevity. It should be emphasized that the snap-through instability induced by the internal pressure can be also exploited to realize the significant change in the resonant frequency.

\begin{figure}[htbp]
	\centering	
	\includegraphics[width=0.49\textwidth]{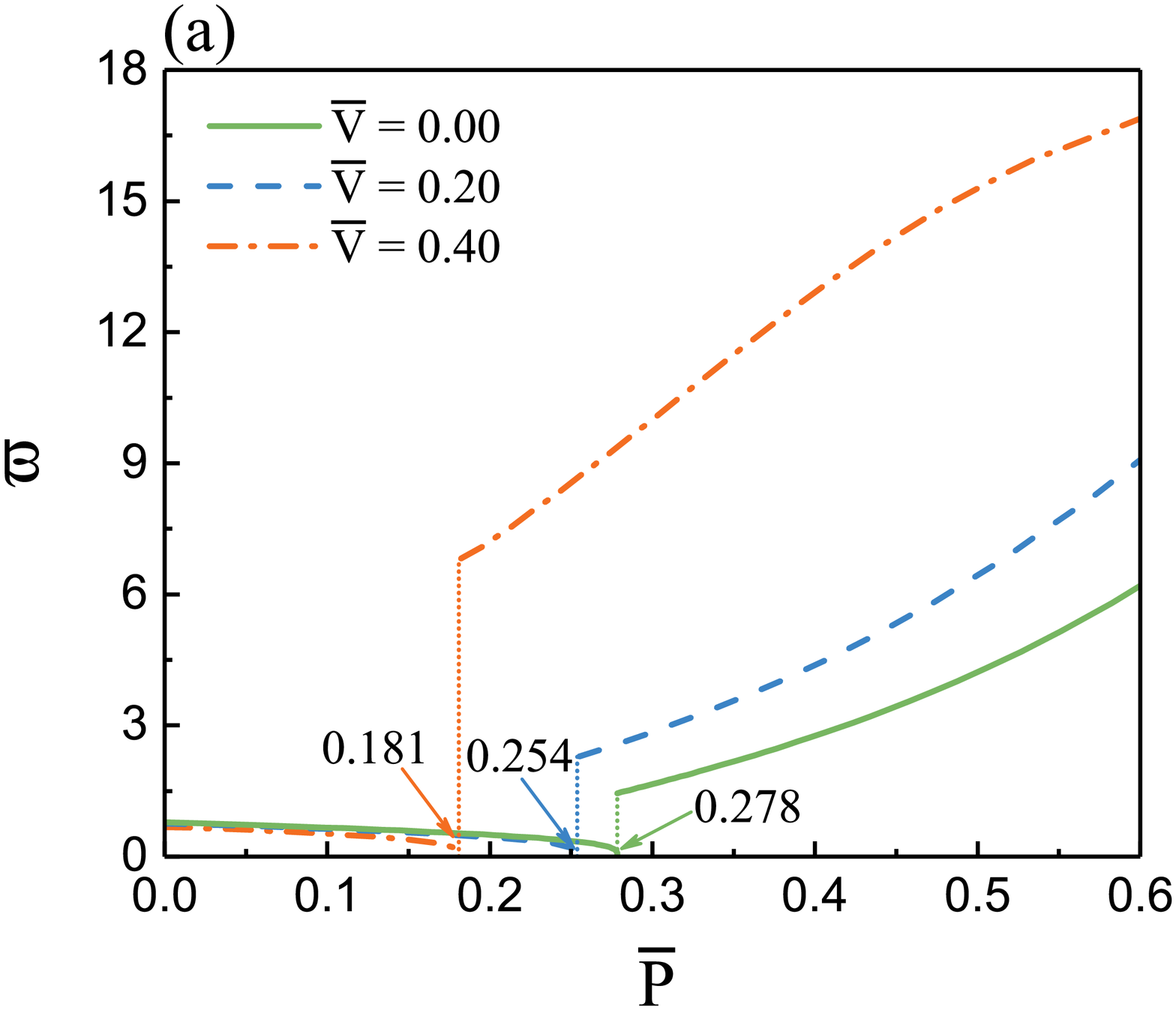}
	\includegraphics[width=0.49\textwidth]{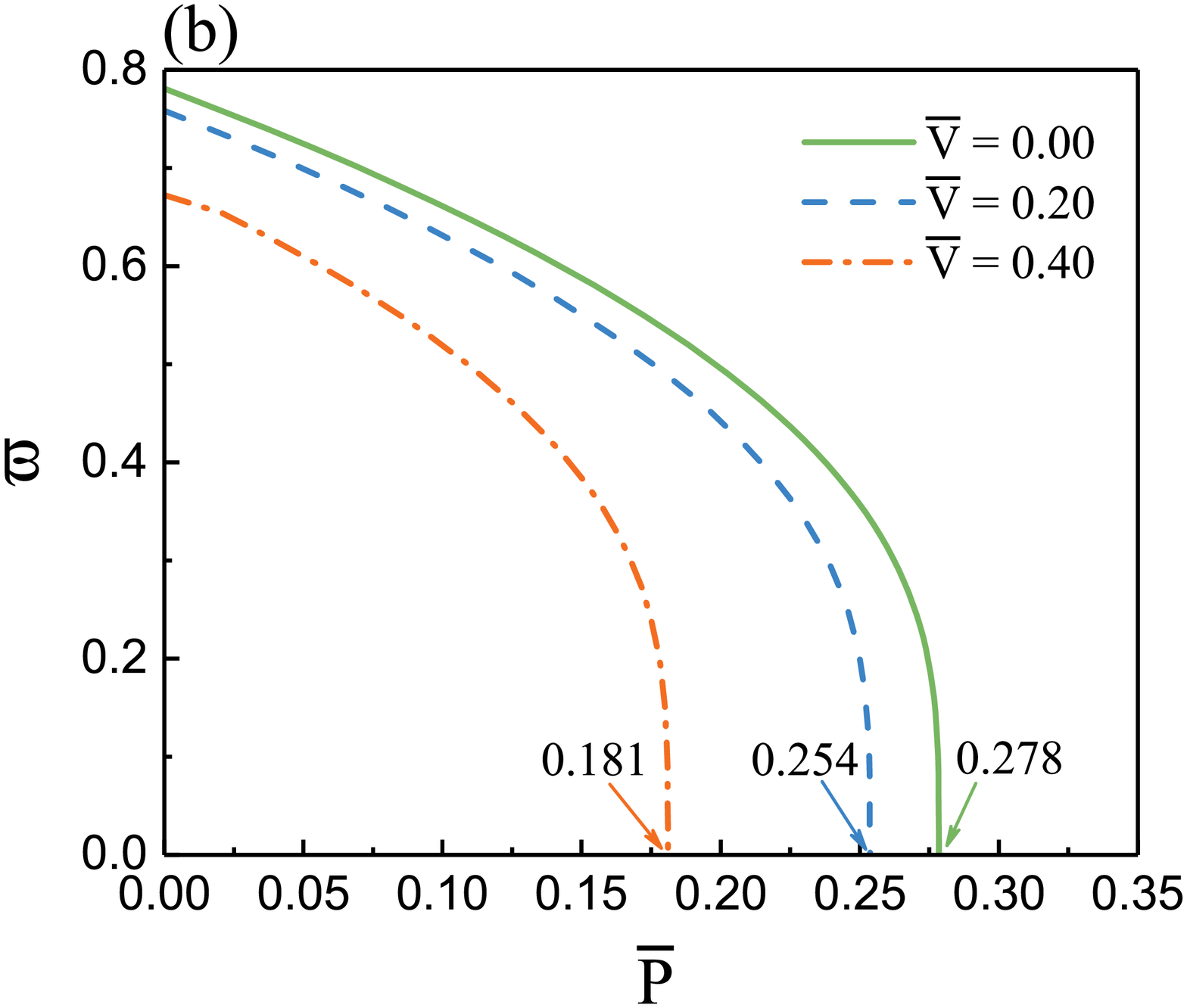}
	\includegraphics[width=0.49\textwidth]{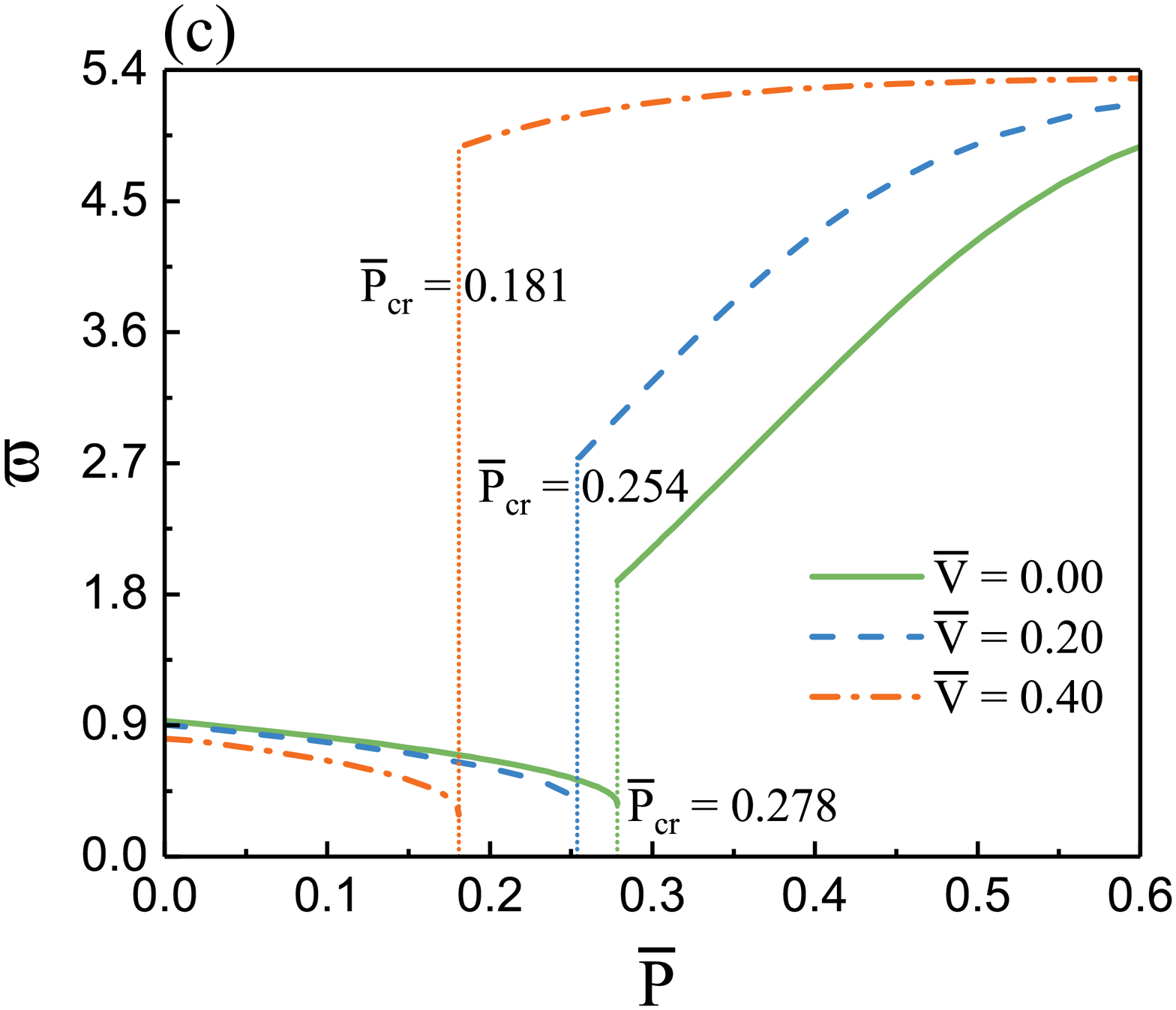}
	\includegraphics[width=0.49\textwidth]{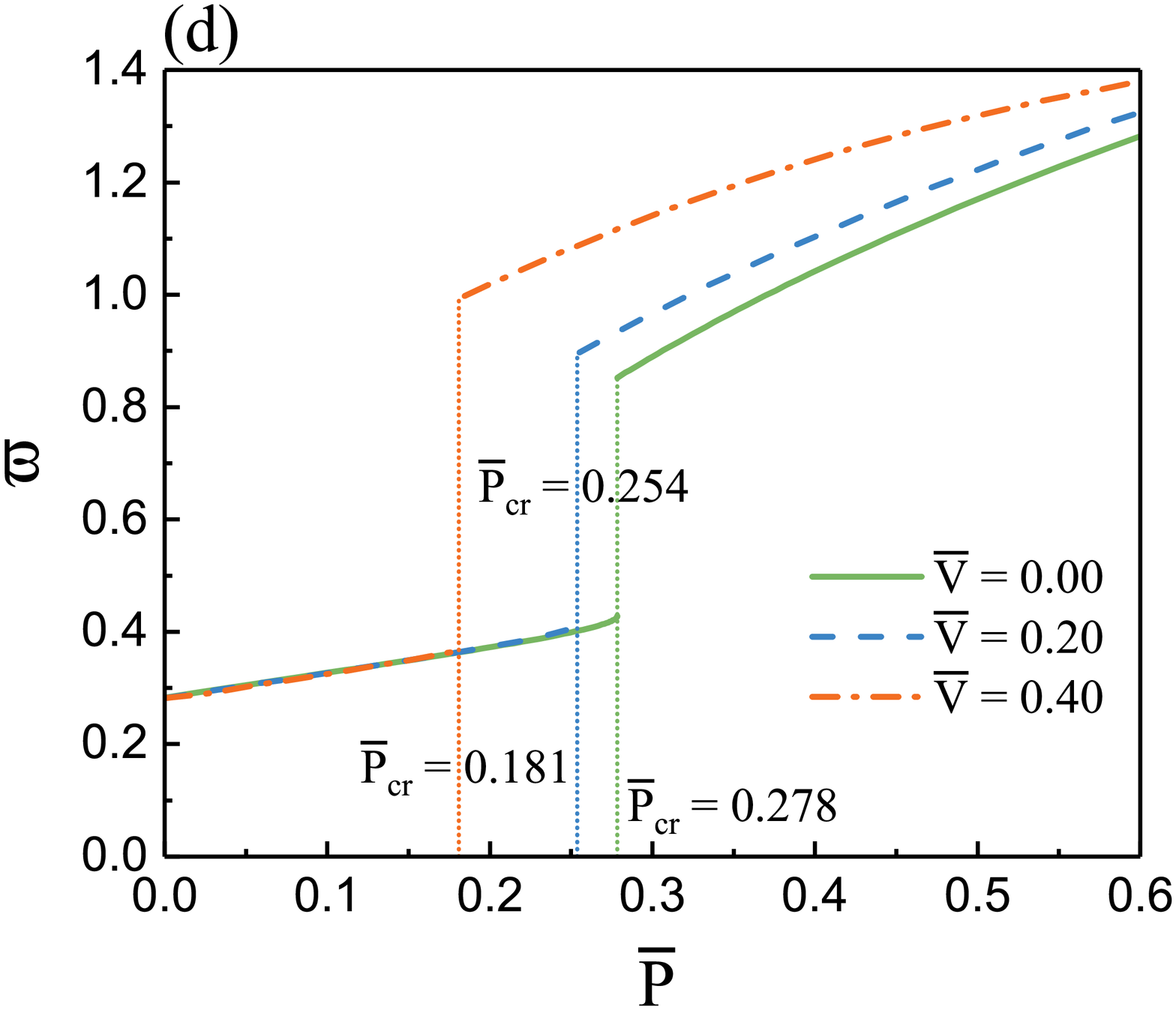}
	\caption{Dimensionless vibration frequencies $\varpi$ of the second class of vibration (spheroidal modes) as a function of the dimensionless internal pressure $\overline{P}$ for a moderately thick SEA balloon with $G=97.2$ under different values of electric voltage: (a) breathing mode ($n=0$) with snap-through instability; (b) breathing mode ($n=0$) without snap-through instability; (c) dipolar mode ($n=1$); (d) quadrupolar mode ($n=2$).}
	\label{Fig7}
\end{figure}

Moreover, for the breathing ($n=0$), dipolar ($n=1$), and quadrupolar modes ($n=2$) of the second class of vibration, we also show the variations of the resonant frequency with the applied internal pressure in Fig.~\ref{Fig7} under three different electric voltages. When the pressure varies from zero to the critical value $\overline{P}_{cr}$, the resonant frequency of the breathing and dipolar modes decrease monotonically, while that of the quadrupolar mode is raised almost linearly. Note that the higher electric voltage results in lower critical value $\overline{P}_{cr}$ (see Fig.~\ref{Fig3}(a)). Since the snap-through instability occurs, the frequency of the breathing mode drops down to zero at $\overline{P}_{cr}$. Furthermore, Fig.~\ref{Fig7} indicates that the snap-through phenomenon caused by the internal pressure can be utilized to realize a sudden jump of the resonant frequency. When the applied internal pressure increases from the critical value $\overline{P}_{cr}$, the frequency of these three modes is lifted up for all the applied electric voltages. In a word, the dependence of the resonant frequency on the internal pressure also provides a possibility to readily adjust the free vibration characteristics of a SEA balloon via tuning the internal pressure.


\subsubsection{Effect of the thickness}

In this subsection, the effect of the SEA balloon thickness on the resonant frequencies of various excited modes will be investigated. For this purpose, we will consider one commercial product `Silicone CF19-2186' by the manufacturer Nusil with $\mu=333$~kPa, $\varepsilon =2.5\times 10^{-11}$~F/m, $\rho=1100$~kg/m$^3$, and $G=46.3$. In general, the vibration frequency of balloon shells depends on the thickness-to-mean radius ratio $H/R_m$, where $R_m=(A+B)/2$ is the mean radius taken as $R_m=5$~cm in the following analysis. Besides, the internal pressure is assumed to be zero.

\begin{figure}[htbp]
	\centering	
	\includegraphics[width=0.484\textwidth]{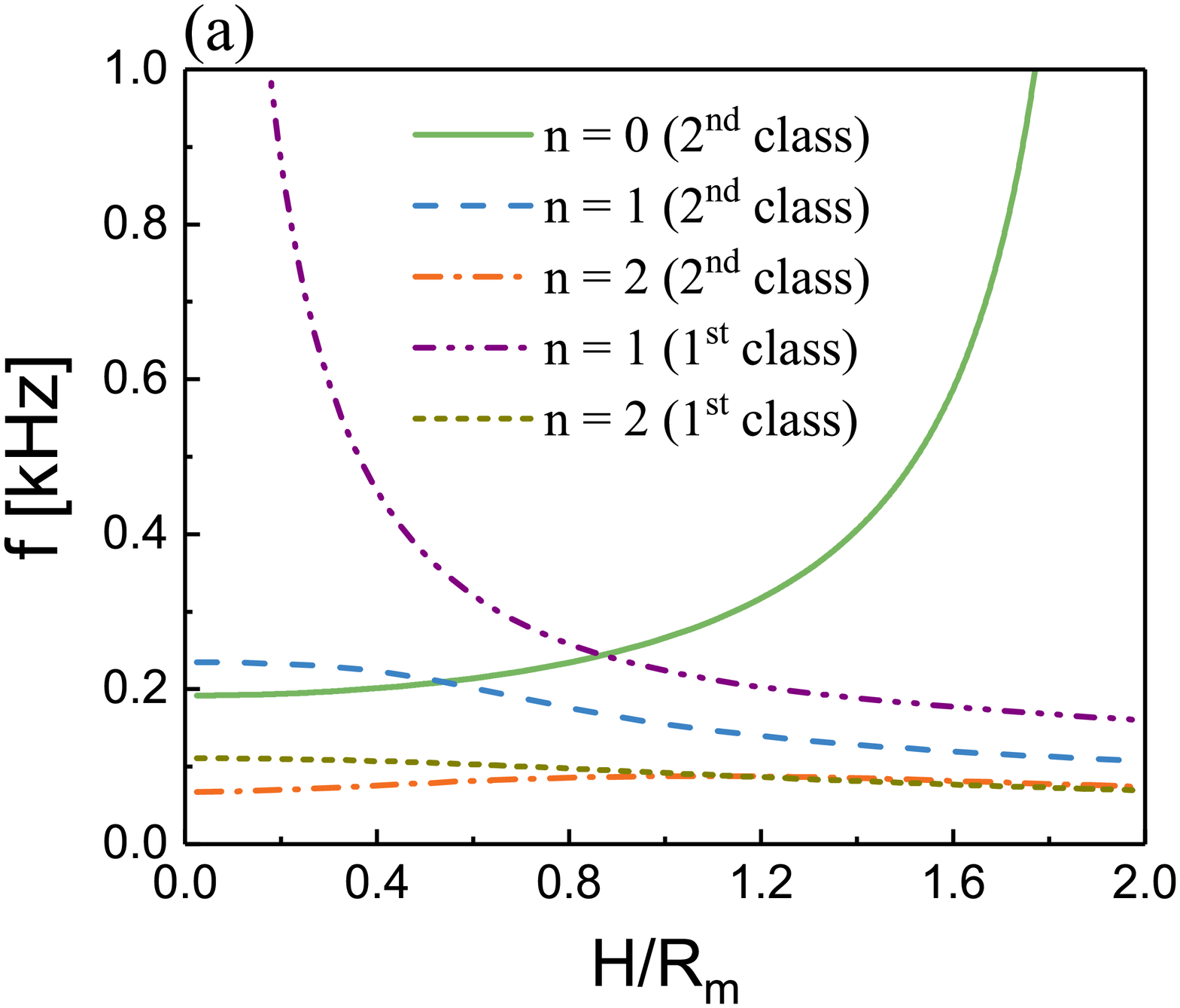}
	\includegraphics[width=0.5\textwidth]{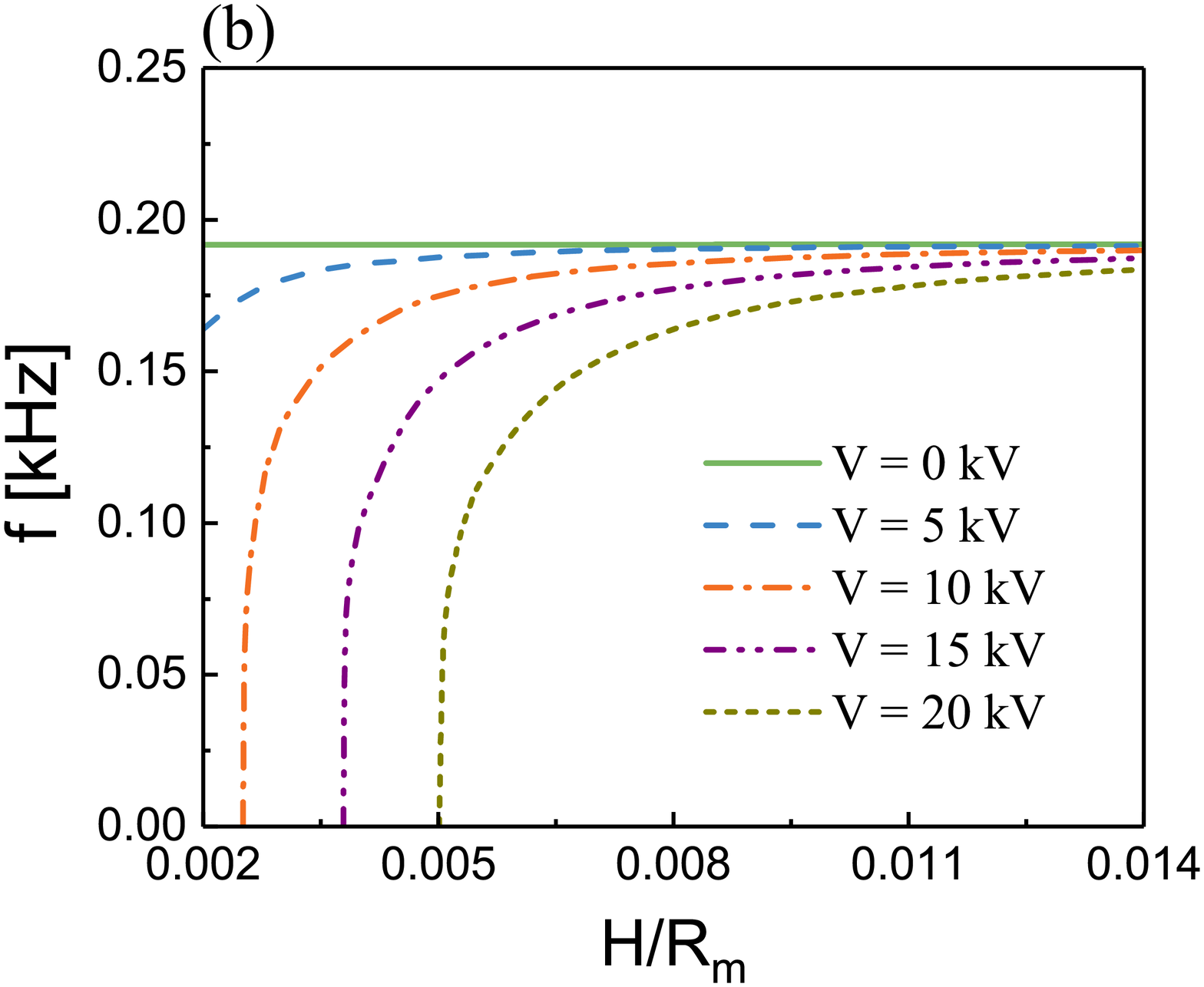}
	\caption{Vibration frequencies $f =\omega/(2\pi)$ versus the thickness-to-mean radius ratio $H/R_m$ of a SEA balloon made of Silicone CF19-2186 by Nusil: (a)  for different excited modes without biasing fields; (b) for the breathing mode ($n=0$) under different applied voltages $V$.}
	\label{Fig8}
\end{figure}

The curves of the lowest vibration frequency $f = \omega/(2\pi)$ versus the thickness-to-mean radius ratio $H/R_m$ are displayed in Fig.~\ref{Fig8}(a) for different excited modes without biasing fields. It can be seen that depending on different vibration modes, the variation trends of $f$ with $H/R_m$ are different. Specifically, for the second class of vibration, as $H/R_m$ increases, the frequency of the breathing mode ($n=0$) increases monotonically, that of the dipolar mode ($n=1$) decreases monotonically, and that of the quadrupolar mode ($n=2$) exhibits a non-monotonic variation (first increases and then reduces). The frequencies of the first two torsional modes ($n=1$ and $n=2$) for the first class of vibration gradually drop down with the increasing $H/R_m$.

Furthermore, for different electric voltages $V$, Fig.~\ref{Fig8}(b) depicts the variation of vibration frequencies $f$ with the thickness-to-mean radius ratio $H/R_m$ for the breathing mode ($n=0$). When $H/R_m$ increases, the vibration frequencies under the action of the nonzero voltage gradually approach that without biasing fields, and thus only the small range of $H/R_m$ is shown. Physically, a higher voltage value (greater than 20~kV) is required to tune the vibration frequency for a larger $H/R_m$. Additionally, corresponding to different nonzero voltages, there exist various critical values of $H/R_m$  such that the vibration frequencies vanish, which means that the snap-instability phenomenon occurs below these thresholds. A similar analysis can be carried out for other excited modes and is not shown here for brevity.

\section{Conclusions}\label{sec7}

In this paper, we conducted a 3D analytical study of the small-amplitude free vibration of a SEA spherical balloon with underlying radially inhomogeneous baising fields. By introducing appropriate incremental potential functions, we obtained two decoupled incremental state equations \eqref{53}-\eqref{56}. Furthermore, two separate frequency equations governing torsional and spheroidal modes (see Eqs.~\eqref{freEQ} and \eqref{freEQ-breath}$_2$) were analytically derived by employing the state-space method (SSM) and imposing the incremental mechanical and electric boundary conditions. Finally, we provided a detailed numerical evaluation to comprehensively elucidate the effects of radial electric voltage and internal pressure on the vibration characteristics of the SEA balloon. What follows is a list of useful conclusions drawn from the present investigation:

\begin{enumerate}[(1)]
	\item The proposed SSM can be utilized to obtain accurate resonant frequencies of the SEA balloon with a high precision.
	
	\item By independently or jointly varying the applied electric voltage and internal pressure, the resonant frequencies of different excited vibration modes for the SEA balloon can be readily adjusted.
	
	\item The snap-through instability induced by both the electric voltage and the internal pressure can be exploited to realize sharp transitions in the resonant frequency.
\end{enumerate}

All these results indicate that biasing fields can be an efficient means to manipulate the 3D free vibrations of SEA structures (not just balloons). The present analytical study provides significant guidance for the accurate prediction of the free vibration characteristics of tunable resonant devices that incorporate the SEA balloon as the tuning element.


\section*{Acknowledgments}


The work was supported by the National Natural Science Foundation of China (Nos. 11872329 and 11621062), the Russian Science Foundation (No. 18-19-00092), and the awarding of Research Fellow at Politecnico di Torino. Partial supports from the Fundamental Research Funds for the Central Universities (No. 2016XZZX001-05) and the Shenzhen Scientific and Technological Fund for R \& D (No. JCYJ20170816172316775) are also acknowledged.
 


\appendix

\section{Non-zero instantaneous electroelastic moduli} \label{AppeA}


For an incompressible isotropic SEA spherical balloon subjected to an internal pressure and a radial electric displacement field, we can employ the incremental theory of nonlinear electroelasticity \citep{dorfmann2010electroelastic} to obtain the non-zero components (the components denoted by the subscripts 1, 2 and 3 correspond to those along the $r$-, $\theta$- and $\varphi$-directions, respectively) of the instantaneous electroelastic moduli tensors ${{\mathcal{A}}_{0}}$, ${{\mathcal{M}}_{0}}$ and ${{\mathcal{R}}_{0}}$ as follows:
\begin{align}
{{\mathcal{A}}_{01111}}&=\frac{2}{{{\lambda }^{8}}}\Big\{ {{\lambda }^{4}}\left( {{\Omega }^{\text{*}}_{1}}+2{{\Omega }^{\text{*}}_{2}}{{\lambda }^{2}} \right)+8{{\lambda }^{2}}\left( {{\Omega }^{\text{*}}_{12}}+{{\Omega }^{\text{*}}_{22}}{{\lambda }^{2}} \right)+2D_{r}^{4}\left( 4{{\Omega }^{\text{*}}_{66}}+4{{\Omega }^{\text{*}}_{56}}{{\lambda }^{4}}+{{\Omega }^{\text{*}}_{55}}{{\lambda }^{8}} \right) \notag \\ 
& +2{{\Omega }^{\text{*}}_{11}}+D_{r}^{2}\left[ 8{{\Omega }^{\text{*}}_{16}}+{{\lambda }^{4}}\left( 4{{\Omega }^{\text{*}}_{15}}+6{{\Omega }^{\text{*}}_{6}}+{{\Omega }^{\text{*}}_{5}}{{\lambda }^{4}} \right)+8{{\lambda }^{2}}\left( 2{{\Omega }^{\text{*}}_{26}}+{{\Omega }^{\text{*}}_{25}}{{\lambda }^{4}} \right) \right]  \hskip -0.2em \Big\}, \notag \\
{{\mathcal{A}}_{02222}}&={{\mathcal{A}}_{03333}}=\frac{2}{{{\lambda }^{4}}}\Big\{ 2{{\Omega }^{\text{*}}_{22}}{{\left( 1+{{\lambda }^{6}} \right)}^{2}}+{{\lambda }^{2}}\Big[ {{\Omega }^{\text{*}}_{2}}\left( 1+{{\lambda }^{6}} \right) \notag \\
&+{{\lambda }^{4}}\left( {{\Omega }^{\text{*}}_{1}}+2{{\Omega }^{\text{*}}_{11}}{{\lambda }^{2}} \right)+4{{\lambda }^{2}}{{\Omega }^{\text{*}}_{12}}\left( 1+{{\lambda }^{6}} \right) \Big] \Big\}, \notag \\ 
{{\mathcal{A}}_{01122}}&={{\mathcal{A}}_{01133}}=\frac{4}{{{\lambda }^{6}}}\Big\{ \left( {{\Omega }^{\text{*}}_{11}}+{{\Omega }^{\text{*}}_{2}} \right){{\lambda }^{4}}+2{{\lambda }^{2}}{{\Omega }^{\text{*}}_{22}}\left( 1+{{\lambda }^{6}} \right)+{{\Omega }^{\text{*}}_{12}}\left( 1+3{{\lambda }^{6}} \right) \notag \\ 
& +D_{r}^{2}\left[ 2{{\Omega }^{\text{*}}_{26}}\left( 1+{{\lambda }^{6}} \right)+{{\lambda }^{4}}\left( 2{{\Omega }^{\text{*}}_{16}}+{{\Omega }^{\text{*}}_{25}}(1+{{\lambda }^{6}})+{{\Omega }^{\text{*}}_{15}}{{\lambda }^{4}} \right) \right]  \hskip -0.2em \Big\}, \notag\\ 
{{\mathcal{A}}_{02233}}&=\frac{4}{{{\lambda }^{4}}}\left\{ \left( {{\Omega }^{\text{*}}_{11}}+{{\Omega }^{\text{*}}_{2}} \right){{\lambda }^{8}}+2{{\lambda }^{4}}{{\Omega }^{\text{*}}_{12}}\left( 1+{{\lambda }^{6}} \right)+{{\Omega }^{\text{*}}_{22}}{{\left( 1+{{\lambda }^{6}} \right)}^{2}} \right\}, \notag\\ 
{{\mathcal{A}}_{01212}}&={{\mathcal{A}}_{01313}}=\frac{2}{{{\lambda }^{4}}}\left\{ {{\Omega }^{\text{*}}_{1}}+{{\Omega }^{\text{*}}_{2}}{{\lambda }^{2}}+D_{r}^{2}\left[ {{\Omega }^{\text{*}}_{6}}\left( 2+{{\lambda }^{6}} \right)+{{\Omega }^{\text{*}}_{5}}{{\lambda }^{4}} \right] \right\},\notag \\
{{\mathcal{A}}_{01221}}&={{\mathcal{A}}_{01331}}=2\left( D_{r}^{2}{{\Omega }^{\text{*}}_{6}}{{\lambda }^{2}}-\frac{{{\Omega }^{\text{*}}_{2}}}{{{\lambda }^{2}}} \right),\quad {{\mathcal{A}}_{02323}}={{\mathcal{A}}_{03232}}=2\left( {{\Omega }^{\text{*}}_{1}}{{\lambda }^{2}}+\frac{{{\Omega }^{\text{*}}_{2}}}{{{\lambda }^{2}}} \right), \notag \\ 
{{\mathcal{A}}_{02121}}&={{\mathcal{A}}_{03131}}=2{{\lambda }^{2}}\left( {{\Omega }^{\text{*}}_{1}}+{{\Omega }^{\text{*}}_{2}}{{\lambda }^{2}}+D_{r}^{2}{{\Omega }^{\text{*}}_{6}} \right),\quad {{\mathcal{A}}_{02332}}=-2{{\Omega }^{\text{*}}_{2}}{{\lambda }^{4}}, \notag \\ 
{{\mathcal{M}}_{0111}}&=\frac{4{{D}_{r}}}{{{\lambda }^{8}}}\Big\{ {{\lambda }^{2}}\left[ 2{{\Omega }^{\text{*}}_{26}}+{{\lambda }^{2}}\left( 2{{\Omega }^{\text{*}}_{6}}+{{\Omega }^{\text{*}}_{15}}+{{\lambda }^{4}}({{\Omega }^{\text{*}}_{5}}+{{\Omega }^{\text{*}}_{14}})+2{{\lambda }^{2}}({{\Omega }^{\text{*}}_{25}}+{{\Omega }^{\text{*}}_{24}}{{\lambda }^{4}}) \right) \right] \notag \\ 
& +{{\Omega }^{\text{*}}_{16}} +D_{r}^{2}\left[ 2{{\Omega }^{\text{*}}_{66}}+3{{\Omega }^{\text{*}}_{56}}{{\lambda }^{4}}+{{\lambda }^{8}}\left( 2{{\Omega }^{\text{*}}_{46}}+{{\Omega }^{\text{*}}_{55}}+{{\Omega }^{\text{*}}_{45}}{{\lambda }^{4}} \right) \right]   \hskip -0.2em \Big\}, \notag \\ 
{{\mathcal{M}}_{0221}}&={{\mathcal{M}}_{0331}}=\frac{4{{D}_{r}}}{{{\lambda }^{6}}}\Big\{ {{\Omega }^{\text{*}}_{26}}\left( 1+{{\lambda }^{6}} \right)+{{\lambda }^{4}}\left[ {{\Omega }^{\text{*}}_{16}}+{{\Omega }^{\text{*}}_{25}}\left( 1+{{\lambda }^{6}} \right) \right] \notag \\
&+{{\lambda }^{8}}\left[ {{\Omega }^{\text{*}}_{24}}\left( 1+{{\lambda }^{6}} \right)+{{\Omega }^{\text{*}}_{15}}+{{\Omega }^{\text{*}}_{14}}{{\lambda }^{4}} \right] \Big\},\quad {{\mathcal{M}}_{0122}}={{\mathcal{M}}_{0133}}=2{{D}_{r}}\left[ {{\Omega }^{\text{*}}_{5}}+\frac{{{\Omega }^{\text{*}}_{6}}\left( {{\lambda }^{6}}+1 \right)}{{{\lambda }^{4}}} \right], \notag \\ 
{{\mathcal{R}}_{011}}&=2\left( {{\Omega }^{\text{*}}_{5}}+{{\Omega }^{\text{*}}_{4}}{{\lambda }^{4}}+\frac{{{\Omega }^{\text{*}}_{6}}}{{{\lambda }^{4}}} \right)+\frac{4D_{r}^{2}}{{{\lambda }^{8}}}\Big\{ {{\Omega }^{\text{*}}_{66}}+{{\lambda }^{4}}\Big[ 2{{\Omega }^{\text{*}}_{56}} \notag \\
&+{{\lambda }^{4}}\left( 2{{\Omega }^{\text{*}}_{46}}+{{\Omega }^{\text{*}}_{55}}+2{{\Omega }^{\text{*}}_{45}}{{\lambda }^{4}}+{{\Omega }^{\text{*}}_{44}}{{\lambda }^{8}} \right) \Big] \Big\}, \quad {{\mathcal{R}}_{022}}={{\mathcal{R}}_{033}}=2\left( {{\Omega }^{\text{*}}_{5}}+{{\Omega }^{\text{*}}_{6}}{{\lambda }^{2}}+\frac{{{\Omega }^{\text{*}}_{4}}}{{{\lambda }^{2}}} \right). \notag 
\end{align}


\section{Derivation of the incremental state equation \eqref{53}}
\label{AppeB}


First, combining Eq.~\eqref{52}$_1$ with Eq.~\eqref{52}$_2$, we have
\begin{equation} \label{B1}
\frac{\partial {{u}_{\theta }}}{\partial \theta }+\frac{1}{\sin \theta }\frac{\partial {{u}_{\varphi }}}{\partial \varphi }+{{u}_{\theta }}\cot \theta =-\nabla _{1}^{2}G,
\end{equation}
where $\nabla _{1}^{2}$ has been defined in Subsec.~\ref{4.2}. Therefore, substitution of Eq.~\eqref{B1} into Eq.~\eqref{50}$_5$ yields
\begin{equation} \label{B2}
{{\nabla }_{2}}w=\nabla _{1}^{2}G-2w.
\end{equation}
Inserting Eq.~\eqref{52}$_{1-3}$ into Eq.~\eqref{51}$_{10}$ and using Eqs.~\eqref{B1} and \eqref{B2} results in
\begin{equation} \label{B3}
{{\nabla }_{2}}\Phi ={{q}_{4}}\left( \nabla _{1}^{2}G-2w \right)-\frac{1}{{{\varepsilon }_{11}}}{{\Delta }_{r}},
\end{equation}
where ${{q}_{4}}=({{{e}_{11}}-{{e}_{12}}})/{{{\varepsilon }_{11}}}$. Next, substituting Eq.~\eqref{52} into Eqs.~\eqref{51}$_{4}$ and \eqref{51}$_{6}$, we have
\begin{equation} \label{B4}
\begin{split}
& -\frac{1}{\sin \theta }\frac{\partial }{\partial \varphi }\left( {{\Sigma }_{1}}-{{c}_{66}}{{\nabla }_{2}}\psi +{{c}_{69}}\psi  \right)-\frac{\partial }{\partial \theta }\left( {{\Sigma }_{2}}-{{c}_{66}}{{\nabla }_{2}}G+{{c}_{69}}w+{{c}_{69}}G+{{e}_{26}}\Phi  \right)=0, \\ 
& \frac{\partial }{\partial \theta }\left( {{\Sigma }_{1}}-{{c}_{66}}{{\nabla }_{2}}\psi +{{c}_{69}}\psi  \right)-\frac{1}{\sin \theta }\frac{\partial }{\partial \varphi }\left( {{\Sigma }_{2}}-{{c}_{66}}{{\nabla }_{2}}G+{{c}_{69}}w+{{c}_{69}}G+{{e}_{26}}\Phi  \right)=0,
\end{split}
\end{equation}
which gives \citep{ding2006elasticity}
\begin{equation} \label{B5}
{{\nabla }_{2}}\psi =\frac{1}{{{c}_{66}}}{{\Sigma }_{1}}+\frac{{{c}_{69}}}{{{c}_{66}}}\psi,\quad {{\nabla }_{2}}G=\frac{1}{{{c}_{66}}}{{\Sigma }_{2}}+\frac{{{c}_{69}}}{{{c}_{66}}}\left( w+G \right)+\frac{{{e}_{26}}}{{{c}_{66}}}\Phi.
\end{equation}

Further, eliminating the incremental Lagrange multiplier $\dot{p}$ with the help of Eq.~\eqref{51}$_{1-3}$ and utilizing Eq.~\eqref{B1}, we can obtain
\begin{align} \label{B6}
{{\Sigma }_{\theta \theta }}&=\left( {{c}_{22}}-{{c}_{12}} \right)\left( \frac{\partial {{u}_{\theta }}}{\partial \theta }+{{u}_{r}} \right)+\left( {{c}_{23}}-{{c}_{12}} \right)\left( \frac{1}{\sin \theta }\frac{\partial {{u}_{\varphi }}}{\partial \varphi }+{{u}_{r}}+{{u}_{\theta }}\cot \theta  \right) \notag \\
&+\left( {{c}_{12}}-{{c}_{11}} \right){{\nabla }_{2}}{{u}_{r}}+\left( {{e}_{12}}-{{e}_{11}} \right){{\nabla }_{2}}\Phi +{{\Sigma }_{rr}},\notag \\ 
{{\Sigma }_{\varphi \varphi }}&=\left( {{c}_{23}}-{{c}_{12}} \right)\left( \frac{\partial {{u}_{\theta }}}{\partial \theta }+{{u}_{r}} \right)+\left( {{c}_{22}}-{{c}_{12}} \right)\left( \frac{1}{\sin \theta }\frac{\partial {{u}_{\varphi }}}{\partial \varphi }+{{u}_{r}}+{{u}_{\theta }}\cot \theta  \right) \notag \\
&+\left( {{c}_{12}}-{{c}_{11}} \right){{\nabla }_{2}}{{u}_{r}}+\left( {{e}_{12}}-{{e}_{11}} \right){{\nabla }_{2}}\Phi +{{\Sigma }_{rr}}, \notag \\ 
{{\Sigma }_{\theta \theta }}&+{{\Sigma }_{\varphi \varphi }}=\left( {{c}_{22}}+{{c}_{23}}-2{{c}_{12}} \right)\left( -\nabla _{1}^{2}G+2w \right) \notag \\
&+2\left( {{c}_{12}}-{{c}_{11}} \right){{\nabla }_{2}}w+2\left( {{e}_{12}}-{{e}_{11}} \right){{\nabla }_{2}}\Phi +2{{\Sigma }_{rr}}.
\end{align}
Inserting Eq.~\eqref{52}$_{1-3}$ into Eq.~\eqref{B6}$_{1,2}$ and Eq.~\eqref{51}$_{5,7,8,9,11,12}$ and after some rearrangements, we can get
\begin{align} \label{B7}
& {{\Sigma }_{\theta \theta }}=\left( {{c}_{22}}-{{c}_{23}} \right)\left( \frac{\cos \theta }{{{\sin }^{2}}\theta }\frac{\partial \psi }{\partial \varphi }-\frac{1}{\sin \theta }\frac{{{\partial }^{2}}\psi }{\partial \theta \partial \varphi } \right)+\left[ {{c}_{22}}+{{c}_{23}}-2{{c}_{12}}+\left( {{c}_{12}}-{{c}_{11}} \right){{\nabla }_{2}} \right]w \notag \\
&\quad+\left( {{e}_{12}}-{{e}_{11}} \right){{\nabla }_{2}}\Phi +{{\Sigma }_{rr}}-\left[ \left( {{c}_{22}}-{{c}_{12}} \right)\frac{{{\partial }^{2}}}{\partial {{\theta }^{2}}}+\left( {{c}_{23}}-{{c}_{12}} \right) \left(\frac{1}{{{\sin }^{2}}\theta }\frac{{{\partial }^{2}}}{\partial {{\varphi }^{2}}}+\cot \theta \frac{\partial }{\partial \theta }\right)\right]G, \notag \\ 
& {{\Sigma }_{\varphi \varphi }}=\left( {{c}_{22}}-{{c}_{23}} \right)\left( \frac{1}{\sin \theta }\frac{{{\partial }^{2}}\psi }{\partial \theta \partial \varphi }-\frac{\cos \theta }{{{\sin }^{2}}\theta }\frac{\partial \psi }{\partial \varphi } \right)+\left[ {{c}_{22}}+{{c}_{23}}-2{{c}_{12}}+\left( {{c}_{12}}-{{c}_{11}} \right){{\nabla }_{2}} \right]w \notag \\
&\quad+\left( {{e}_{12}}-{{e}_{11}} \right){{\nabla }_{2}}\Phi +{{\Sigma }_{rr}}-\left[ \left( {{c}_{23}}-{{c}_{12}} \right)\frac{{{\partial }^{2}}}{\partial {{\theta }^{2}}}+\left( {{c}_{22}}-{{c}_{12}} \right) \left(\frac{1}{{{\sin }^{2}}\theta }\frac{{{\partial }^{2}}}{\partial {{\varphi }^{2}}}+\cot \theta \frac{\partial }{\partial \theta }\right)\right]G, \notag \\ 
& {{\Sigma }_{\theta r}}=-\frac{1}{\sin \theta }\frac{\partial }{\partial \varphi }\left( {{c}_{69}}{{\nabla }_{2}}\psi -{{c}_{99}}\psi  \right)-\frac{\partial }{\partial \theta }\left( {{c}_{69}}{{\nabla }_{2}}G-{{c}_{99}}w-{{c}_{99}}G-{{e}_{26}}\Phi  \right), \notag \\ 
& {{\Sigma }_{\varphi r}}=\frac{\partial }{\partial \theta }\left( {{c}_{69}}{{\nabla }_{2}}\psi -{{c}_{99}}\psi  \right)-\frac{1}{\sin \theta }\frac{\partial }{\partial \varphi }\left( {{c}_{69}}{{\nabla }_{2}}G-{{c}_{99}}w-{{c}_{99}}G-{{e}_{26}}\Phi  \right), \notag \\ 
& {{\Sigma }_{\theta \varphi }}=\left( {{c}_{44}}\frac{{{\partial }^{2}}\psi }{\partial {{\theta }^{2}}}-{{c}_{47}}\cot \theta \frac{\partial \psi }{\partial \theta }-{{c}_{47}}\frac{1}{{{\sin }^{2}}\theta }\frac{{{\partial }^{2}}\psi }{\partial {{\varphi }^{2}}} \right)+\left( {{c}_{44}}+{{c}_{47}} \right)\left( \frac{\cot \theta }{\sin \theta }\frac{\partial G}{\partial \varphi }-\frac{1}{\sin \theta }\frac{{{\partial }^{2}}G}{\partial \varphi \partial \theta } \right), \notag \\ 
& {{\Sigma }_{\varphi \theta }}=\left( {{c}_{47}}\frac{{{\partial }^{2}}\psi }{\partial {{\theta }^{2}}}-{{c}_{44}}\cot \theta \frac{\partial \psi }{\partial \theta }-{{c}_{44}}\frac{1}{{{\sin }^{2}}\theta }\frac{{{\partial }^{2}}\psi }{\partial {{\varphi }^{2}}} \right)+\left( {{c}_{44}}+{{c}_{47}} \right)\left( \frac{\cot \theta }{\sin \theta }\frac{\partial G}{\partial \varphi }-\frac{1}{\sin \theta }\frac{{{\partial }^{2}}G}{\partial \varphi \partial \theta } \right), \notag \\ 
& {{\Delta }_{\theta }}=\frac{\partial }{\partial \theta }\left( {{e}_{26}}w-{{e}_{26}}{{\nabla }_{2}}G+{{e}_{26}}G-{{\varepsilon }_{22}}\Phi  \right)-\frac{1}{\sin \theta }\frac{\partial }{\partial \varphi }\left( {{e}_{26}}{{\nabla }_{2}}\psi -{{e}_{26}}\psi  \right), \notag \\ 
& {{\Delta }_{\varphi }}=\frac{\partial }{\partial \theta }\left( {{e}_{26}}{{\nabla }_{2}}\psi -{{e}_{26}}\psi  \right)+\frac{1}{\sin \theta }\frac{\partial }{\partial \varphi }\left( {{e}_{26}}w-{{e}_{26}}{{\nabla }_{2}}G+{{e}_{26}}G-{{\varepsilon }_{22}}\Phi  \right).  
\end{align}

Thus, substituting Eqs.~\eqref{B6}$_{3}$ and \eqref{B7}$_{3,4}$ into \eqref{50}$_{1}$ and employing Eqs.~\eqref{B2}, \eqref{B3} and \eqref{B5}$_{2}$ provides
\begin{equation} \label{B8}
\begin{split}
{{\nabla }_{2}}{{\Sigma }_{rr}}&=\frac{{{c}_{69}}}{{{c}_{66}}}\nabla _{1}^{2}{{\Sigma }_{2}}+{{\Sigma }_{rr}}+\left( {{q}_{1}}+{{q}_{2}}+2{{q}_{3}} \right)\nabla _{1}^{2}G\\
&+\left[ {{q}_{1}}\nabla _{1}^{2}-2\left( {{q}_{2}}+2{{q}_{3}} \right)+\rho {{r}^{2}}\frac{{{\partial }^{2}}}{\partial {{t}^{2}}} \right]w+2{{q}_{4}}{{\Delta }_{r}}+{{q}_{5}}\nabla _{1}^{2}\Phi,
\end{split}
\end{equation}
where the relation ${{c}_{22}}-{{c}_{23}}-{{c}_{47}}={{c}_{44}}$ has been used in deriving Eq.~\eqref{B8} and we have
\begin{equation} \label{B9}
\begin{split}
&{{q}_{1}}=\frac{c_{69}^{2}}{{{c}_{66}}}-{{c}_{99}},\quad {{q}_{2}}={{c}_{44}}+{{c}_{47}},\quad {{q}_{4}}=\frac{{{e}_{11}}-{{e}_{12}}}{{{\varepsilon }_{11}}},\\
&{{q}_{3}}=2{{c}_{12}}-{{c}_{11}}-{{c}_{22}}-\frac{{{\left( {{e}_{11}}-{{e}_{12}} \right)}^{2}}}{{{\varepsilon }_{11}}},\quad {{q}_{5}}=\left( \frac{{{c}_{69}}}{{{c}_{66}}}-1 \right){{e}_{26}}, 
\end{split}
\end{equation}
Similarly, we can substitute Eq.~\eqref{B7}$_{7,8}$ into \eqref{50}$_{4}$ and use Eq.~\eqref{B5}$_{2}$ to obtain
\begin{equation} \label{B10}
{{\nabla }_{2}}{{\Delta }_{r}}=\nabla _{1}^{2}\left[ {{q}_{5}}\left( w+G \right)+\frac{{{e}_{26}}}{{{c}_{66}}}{{\Sigma }_{2}}+{{q}_{6}}\Phi  \right]-{{\Delta }_{r}},
\end{equation}
where ${{q}_{6}}={e_{26}^{2}}/{{{c}_{66}}}+{{\varepsilon }_{22}}$.

Finally, inserting Eqs.~\eqref{52}$_{2,5}$ and \eqref{B7}$_{2,4-6}$ into Eq.~\eqref{50}$_3$ and through some lengthy mathematical manipulations, we obtain
\begin{equation} \label{B11}
\frac{\partial {{L}_{1}}}{\partial \theta }-\frac{1}{\sin \theta }\frac{\partial {{L}_{2}}}{\partial \varphi }=0,
\end{equation}
where
\begin{equation} \label{B12}
\begin{split}
{{L}_{1}}&={{\nabla }_{2}}{{\Sigma }_{1}}+{{\Sigma }_{1}}+{{c}_{69}}{{\nabla }_{2}}\psi +\left( {{c}_{44}}\nabla _{1}^{2}+{{c}_{44}}+{{c}_{47}}-{{c}_{99}} \right)\psi -\rho {{r}^{2}}\frac{{{\partial }^{2}}\psi }{\partial {{t}^{2}}}, \\ 
{{L}_{2}}&={{\nabla }_{2}}{{\Sigma }_{2}}+{{\Sigma }_{2}}-\left[ {{c}_{23}}+{{c}_{22}}-2{{c}_{12}}+{{c}_{99}}+\left( {{c}_{12}}-{{c}_{11}} \right){{\nabla }_{2}} \right]w-\left( {{e}_{12}}-{{e}_{11}} \right){{\nabla }_{2}}\Phi \\ 
&-{{e}_{26}}\Phi-{{\Sigma }_{rr}}+{{c}_{69}}{{\nabla }_{2}}G+\left[ \left( {{c}_{22}}-{{c}_{12}} \right)\nabla _{1}^{2}+{{c}_{44}}+{{c}_{47}}-{{c}_{99}} \right]G-\rho {{r}^{2}}\frac{{{\partial }^{2}}G}{\partial {{t}^{2}}}.
\end{split} 
\end{equation}
Note that the expressions $\nabla _{1}^{2}$ and ${{c}_{22}}-{{c}_{23}}-{{c}_{47}}={{c}_{44}}$ have been exploited in deriving Eq.~\eqref{B11}, along with the following relation:
\begin{align} \label{B13}
\frac{\partial }{\partial \theta }\nabla _{1}^{2}=\frac{{{\partial }^{3}}}{\partial {{\theta }^{3}}}+\frac{1}{{{\sin }^{2}}\theta }\frac{{{\partial }^{3}}}{\partial {{\varphi }^{2}}\partial \theta }-\frac{2\cos \theta }{{{\sin }^{3}}\theta }\frac{{{\partial }^{2}}}{\partial {{\varphi }^{2}}}+\cot \theta \frac{{{\partial }^{2}}}{\partial {{\theta }^{2}}}-\frac{1}{{{\sin }^{2}}\theta }\frac{\partial }{\partial \theta }.
\end{align}
In the same way, applying Eqs.~\eqref{52}$_{1,4}$ and \eqref{B7}$_{1-3,6}$ into Eq.~\eqref{50}$_2$ and carrying out some lengthy mathematical derivations leads to
\begin{equation} \label{B14}
-\frac{1}{\sin \theta }\frac{\partial {{L}_{1}}}{\partial \varphi }-\frac{\partial {{L}_{2}}}{\partial \theta }=0,
\end{equation}
which, combined with Eq.~\eqref{B11}, yields \citep{ding2006elasticity}
\begin{equation} \label{B15}
{{L}_{1}}=0,\quad {{L}_{2}}=0,
\end{equation}
Hence, from Eqs.~\eqref{B12} and \eqref{B15}, we get
\begin{equation} \label{B16}
{{\nabla }_{2}}{{\Sigma }_{1}}=\left( \rho {{r}^{2}}\frac{{{\partial }^{2}}}{\partial {{t}^{2}}}-{{c}_{44}}\nabla _{1}^{2}-{{q}_{1}}-{{q}_{2}} \right)\psi -{{q}_{7}}{{\Sigma }_{1}},
\end{equation}
and
\begin{equation} \label{B17}
\begin{split}
{{\nabla }_{2}}{{\Sigma }_{2}}&={{\Sigma }_{rr}}-\left( {{q}_{1}}+{{q}_{2}}+2{{q}_{3}} \right)w+\left( {{q}_{3}}\nabla _{1}^{2}+\rho {{r}^{2}}\frac{{{\partial }^{2}}}{\partial {{t}^{2}}}-{{q}_{1}}-{{q}_{2}} \right)G\\
&+{{q}_{4}}{{\Delta }_{r}}-{{q}_{5}}\Phi -{{q}_{7}}{{\Sigma }_{2}}. 
\end{split}
\end{equation}

For the incremental state vector $\mathbf{Y_1}={{\left[ {{\Sigma }_{1}},\psi  \right]}^{\text{T}}}$ defined in Eq.~\eqref{54}$_1$, Eqs.~\eqref{B5}$_1$ and \eqref{B16} can be rewritten into the form of incremental state equation \eqref{53} with $k=1$, while for another incremental state vector $\mathbf{Y_2}={{\left[ {{\Sigma }_{rr}},{{\Sigma }_{2}},G,w,{{\Delta }_{r}},\Phi  \right]}^{\text{T}}}$ given in Eq.~\eqref{54}$_2$, we can rewrite Eqs.~\eqref{B2}, \eqref{B3}, \eqref{B5}$_2$, \eqref{B8}, \eqref{B10} and \eqref{B17} into Eq.~\eqref{53} with $k=2$.



\section*{References}

\bibliographystyle{elsarticle-harv.bst}
\nocite{*}
\bibliography{Soft_Dielectric_Spherical_Shell.bib}







\end{document}